\documentclass[authoryear,sort&compress,final,3p,times,english]{elsarticle}
\usepackage{lineno}
\usepackage{amsmath}
\usepackage{hyperref}
\usepackage{cleveref}
\usepackage{amsthm}
\usepackage{amsfonts}
\usepackage{natbib}
\usepackage{bibunits} 
\usepackage[bf]{caption}
\usepackage{caption}
\usepackage[T1]{fontenc}
\usepackage[utf8]{inputenc}
\usepackage{babel}
\usepackage[export]{adjustbox}
\usepackage{a4,graphicx}
\usepackage{subfig}
\usepackage{float}
\usepackage{array}
\usepackage{setspace}
\usepackage{geometry}
\usepackage{array}
\usepackage{tikz}
\usepackage{longtable}
\usepackage{xcolor}
\usepackage{color}
\usepackage{tabularx}
\usepackage{cuted}
\usepackage{scrextend}
\usepackage{hyperref}
\usepackage{graphicx}
\usepackage{xtab}
\usepackage{footnote}
\usetikzlibrary{shapes,shapes.geometric,arrows,fit,calc,positioning,automata}
\usetikzlibrary{automata,positioning}
\usepackage{varwidth}
\usepackage{mathrsfs}
\usepackage{epstopdf}
\usepackage[toc,page]{appendix}

\usepackage{algorithm}
\usepackage{algorithmicx}
\usepackage{multicol}
\usepackage{arydshln}
\usepackage{caption}
\usepackage{mathtools}
\usepackage{algpseudocode}
\usepackage{hhline}
\usepackage[inline]{enumitem}  
\usepackage{pgfplots}
\hypersetup{colorlinks=true,linkcolor=green,filecolor=blue,urlcolor=blue}
\makeatletter
\def\ps@pprintTitle{%
\let\@oddhead\@empty
\let\@evenhead\@empty
\def\@oddfoot{\reset@font\hfil\thepage\hfil}%
\let\@evenfoot\@oddfoot}

\def\R{\mathbb R}

\def\R{\mathbb R}
\def\1{\mathbf 1}
\def\x{\mathbf x}
\def\y{\mathbf y}

\DeclarePairedDelimiter\abs{\lvert}{\rvert}

\DeclareFontEncoding{LS1}{}{}
\DeclareFontEncoding{LS2}{}{\noaccents@}
\DeclareFontSubstitution{LS1}{stix}{m}{n}
\DeclareFontSubstitution{LS2}{stix}{m}{n}
\DeclareMathAlphabet\mathscr{LS1}{stixscr}{m}{n}
\SetMathAlphabet\mathscr{bold}{LS1}{stixscr}{b}{n}
\DeclareMathAlphabet\mathcal{LS2}{stixcal}{m}{n}
\SetMathAlphabet\mathcal{bold}{LS2}{stixcal}{b}{n}
\DeclareMathOperator*{\argmin}{arg\,min}
\DeclareMathOperator*{\argmax}{arg\,max}
\DeclareFontFamily{U}{BOONDOX-calo}{\skewchar\font=45 }
\DeclareFontShape{U}{BOONDOX-calo}{m}{n}{ <-> s*[1.05] BOONDOX-r-calo}{}
\DeclareFontShape{U}{BOONDOX-calo}{b}{n}{ <-> s*[1.05] BOONDOX-b-calo}{}
\DeclareMathAlphabet{\mathcalboondox}{U}{BOONDOX-calo}{m}{n}
\SetMathAlphabet{\mathcalboondox}{bold}{U}{BOONDOX-calo}{b}{n}
\DeclareMathAlphabet{\mathbcalboondox}{U}{BOONDOX-calo}{b}{n}
\DeclareMathOperator{\tr}{tr}

\newcounter{algsubstate}
\renewcommand{\thealgsubstate}{\alph{algsubstate}}

\algrenewcommand\algorithmicforall{\textbf{foreach}}
\algrenewcommand\algorithmicindent{.8em}
\makesavenoteenv{tabular}

\newcommand{\comm}[1]{}

\makeatletter
\newcommand{\inlineitem}[1][]{
\ifnum\enit@type=\tw@
{\descriptionlabel{#1}}
\hspace{\labelsep}
\else
\ifnum\enit@type=\z@
\refstepcounter{\@listctr}\fi
\quad\@itemlabel\hspace{\labelsep}
\fi}

\makeatother
\begin{document}
\begin{frontmatter}
\title{\textcolor{blue}{Spatial point process via regularisation  modelling of ambulance call risk}}
\author[1]{Fekadu L. Bayisa}
\author[1]{Markus \AA dahl}
\author[1]{Patrik Ryd\'{e}n}  
\author[1,2]{Ottmar Cronie$^{*}$}
\address[1]{Department of Mathematics and Mathematical Statistics, Ume{\aa} University, 901 87, Ume{\aa}, Sweden}
\address[2]{Biostatistics, School of Public Health and Community Medicine, Institute of Medicine, University of Gothenburg, Sweden}
\cortext[author] {Corresponding author.\\\textit{E-mail address:} ottmar.cronie@gu.se}	

\begin{abstract}
This study investigates the spatial distribution of ambulance/emergency alarm call events in order to identify spatial covariates associated with the events and discern hotspot regions of the events. The study, which focuses on the Swedish municipality of Skellefte{\aa}, is motivated by the problem of developing optimal dispatching strategies for prehospital resources such as ambulances. The dataset at hand is a large-scale multivariate spatial point pattern of call events stretching between the years 2014--2018. For each event, we have recordings of the spatial location of the call as well as marks containing the associated priority level, given  by 1 (highest priority) or 2, and sex labels, given by female or male. To achieve our goals, we begin by modeling the spatially varying call occurrence risk as an intensity function of a (multivariate) inhomogeneous spatial Poisson process that we assume is a log-linear function of some underlying spatial covariates. The spatial covariates used in this study are related to road network coverage, population density, and the socio-economic status of the population in Skellefte{\aa}. Since mobility is clearly a factor that has a large impact on where people are in need of an ambulance, and since none of our spatial covariates quantify human mobility patterns, we here take a pragmatic approach where, in addition to other spatial covariates, we include a non-parametric intensity estimate of the events as a covariate in the intensity function.  A new heuristic algorithm has been developed to select an optimal estimate of the kernel bandwidth in order to obtain the non-parametric intensity estimate of the events and to generate other covariates. Since we consider a large number of spatial covariates as well as their products (the first-order interaction terms), and since some of them may be strongly correlated, lasso-like elastic-net regularisation has been used in the log-likelihood intensity modeling to perform variable selection and reduce variance inflation from overfitting and bias from underfitting. As a result of the variable selection, the fitted model structure contains individual covariates of both road network and demographic types. We discovered that hotspot regions of calls have been observed along dense parts of the road network in Skellefte{\aa}. Furthermore, a mean absolute error evaluation of the proposed model to generate the intensity of emergency alarm/ambulance call events indicates that the estimated model is stable and can be used to generate a reliable intensity estimate over the region, which can be used as an input in the problem of designing prehospital resource dispatching strategies.
\end{abstract}

\end{frontmatter}

\section{Introduction}
In many (if not most) countries/regions, the prehospital resources available, e.g.~ambulances, are scarce, and this clearly affects the availability and general quality of the prehospital service in question. For example, a small decrease in the median response time, which could be the effect of the addition of an ambulance, could save a significant number of lives annually. However, simply adding additional ambulances to a given region might not necessarily yield the expected outcome, i.e.~drastically decreased response times, but instead the main challenge might be one of resource optimisation. More specifically, it may very well be the case that the existing resources are not utilised optimally; note that, given an expected performance outcome in terms of the response time distribution, one might have to both add and optimise prehospital resources. Moreover, in most/all cases, the addition of a new ambulance comes at a relatively high cost, so simply adding ambulances until the expected outcome has been achieved is essentially never a viable option.  Such optimisation can be quite challenging and extensive; in 2018, in Sweden, approximately 660 ambulances responded to roughly 1.2 million {\em ambulance/emergency alarm calls}, i.e.~events defined by the deployment of an ambulance. In addition, the deployment of ambulances in 2018 cost more than 4 billion Swedish krona. \\

Carrying out resource optimisation within a prehospital care organisation/system, which to a large degree boils down to placing ambulance stations at locations where, e.g.~the median response time is minimised annually, requires knowledge/understanding of where (and when) calls tend to occur spatially (and temporally).  In essence, minimising the response times of ambulances is crucial for obtaining the desired clinical outcomes following ambulance calls  \citep{blackwell2002response,pell2001effect,o2011role}.  The call occurrence risk can be influenced by various underlying factors, such as demographic factors. More specifically, within the spatial context, if one is able to exploit different covariates/predictors to model the expected number of calls occurring within a given region, within an arbitrary period of time, then one can obtain an understanding of which spatial population characteristics (e.g.~age distribution in a given subregion) that are driving the risk of an ambulance call occurring at a given spatial location. In addition, such a model may be extended to be exploited for predictive purposes, most notably within a spatio-temporal context.  One can easily imagine that the distribution of calls in Sweden is both complex, dynamical, and heterogeneously distributed within the spatial study region. \\

This work is aimed at describing the spatial dynamics of ambulance calls in the municipality of Skellefte{\aa}, Sweden. The main focus of our study is to generate a risk-map for the ambulance calls, which is modelled by means of different spatial covariates, and to identify hotspot regions in Skellefte{\aa}, which can play a key role in designing optimal dispatching strategies for prehospital resources. It also focuses on identifying spatial covariates that tend to influence the occurrence of call events. Hence, we aim for a model with good predictive performance. \\

Note that, a priori, we do not know how many calls there will be within the spatial region studied and within the timeframe considered, which here is given by the years 2014--2018. In addition, we have access to the exact (GPS) locations of the events, as well as additional information, so called marks, attached to each event. Such datasets are commonly referred to as {\em (marked) point patterns}, and their natural modelling framework is that of {\em (marked) spatial point processes} \citep{baddeley2015spatial, moller2003statistical, diggle2013statistical,van2000markov}, which may be thought of as generalised random samples with the properties that the number of points/elements in the sample is allowed to be random and the points are allowed to be dependent. When the marks are discrete, one commonly speaks of a {\em multivariate/multitype} point process.  Here, in our specific dataset, each event has two mark components attached to each spatial location: the priority/risk level (1 or 2) of the call and the gender (male or female) of the patient associated with the call.\\

The properties of a univariate point process are most commonly characterised through its intensity function, which essentially reflects the probability of the point process having a point at a given location in the study region. Formally, it is defined as the density function of the first-order expectation measure over the spatial domain. In most applications, it is unrealistic to assume that the underlying point process is homogeneous, i.e.~the intensity function is constant, and there are various approaches available to deal with the spatial inhomogeneity of the points/events \citep{baddeley2015spatial,diggle2013statistical}.   Quite often, one has access to a collection of spatial covariates, which may be used to model the spatially varying intensity function. In particular, it is commonly assumed that the intensity has a log-linear form, i.e.~the log-intensity is a linear combination of the covariates. Here, the common practice is to first model the intensity function using a Poisson process log-likelihood function, which is a closed-form function of only the (assumed) intensity function.  In the case of a general point process,  this is commonly  referred to as a composite likelihood estimation and,  although a Poisson process is a point process with independent points/events and Poisson distributed counts within any subregion, the indicated intensity estimation approach still has good large sample properties for general point processes \citep{coeurjolly2019understanding}.  Once an intensity function estimate is obtained, one would then proceed with analysing and modelling possible spatial interaction, i.e.~dependence between the events \citep{baddeley2000non, baddeley2015spatial, van2011aj, cronie2016summary}.  It should here be emphasised that observed clumps of points in a point pattern may be the result of either inhomogeneity, spatial interaction (clustering/aggregation/attraction) or both.  Since our main interest here is generating a spatial risk map for the observed events, i.e.~the ambulance calls, which can be exploited for predictive purposes and also used as input in the problem of designing an optimal  dispatching strategy of ambulances, we will solely focus on the former, i.e. modelling the intensity function as a log-linear function of a collection of spatial covariates.  This allows us to address one of the main objectives of this work, which is to identify hotspot regions of the events. A further main objective of this work is to select covariates governing the intensity function. Here the list of covariates includes population density, shortest distance to road networks, line density of road networks, line density of densely populated regions, bench mark estimated intensity, proportion of population by age category, proportion of population by gender, proportion of population (aged 20 + years)  by income status,  proportion of household (aged 20 + years) by economic standard, proportion of population (aged 25-64 years) by education level, proportion of population by Swedish or foreign background, and proportion of population (aged 20-64 years) by employment status.  Aside from including the individual covariates in our model framework, we also include the cross-terms given by the products of the individual covariates, since there may potentially be interactive effects of the covariates. This results in a high-dimensional data setting (a total of nine hundred eighty-nine covariates), with possibly hard to interpret cross-terms, and it is clearly a challenge to identify which covariates sensibly explain the actual intensity of the events, i.e.~which covariates should actually be included in the final model. As a solution, one may apply regularisation when fitting the model, which, aside from carrying out variable selection, also reduces variance inflation from overfitting and bias from underfitting. In addition, we want to adjust for the fact that the demographic covariates we have access to only reflect where different demographic groups live, and not how they move around. Since we do not have access to any human mobility covariates, we have to be pragmatic and use some proxy for such covariates. Our solution is to additionally include a non-parametric intensity estimate of the spatial locations of the events as an additional covariate. In other words, these added covariates would (to some degree) represent the portion of the intensity function that the original spatial covariates could not explain.  Here we have another benefit of the regularisation: having accounted for the other covariates, if this added covariate has little/no influence on the intensity function, then the regularisation would indicate this. \\
  
Since many of the spatial covariates which we deal with are most likely strongly correlated, variable selection in the modelling of the spatially varying intensity function is most likely necessary to overcome the use of correlated covariates in the modelling of  ambulance call events. \citet{tibshirani1996regression} introduced a penalised likelihood procedure, which has been a cornerstone in the development of variable selection methods via regularisation (or penalisation). The idea of \citet{tibshirani1996regression} is to add a least absolute shrinkage and selection operator (lasso) penalty to the loss function, often the likelihood function, to shrink small coefficients of covariates to zero while retaining covariates with large coefficients in the model, thus leaving us with a sparse model with highly influential covariates. Hence, the approach simultaneously performs variable selection and parameter estimation. A plethora of regularisation methods, such as elastic-net \citep{zou2005regularization} and adaptive lasso \citep{zou2006adaptive}, have been developed subsequent to the work of \citet{tibshirani1996regression}. A predecessor of the lasso penalty, the ridge penalty, which is an $\ell_2$-penalty, may be combined linearly with the lasso $\ell_1$-penalty to obtain elastic-net regularisation, which is used to select variables and shrinks the coefficients of correlated variables to each other. Moreover, in adaptive lasso regularisation, adaptive weights are used in the penalisation of the coefficients of the variables. It should be emphasised that these shrinkage/regularisation methods have the effect of balancing estimation bias and variance, which is an additional motivation for their employment. Turning to the point process context, also here variable selection has been developed to reduce variance inflation from overfitting and bias from underfitting. Considering Poisson process likelihood estimation together with a lasso penalty, \citet{renner2013equivalence} has introduced a maximum entropy approach for modelling the spatial distribution of a specie.  \citet{thurman2014variable} considered an adaptive lasso penalty to select variables in Poisson point process modelling. Regularisation methods such as lasso, adaptive lasso, and elastic-net  have also been considered in the context of a wide range of inhomogeneous spatial point processes models by \citet{yue2015variable}. \citet{choiruddin2018convex} further extended the above works to include a larger range of models and penalties. We here essentially follow the track of \citet{yue2015variable}. \\

With regard to the estimation of the regularised models, \citet{efron2004least} have developed the least angle regression to estimate the entire lasso regularisation paths. According to \cite{friedman2007pathwise} and \cite{friedman2010regularization}, in comparison with the least angle regression algorithm, the cyclical coordinate descent algorithm computes the regularisation paths of different regularisation methods with lower computational costs. In this study,  the cyclical coordinate descent method has been used to estimate the entire regularisation paths since it is computationally fast on large datasets \citep{fercoq2016optimization}. The general idea of the coordinate descent algorithm is that the objective function is optimised with respect to a single parameter at a time, and  the optimisation of the objective function is iteratively carried out for all parameters until a convergence criterion is fulfilled. In this work, the objective function represents a regularised quadratic approximation to the log-likelihood function of an inhomogeneous spatial Poisson process. \\

Two approaches have been used to evaluate the performance of the proposed model. The first approach involves training the proposed model on the whole dataset.   We treat the trained (or estimated)  intensity function on the whole dataset  as the true intensity function of the call events. Based on the estimated intensity function,  undersampling  i.e. 70\% of the whole dataset, has been used to generate a hundred datasets, which are  then used to generate a hundred intensity images. Using quantiles and mean absolute errors between pixel-by-pixel differences of the estimated image (i.e. image based on the whole dataset) and the hundred estimated intensity images (i.e. the intensity images based on the hundred undersampled datasets) can be used to evaluate the stability of the proposed model in estimating the intensity function of the emergency alarm call events. The second approach is to visually  evaluate the performance of the proposed model. We train the proposed model on 70\% and  on the remaining 30\% of the whole dataset.  Then, we compare the patterns of hotspot regions in the estimated intensity images obtained from the two datasets.  If the patterns of hotspot regions and the spatial locations in the respective datasets seemingly coincide with each other, then the proposed model is more likely applicable for the modelling of the call events. \\

In summary, the aim of this article is to explore the space-dynamics of ambulance calls and to identify spatial covariates associated with the call events. The result of this work will be used as input in designing optimal dispatching strategies for prehospital resources such as ambulances. \\

The structure of the article is as follows. Section \ref{data} and  \ref{Statisticalmethod} provide an overview of the data and statistical methods used in this work. Section  \ref{s:CreatedCovariates}  and \ref{Results} present the created spatial features for modelling emergency alarm call events and the results of the study. We evaluate the fitted model in section \ref{Evaluation} and discuss the implication of the results and provide a precise summary of the work in section \ref{Discn}.

\section{Data}\label{data}
Given a spatial/geographical region $W$, which we assume to be a subset of $\R^2$, by an {\em event} we will mean a location (GPS position) in $W$ to which an ambulance is dispatched during a given time period and the total collection of events will be referred to as a set of {\em ambulance/emergency call dataset}.  Here, the available information associated to an ambulance call dataset is the collection of locations  $\{x_1,\ldots,x_n\}\subseteq W$, $n\geq0$, as well a {\em mark} $m_i$ which is attached to each location $x_i$, $i=1,\ldots,n$. As we a priori do not know $n$, i.e.~how many calls there will be within $W$ during the period in question, such a dataset $\y=\{(x_1,m_1),\ldots,(x_n,m_n)\}$ is most naturally classified as a {\em marked point pattern} \citep{baddeley2015spatial}. Note that when the marks are discrete, we usually refer to the point pattern as {\em multivariate/multitype} rather than as {\em marked}. 

\subsection{Spatial ambulance call dataset}
Turning to the dataset at hand, $W$ will represent the Swedish municipality of Skellefte{\aa}, and the time period under consideration is given by the years 2014--2018. Our dataset consists of 14,919 events where the mark structure is such that for each event, and thereby the person associated with the ambulance call in question, there is a recording of the event's priority label (1 indicates the highest severity, implying turned on sirens, and 2 is a lower severity level) and a recording of the sex/gender of the person (female or male). Among the 14,919 events, 7,204 and 7,715 of them have priority labels 1 and 2, respectively.  Unfortunately, a missing data issue is present here: 5,236 and 5,238 of the events are recorded as male and female, respectively, and the remaining events do not have any sex recording for the person related to the event, so they are labelled as "missing".  Since our interest in the sex label lies mostly in highlighting possible structural differences related to the different covariates at hand, we have decided to proceed by studying the data as two separate marked point patterns: one where the marks are given by the priority labels and one where the marks are given by the sex labels.  It should be acknowledged that there are other ways of dealing with this issue. \\

\begin{figure}[!htpb]
\centering
\includegraphics[width=0.4\textwidth, height=0.35\linewidth]{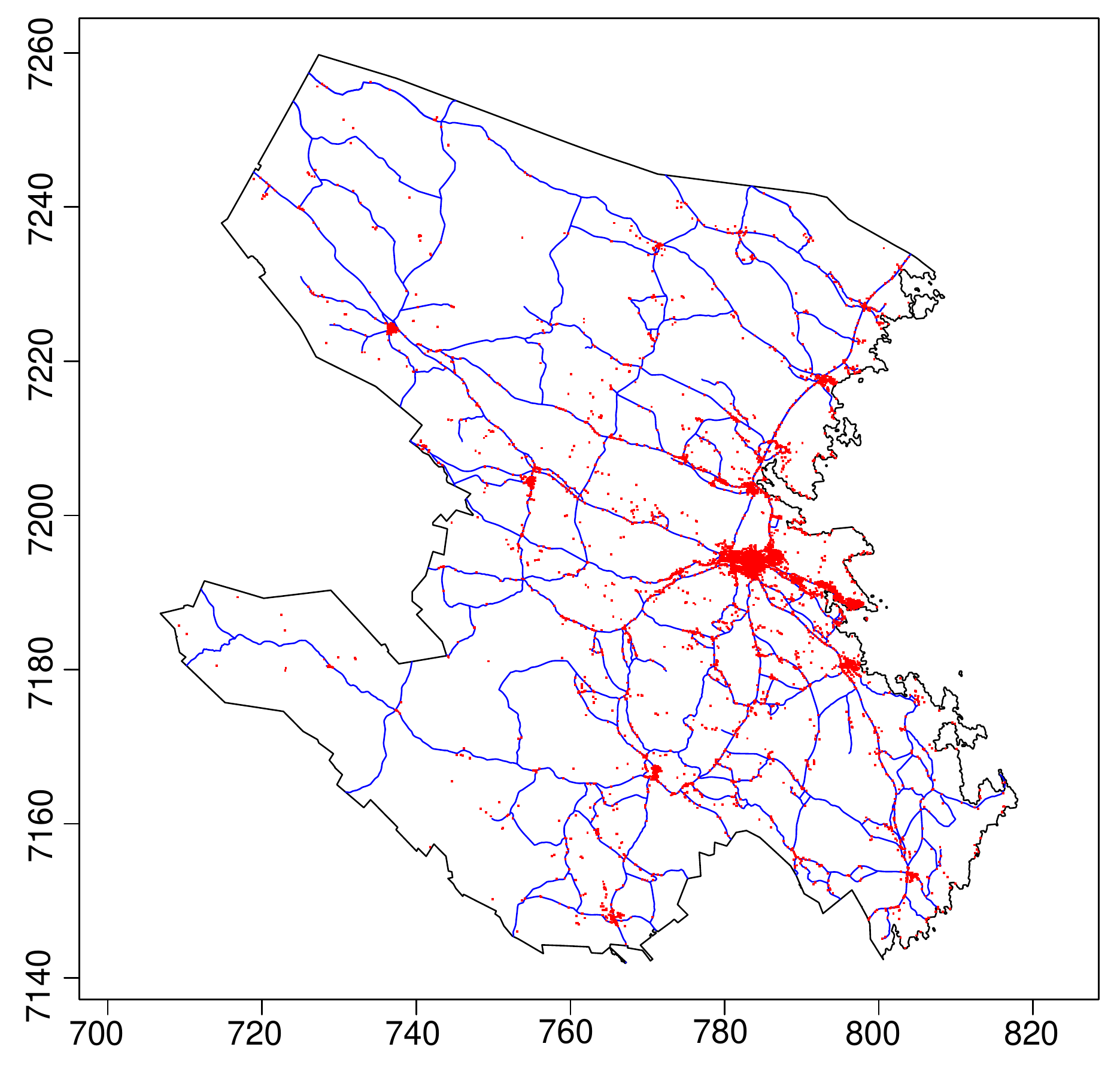}
\caption{Ambulance emergency alarm call locations and the main road network of Skellefte{\aa}.}
\label{Or213}
\end{figure}

Figure \ref{Or213} demonstrates the locations of all calls as well as the main road network of Skellefte{\aa}, Sweden. The figure highlights that the call locations are unevenly distributed over the study region and they tend to lie along the road network, which will make the statistical modelling quite challenging.  For data privacy reasons,  the axis labels of the figure have been scaled (i.e.~divided) by a factor of 1000. 

\subsection{Spatial covariates}
In order to properly model the ambulance call risk, we also need a range of spatial covariates.  To begin with,  a closer look at Figure \ref{Or213} and the relation between the road network and the call locations justifies the inclusion of road network related covariates. In addition, demographic spatial covariates should also play a role here, given that different demographic zones have different behavioural patterns.  The demographic spatial covariates considered have been supplied by Statistics Sweden (SCB), and the road network related covariates considered have been provided by the Swedish Transport Administration (Trafikverket). \\

Here, we will distinguish between two categories of covariates. The first is the collection of 'original' covariates, which are raw covariates retrieved from SOS alarm, Statistics Sweden (SCB), and Trafikverket. These covariates will, in turn, be used to generate a collection of new covariates, which we will refer to as 'created' covariates. In Section \ref{s:CreatedCovariates}, we outline the construction of the created covariates in detail.\\

All covariates considered have been based on the year 2018 and in their final forms they are given as functions of the form $z_i(s)$, $s\in W$.  Moreover, in addition to including each individual covariate $z_i$ in the analysis, we also include each interaction term $z_{ij}(s)=z_i(s)z_j(s)$, $s\in W$, which makes sense because we suspect that many of the covariates interact in one way or another.  We have a total of 43 individual spatial covariates and, all in all, we consider a total of  989 $(={43 \choose 2} + 2\times 43)$ covariates as candidates to be included in the modelling of the ambulance call risk. 

\subsubsection{Demographic spatial covariates}
Each demographic spatial covariate $z_i(s),$ $s\in W,$ which has been sampled on the 31st of December 2018, is piecewise constant and its value changes depending on which DeSO zone the location $s$ belongs to. The DeSO zones, which we use to define different demographic zones, partition Sweden into 5984 smaller spatial sub-regions, which do not overlap with the borders of any of the country's 290 municipalities and are encoded based on e.g.~how rural a DeSO zone is.  Below follows a short description of these covariates, where some have been graphically illustrated in Figure \ref{Covariates}:
\begin{itemize}
\item {\bf Population density:} Per DeSO, this gives us the ratio of the total population size of the DeSO to the area of the DeSO.
\item {\bf Population by age (counts):} The number of individuals of a given age category living in a given DeSO zone. There are a total of 17 such covariates, reflecting the following age categories (ages in years):  0-5, 6-9, 10-15, 16-19, 20-24, 25-29, 30-34, 35-39, 40-44, 45-49, 50-54, 55-59, 60-64, 65-69, 70-74, 75-79 and 80+.
\item {\bf Population by sex (counts):} The number of individuals of a given sex classification living in a given DeSO zone. There are 2 sex related covariates: women and men.
\item {\bf Population by Swedish/non-Swedish background (counts):} The number of individuals of a given Swedish/non-Swedish classification living in a given DeSO zone. There are 2 such covariates: Swedish background and non-Swedish background (the latter includes people born outside Sweden as well as people born in Sweden with both parents born outside Sweden).
\item {\bf Population 20-64 years of age by occupation (counts):} The number of individuals of a given occupational status living in a given DeSO zone. There are a total of 2 such covariates: (gainfully) employed and unemployed. 
\item {\bf Population 25-64 years of age by education level (counts):} The number of individuals of a given maximal education level category living in a given DeSO zone. There are a total of 4 such covariates: no secondary education, secondary education, post-secondary education of at most three years, and post-secondary education of more than three years (including doctoral degrees).
\item {\bf Population 20+ years of age by accumulated income (counts):} The number of individuals of a given income level category living in a given DeSO zone. There are a total of 2 such covariates:  below the national median income and above the national median income. 
\item {\bf Population 20+ years of age by economical standard (counts):} The number of individuals of a given economical standard category living in a given DeSO zone. There are a total of  2 such covariates: below the national median economical standard and above the national median economical standard. 
\end{itemize}
\subsubsection{Road network related covariates}
The road network data at hand have been broken down into sub-networks, which reflect the complete road network, the main road network, and the densely populated areas; a graphical illustration of these can be found in Figure \ref{BandWidthLineDensity}. Each network is included in the analysis with the aim of indicating a different level of call activity: The complete road network indicates the overall spatial region where calls tend to occur, the main road network indicates the parts of the complete network where there is a reasonable amount of activity, and the densely populated area indicates on which parts of the complete road network most people live. \\
 
We would further like to adjust for the fact that the covariates above mainly reflect where different people (of different demographic groups) live, and not how they move about. Unfortunately, we do not have access to explicit mobility covariates such as aggregated movement patterns in the population, and, as one may guess, people do not only need access to ambulances when they are at home. A partial solution to this, as we see it, is to consider an additional covariate, namely a spatial point pattern of bus stop locations. The idea is that bus stops reflect where there is a large amount of human mobility/activity. \\

It may be noted that the above original covariates are not functions of the form $z_i(s)$, $s\in W$, but rather line segment patterns and point patterns \citep{baddeley2015spatial}. Our approach to including them in the analysis is to let them give rise to a set of created covariates, which are described in detail in Section \ref{s:LineDensity}.

\subsubsection{The spatial domain}
We finally emphasis that the $x$- and $y$-coordinates of the spatial domain $W\subseteq\R^2$ will be included as covariates. These are intended to explain residual and explicit spatial variation, having adjusted for the full range of spatial covariates.

\section{Statistical methods}\label{Statisticalmethod}

As emphasised in Section \ref{data}, in particular through Figure \ref{Or213}, the ambulance calls are mainly located on/close to the underlying road network. One way to deal with the analysis is to project all events to a linear network representation of the road network and proceed with a linear network point pattern analysis \citep{baddeley2015spatial}. However, we here avoid such projections and instead choose to treat the spatial domain, and thereby the data point pattern, as Euclidean. As a result, we will instead introduce road network generated covariates which control that the fitted model generates events close to the underlying road network; details on the construction of such covariates can be found in Section \ref{s:LineDensity}.

\subsection{Point process preliminaries}\label{Ppp}

In application areas such as environmental science, epidemiology, ecology, etc., aside from the spatial locations of the events, additional information about the events may be available, which can be associated to the locations of the events. Such information pieces are referred to as marks, and by including them in the analysis, we can often obtain more realistic spatial point process models for the events -- note that, in contrast, a covariate reflects information which is known throughout the spatial domain before realisation of the events. For instance, as we saw above, it is both common and practically appropriate in emergency medical services to document the priority levels, the gender, the incident time, etc.~for a call/patient; recall that by spatial points of the events we mean the set of spatial locations of the patients to which ambulances have been dispatched or the set of spatial locations of patients from which calls have been made to the dispatcher/emergency alarm center.  \\

As previously indicated, marked point patterns are modelled by {\em marked point processes} \citep{baddeley2015spatial, moller2003statistical, diggle2013statistical,van2000markov}. Given a spatial domain $W\subseteq\R^2$ and a mark space\footnote{Formally, $\mathcal{M}$ is assumed to be a Polish space.} $\mathcal{M}$, a point process: 
$$
Y=\{(x_i,m_i)\}_{i=1}^{N}\subseteq W\times \mathcal{M}, 
\qquad
0\leq N \leq\infty,
$$ 
is a random subset such that, with probability 1, $Y\cap A\times B$ has finite cardinality $Y(A\times B)$ for any\footnote{Throughout, any set under consideration is a Borel set.} $A\times B\subseteq W\times \mathcal{M}$; this is referred to as local finiteness.  If we additionally have that $Y_g=\{x_i\}_{i=1}^{N}\subseteq W$ is a well-defined point process on $W$ (locally finite) in its own right, then we say that $Y$ is a marked point process. Note that each $(x_i,m_i)$ is a random variable and note in particular that if $W$ is bounded then $Y$ is automatically a marked point process. 

When the mark space is discrete, e.g.~$\mathcal{M}= \left\lbrace j: j = 1, 2, \ldots,k,\; k>1\right\rbrace $, we say that $Y$ is multitype and we note that we may split $Y$ into the marginal (purely spatial) point processes: 
$$
Y_j=\{x_i:(x_i,m_i)\in Y, m_i=j\}, 
\qquad
j=1, 2, \ldots,k.
$$ 
This collection may formally be represented by the vector $(Y_1, Y_2, \ldots,Y_k)$, which is referred to be a {\em multivariate} point process, and one commonly uses the two notions interchangeably. 

\subsection{Spatial intensity functions}
Our main interest here is to create a set of "heat maps" which describe the risk of a call occurring at a given spatial location $x\in W$, given an associated mark. This is accomplished by assuming that our data are generated by a multivariate point process $Y\subseteq W\times\{1, 2, \ldots,k\}$, $k>1$ and then modelling the {\em spatial intensity function} of each of the component $Y_j, j = 1, 2, \ldots, k.$ Formally, the spatial intensity function of $Y$ is defined as the function $\rho_Y$ satisfying\footnote{We assume that $\rho$ is the Radon-Nikodym derivative of the Borel measure $(A\times B)\mapsto E[Y(A\times B)]$ w.r.t.~the product measure given by the product of a Lebesgue measure and a counting measure.}
$$
E[Y(A\times B)]= \sum_{j=1}^k \1\{j\in B\}\int_{A} \rho_Y(x,j)dx, 
\quad 
A\times B\subseteq W\times \{1, 2, \ldots,k\},
$$
where $\1\{\cdot\}$ denotes the indicator function. Heuristically, $\rho_Y(x,j)dx$ may be interpreted as the probability that $Y$ has an event with mark $j$ in an infinitesimal neighbourhood of $x$ with size $dx$.  Since $E[Y(A\times\{j\})]=E[Y_j(A)]$, it follows that the spatial intensity function of $Y_j$ satisfies 
$$
\rho_j(x)=\rho_Y(x,j), 
\quad
x\in W, j=1, 2, \ldots,k. 
$$
Hence, by modelling the marginal processes $\rho_j$ separately, we obtain a model for $\rho_Y$. Letting $X$ denote an arbitrary $Y_j$, $j=1, 2, \ldots,k$, we see that its spatial intensity function $\rho$ satisfies
$$
E[X(A)] = \int_{A} \rho(x)dx, 
\quad 
A\subseteq W,
$$
or equivalently,
$
\rho(x) = \lim_{\abs{dx}\to 0} E\left[X\left(dx\right)\right]/\abs{dx}. 
$ 
If a point process has constant intensity function then we say that it is homogeneous, otherwise we refer to it as inhomogeneous. In a similar fashion, we may define higher-order intensity functions $\rho_n(x_1, x_2, \ldots,x_n)$, $x_1, x_2\ldots,x_n\in W$, $n\geq2$ as $\rho_n(x_1, x_2, \ldots,x_n)=E[X(dx_1)X(dx_2)\cdots X(dx_n)]=P(X(dx_1)=1, X(dx_2)=1, \ldots,X(dx_n)=1)$ for disjoint infinitesimal neighbourhoods $dx_1, dx_2, \ldots,dx_n\subseteq W$ of $x_1, x_1, \ldots,x_n$. If $x_i=x_j$ for any $i\neq j$, then $\rho_n(x_1, x_2, \ldots,x_n)=0$.

\subsection{Poisson processes}
Poisson processes, for which $\rho_n(x_1, x_2, \ldots,x_n)=\rho(x_1)\rho(x_2)\cdots\rho(x_n)$, $n\geq1$ and $X(A)$ is Poisson distributed with mean $\int_A\rho(x)dx$ for any $A\subseteq W$, are used as baseline models for the case of complete spatial randomness, i.e.,~ the case where there is no spatial interaction/dependence present. Poisson processes are completely governed by their spatially varying intensity functions and may be viewed as generalisations of random samples to the case where the size of the sample is random.

\subsection{Parametric spatial intensity function modelling}
We have observed a high degree of inhomogeneity in the ambulance call locations and we believe that a large portion of this inhomogeneity can be attributed to (some of) the spatial covariates which we have access to. Hence, as a starting point, for a given mark, we will consider parametric modelling of the call locations.\\

Consider a family of spatial intensity functions $\rho_{\boldsymbol\theta}(x)$, $x\in W$, which depends on spatial covariates through a parameter vector $\boldsymbol\theta\in\mathbb{R}^{K+1}$.  A common and convenient approach when modelling the intensity function of an arbitrary point process is to proceed as if we are considering a Poisson process, which is commonly referred to as composite likelihood estimation and is often motivated by good large sample properties \citep{coeurjolly2019understanding}.  Accordingly, suppose that a point pattern $\x = \left\lbrace x_{1}, x_{2}, \cdots, x_{n} \right\rbrace \subseteq W$ represents a realisation of a spatial point process $X$ which is observed within a bounded study region $W$. If $X$ is an inhomogeneous Poisson point process, then the associated log-likelihood function is given by
\begin{eqnarray}
 \label{ogeetti3}  
\log \mathcal{L}\left(\boldsymbol\theta\mid \mathbf{x} \right) =  \sum_{i=1}^{n}\log\rho_{\boldsymbol\theta}\left(x_{i}\right) - \int_{W}\rho_{\boldsymbol\theta}\left(x\right)dx .
\end{eqnarray}
Using any quadrature rule, the integral in \eqref{ogeetti3} can be approximated by a finite sum 
\begin{eqnarray}\label{Ogeettikoo}
	\int_{W}\rho_{\boldsymbol\theta}\left(x\right)dx
	\approx 
	\sum_{j=1}^{m} \rho_{\boldsymbol\theta}(s_{j})w_{j},
\end{eqnarray}
where  the positive numbers $w_{j}$,  $j = 1, 2, \ldots, m$, are quadrature weights summing to the area $\abs{W}$ and $s_{j}\in W$, $j = 1, 2, \ldots, m,$ are quadrature points.   Following this approximation, the approximated log-likelihood function may be expressed as
\begin{eqnarray}\label{logl}
	\log \mathcal{L}\left(\boldsymbol\theta\mid \mathbf{x} \right) 
	\approx 
	\sum_{i=1}^{n}\log\rho_{\boldsymbol\theta}\left(x_{i}\right) - \sum_{j=1}^{m} \rho_{\boldsymbol\theta}(s_{j})w_{j},
\end{eqnarray}
where $\textbf{s} = \{ s_1,\ldots, s_m\} \subseteq W$ represents the union of the observed spatial locations of events $\mathbf{x} = \{ x_1,\ldots, x_n\}$ and the set of dummy points $\textbf{s}\setminus\mathbf{x} = \{v_1,\ldots,v_q\}$, $q = m-n$. Here we assume that $m$ is much larger than $n$ for a better approximation of the log-likelihood function $\log \mathcal{L}\left(\boldsymbol\theta\mid \mathbf{x} \right)$. The approximated log-likelihood function in expression \eqref{logl} can then be rewritten as follows: 
\begin{eqnarray}
\label{Ogeettikoo2}
	\log \mathcal{L}\left(\boldsymbol\theta\mid \mathbf{x} \right) 
	\approx  
	\sum_{j=1}^{m}\left( y_{j}\log\rho_{\boldsymbol\theta}(s_{j}) -  \rho_{\boldsymbol\theta}(s_{j})\right) w_{j}, 
\end{eqnarray}
where  
\begin{eqnarray}
\label{indicator}
	y_{j} = \frac{a_{j}}{ w_{j}} 
	\quad 
	\text{and}
	\quad
	a_{j} = \1\{s_{j}\in \mathbf{x}\}.
\end{eqnarray}
Note that $a_j=0$ means that $s_{j}$ is a dummy point. Exploiting the approximation in equation \eqref{Ogeettikoo}, a large number of dummy points are required to obtain accurate parameter estimation using equation \eqref{Ogeettikoo2}. \cite{waagepetersen2008estimating} proposed two ways of obtaining dummy points and  quadrature weights. With regard to the quadrature weights, the first method is a grid approach in which the observation window $W$ is partitioned into a collection of rectangular tiles. The quadrature weight for a quadrature point $s\in \textbf{s}$ falling in a tile $R$ is the area of $R$ divided by the total number of quadrature points falling in $R$. This approach is advantageous since the computation of the quadrature weights is easy. The second approach is the Dirichlet approach \citep{okabe2009spatial} in which the quadrature weights are the areas of the tiles of the Dirichlet/Voronoi tessellation generated by the quadrature points in $\textbf{s}$. With regard to the dummy points, \cite{waagepetersen2008estimating} proposed two ways of generating dummy points. The first approach is to use stratified dummy points combined with grid-type weights while the second approach is to exploit binomial dummy points with the Dirichlet-type weights.  \\

According to \cite{baddeley2000practical} and \cite{thurman2015regularized}, a computationally cheaper approach to generate dummy points and compute quadrature weights is to partition the study region $W$ into tiles $R$ of equal area.  To generate dummy points, we place one dummy point exactly in each tile either systematically or randomly.  It follows that the quadrature weights for quadrature points $s_{j}$ can be set to $w_{j} = \Delta/E_{j}$,  where $\Delta$ is the area of each tile and $E_{j}$ is the number of events and dummy points in the same tile as point $s_{j}$. \\

Modelling spatial intensity functions parametrically, in particular modelling based on spatial covariates, we often assume that $\rho_{\boldsymbol\theta}$ has a log-linear form. More specifically, letting $\boldsymbol\beta = \left(\beta_{1}, \beta_{2}, \ldots, \beta_{K}\right)$ and $\boldsymbol{\theta} = \left(\beta_{0}, \boldsymbol\beta\right)$,  we assume that
\begin{eqnarray}\label{candidate}
\rho_{\boldsymbol\theta}(x) = \exp\left\lbrace \beta_{0} + \mathbf{z}(x)\boldsymbol\beta'\right\rbrace,
\end{eqnarray}
where  $\mathbf{z}(x)= (z_{1}(x),  z_{2}(x), \cdots, z_{K}(x))$ is a vector of spatial covariates at location $x\in W$. Combining equations \eqref{Ogeettikoo2} and \eqref{candidate}, we thus find that the log-likelihood function $\log \mathcal{L}\left(\boldsymbol\theta\mid \mathbf{x}\right)$ can be approximated by  
\begin{eqnarray}
\label{werqqq}
\log \mathcal{L}\left(\boldsymbol\theta\mid \mathbf{x}\right)  \approx  \sum_{j=1}^{m}\left(  y_{j}\left(\beta_{0} + \mathbf{z}(s_{j})\boldsymbol\beta'\right)  - \exp\left\lbrace \beta_{0} + \mathbf{z}(s_{j})\boldsymbol\beta'\right\rbrace\right) w_{j}. 
\end{eqnarray}
It follows that the expression on the right-hand side of the approximation sign in equation \eqref{werqqq} is a weighted log-likelihood function of independent Poisson random variables $Y_{j}, j = 1, 2, \cdots, m$. That is, for $j = 1, 2, \cdots, m$,  $y_{j}$ are the observations, $\rho_{\boldsymbol\theta}(s_{j}) = \exp\{\beta_{0} + \mathbf{z}(s_{j})\boldsymbol\beta'\}$  are the intensities of the Poisson distributions,  and $w_{j}$ are the weights.
Thus, the weighted log-likelihood function in equation \eqref{werqqq} can be maximised using standard software for fitting generalised linear models \citep{McCullagh1989generalised}. 

\subsection{Variable selection: Elastic-net regularisation}\label{s:Regularisation}

Incorporating regularisation into the log-likelihood function in equation \eqref{werqqq} can help to simultaneously select variables and estimate the parameters of the model. A penalised log-likelihood function   based on equation \eqref{werqqq} may be given by
\begin{eqnarray}
\label{WaaqayyoItalyEegi}
\mathcal{L}_{p}\left(\boldsymbol\theta\right) 
\approx
 \frac{1}{m}\log \mathcal{L}\left(\boldsymbol\theta\mid \mathbf{x}\right)  + \lambda R(\boldsymbol\beta), 
\end{eqnarray}
where $R(\boldsymbol\beta)$ is a regularisation method or penalty function, and $\lambda\ge 0$ is a tuning or smoothing parameter determining the strength of the penalty, or the amount of shrinkage. Several penalties such as garrote \citep{breiman1995better}, least absolute shrinkage and selection operator (lasso), elastic-net and fused lasso \citep{tibshirani2005sparsity}, group lasso \citep{yuan2006model}, Berhu penalty \citep{owen2007robust},  adaptive lasso \citep{zou2006adaptive}, and  LAD-lasso \citep{wang2007robust} have been developed for penalised regression modelling. The most common regularisation methods are lasso, ridge regression, elastic-net, and adaptive lasso. In short, lasso has a tendency to shrink several coefficients to zero, leaving only the most influential ones in the model, while ridge regression shrinks the coefficients of correlated covariates towards each other to borrow strength from each other \citep{friedman2010regularization}. The elastic-net penalty provides a mix between the ridge and the lasso penalty, and it is useful in cases where there are many correlated covariates or when the number of covariates exceeds the size of observations. The elastic-net regularisation penalty has the form
\begin{align}\label{Regu}
R(\boldsymbol\beta) = \sum_{k=1}^{K}\left\lbrace \frac{1}{2}\left(1-\alpha\right) \beta^{2}_{k} + \alpha\abs{\beta_{k}}\right\rbrace,
\end{align}
where the elastic-net parameter $\alpha \in [0, 1]$ turns \eqref{Regu} into a ridge penalty if $\alpha = 0$ and a lasso penalty if $\alpha = 1$. If $\alpha = 1-\epsilon$ for small $\epsilon>0$, then elastic-net performs like lasso but it avoids unstable behaviour due to extreme correlation \citep{yue2015variable}; 
empirical studies have indicated that elastic-net technique tends to outperform lasso on data with highly correlated features \citep{comber2018geographically, friedman2010regularization}. 
\cite{zou2006adaptive} proposed an adaptive lasso to address the shortcomings of lasso, such as biased estimates of large coefficients and conflict between optimal prediction and consistent variable selection.  According to \cite{kramer2009regularized}, however, the performance of  adaptive lasso is poor in the presence of highly correlated variables. 
Hence, our way forward here is to consider elastic-net penalisation and substituting equation \eqref{Regu} into equation \eqref{WaaqayyoItalyEegi}, the optimization problem of the elastic-net penalization of the log-likelihood function can be summarized as 
\begin{align}\label{waaqayyoo234}
\operatorname*{argmin}_{\boldsymbol\theta\in \mathbb{R}^{K+1} }
\mathcal{L}_{p}\left(\boldsymbol\theta\right) 
\approx 
\operatorname*{argmin}_{\boldsymbol\theta\in \mathbb{R}^{K+1} }
\left\lbrace
- \frac{1}{m}\log\mathcal{L}\left(\boldsymbol\theta\right) + \lambda \sum_{k=1}^{K}\left\lbrace \frac{1}{2}\left(1-\alpha\right) \beta^{2}_{k} + \alpha\abs{\beta_{k}}\right\rbrace\right\rbrace.  
\end{align}

\subsubsection{Optimisation methods}\label{Optimisation}
To deal with the optimisation problem in \eqref{waaqayyoo234}, we carry out the optimisation using a cyclical coordinate descent method, which optimises a target function/optimisation problem with respect to a single parameter at a time and iteratively cycles through all parameters until a convergence criterion is reached. Here, we present the coordinate descent algorithm for solving the regularized log-likelihood function with the elastic-net penalty. \\

Let $f\left(\boldsymbol\theta\right)$ be the approximated log-likelihood function \eqref{werqqq}, i.e.,
\begin{eqnarray*}
f(\boldsymbol\theta) 
= 
f(\beta_{0}, \boldsymbol\beta) 
= \sum_{j=1}^{m} w_{j}\left(y_{j}\left(\beta_{0} + \mathbf{z}(s_{j})\boldsymbol\beta'\right)  - \exp\left\lbrace \beta_{0} + \mathbf{z}(s_{j})\boldsymbol\beta'\right\rbrace\right).
\end{eqnarray*}
Let $r$ denote the step number in the optimisation algorithm, and $\boldsymbol\theta^{(r-1)} =  (\beta_{0}^{(r-1)},  \boldsymbol\beta^{(r-1)})$ represent  the current estimates of the parameters.
A quadratic approximation of $f\left(\boldsymbol\theta\right)$ at the point $\boldsymbol\theta^{(r-1)}$ is given by
\begin{eqnarray}\label{werqqqqweq}
f(\boldsymbol\theta)  
\approx 
f\left(\boldsymbol\theta^{(r-1)}\right) + \frac{d f\left(\boldsymbol\theta^{(r-1)}\right) }{d\boldsymbol\theta}\left( \boldsymbol\theta - \boldsymbol\theta^{(r-1)}\right)' + \frac{1}{2}\left( \boldsymbol\theta - \boldsymbol\theta^{(r-1)}\right)\frac{d^{2}f\left(\boldsymbol\theta^{(r-1)}\right) }{d\boldsymbol\theta d\boldsymbol\theta'}\left( \boldsymbol\theta - \boldsymbol\theta^{(r-1)}\right)',
\end{eqnarray}
where the first and the second-order derivatives of the function $f$ with respect to $\boldsymbol\theta$ are given by  
\begin{eqnarray}
\frac{d f(\boldsymbol\theta) }{d\boldsymbol\theta}
&=& 
\sum_{j=1}^{m} w_{j}\left(  y_{j}\mathbf{\bar{z}}(s_{j})  - \exp\left\lbrace \beta_{0} + \mathbf{z}(s_{j})\boldsymbol\beta'\right\rbrace\mathbf{\bar{z}}(s_{j})\right),
\label{werqqq2341}
\\
\frac{d^{2} f(\boldsymbol\theta) }{d\boldsymbol\theta d\boldsymbol\theta'}
&=& 
- \sum_{j=1}^{m} w_{j} \exp\left\lbrace \beta_{0} + \mathbf{z}(s_{j})\boldsymbol\beta'\right\rbrace\mathbf{\bar{z}}(s_{j})'\mathbf{\bar{z}}(s_{j}),
\label{werqqq2342}
\end{eqnarray}
and $\mathbf{\bar{z}}(s_{j}) = (1,  \mathbf{z}(s_{j}))$. 
Hence, a quadratic approximation of the approximated log-likelihood function in \eqref{werqqq} can be obtained through equations \eqref{werqqq2341},  \eqref{werqqq2342}, and \eqref{werqqqqweq}, i.e.,
\begin{eqnarray}
\label{addafacaane2}
\mathcal{L}_{q}\left(\boldsymbol\theta\right) = -\frac{1}{2} \sum_{j=1}^{m}u_{j} \left(y^{*}_{j} - \beta_{0} - \mathbf{z}(s_{j})\boldsymbol\beta'\right)^{2} + C\left(\boldsymbol\theta^{(r-1)}\right), 
\end{eqnarray}
where $C(\boldsymbol\theta^{(r-1)})$ is a constant function of $\boldsymbol\theta$ and the remaining variables in equation \eqref{addafacaane2} are given by 
\begin{eqnarray}\label{addafacaane3452}
y^{*}_{j} = \beta^{(r-1)}_{0} + \mathbf{z}(s_{j}){\boldsymbol\beta^{(r-1)}}'+ \frac{y_{j} }{\exp\left\lbrace \beta^{(r-1)}_{0} + \mathbf{z}(s_{j}){\boldsymbol\beta^{(r-1)}}'\right\rbrace} - 1 
\quad \text{and} \quad
u_{j} =  w_{j}\exp\left\lbrace \beta^{(r-1)}_{0} + \mathbf{z}(s_{j}){\boldsymbol\beta^{(r-1)}}'\right\rbrace.
\end{eqnarray}
As can be seen from equations \eqref{addafacaane2} and \eqref{addafacaane3452},  the variable $y^{*}_{j}$ is the working response variable while $u_{j}$ is the updated weight.  Replacing the log-likelihood function  $\log\mathcal{L}\left(\boldsymbol\theta\right)$ in equation \eqref{waaqayyoo234} by the quadratic approximation $\mathcal{L}_{q}\left(\boldsymbol\theta\right)$, 
the optimisation problem of the regularised quadratic approximation of the log-likelihood function becomes
\begin{eqnarray}
\label{addafacaane3}
\displaystyle\operatorname*{argmin}_{\boldsymbol\theta\in \mathbb{R}^{K+1} } \mathcal{L}_{qp}\left(\boldsymbol\theta\right)  = \displaystyle\operatorname*{argmin}_{\boldsymbol\theta\in \mathbb{R}^{K+1} } \left\lbrace -\frac{1}{m}\mathcal{L}_{q}\left(\boldsymbol\theta\right) + \lambda \displaystyle\sum_{k=1}^{K}\left\lbrace \frac{1}{2}\left(1-\alpha\right) \beta^{2}_{k} + \alpha\abs{\beta_{k}}\right\rbrace\right\rbrace.  
\end{eqnarray}
The optimisation problem in equation \eqref{addafacaane3} can be solved by the coordinate descent algorithm. 
More specifically, for any pre-specified value of the tuning parameter $\lambda$, in iteration $r = 1, 2, \ldots$, the coordinate descent algorithm partially optimises the optimisation problem with respect to $\beta_{k}$, given the estimates $\beta_{0}^{(r-1)}$ and $\beta_{h}^{(r-1)}$, $h\in\{1,\ldots,K\}\setminus\{k\}$. 
Explicitly, the optimisation can be described by 
\begin{eqnarray}\label{addafacaane4}
\displaystyle\operatorname*{argmin}_{\boldsymbol\theta\in \mathbb{R}^{K+1} } \mathcal{L}_{qp}\left(\boldsymbol\theta\right) \approx  \displaystyle\operatorname*{argmin}_{\beta_{k}\in \mathbb{R}} \mathcal{L}_{qp}\left(\beta^{(r-1)}_{0}, \beta^{(r-1)}_{1}, \cdots, \beta^{(r-1)}_{k-1}, \beta_{k},  \beta^{(r-1)}_{k+1}, \cdots, \beta^{(r-1)}_{K}\right).
\end{eqnarray}
According to \cite{friedman2007pathwise}, there are closed form coordinate-wise updates to estimate the parameters of the optimisation problem. Letting $\beta_{k}\geq 0$, the first-order derivative of $\mathcal{L}_{qp}\left(\boldsymbol\theta\right)$ in equation \eqref{addafacaane4} with respect to $\beta_{k}$ is given by  
\begin{eqnarray}\label{addafww2}
\frac{d\mathcal{L}_{qp}\left(\boldsymbol\theta\right)}{d\beta_{k}} = -\frac{1}{m} \sum_{j=1}^{m}u_{j}z_{k}(s_{j}) \left(y^{*}_{j} - \tilde{y}^{(k)}_{j}\right) +  \frac{1}{m}\sum_{j=1}^{m}u_{j}z^{2}_{k}(s_{j})\beta_{k} + \lambda(1-\alpha)\beta_{k} + \lambda\alpha, 
\end{eqnarray}
where $\tilde{y}^{(k)}_{j} = \beta^{(r-1)}_{0} + \sum_{h\ne k}^{K}\beta_{h}^{(r-1)}z_{h}(s_{j})$ is the fitted value excluding the covariate $z_{k}(s_{j})$. Similarly,  the first-order derivative of $\mathcal{L}_{qp}\left(\boldsymbol\theta\right)$ for the case $\beta_{k} < 0$ can easily be obtained.
It follows that the coordinate-wise updates for parameter estimation in the elastic-net penalisation can be obtained by  
\begin{eqnarray}\label{addafacaane5}
\beta^{(r)}_{k} = 
\frac{S\left(\frac{1}{m}\displaystyle\sum_{j=1}^{m} u_{j} z_{k}(s_{j})\left(y^{*}_{j}-\tilde{y}^{(k)}_{j} \right), \lambda\alpha\right) }{\frac{1}{m}\displaystyle\sum_{j=1}^{m} u_{j} z^{2}_{k}(s_{j})+\lambda\left(1-\alpha\right)}, \quad r = 1, 2, \ldots, \quad k = 1, 2, \ldots, K, 
\end{eqnarray}
where  $S(z, \vartheta)$ is the soft-thresholding operator given by
\begin{eqnarray*}  
S(z, \vartheta) = \text{sign}\left(z\right) \left(\abs{z}-\vartheta \right)_{+} = 
\begin{cases}
z-\vartheta, & \text{if } z > 0 \text{ and } \vartheta < \abs{z} ,\\
0, & \text{if } \vartheta \ge \abs{z},\\
z + \vartheta, & \text{if } z < 0 \text{ and }\vartheta < \abs{z}.
\end{cases}
\end{eqnarray*}
The intercept parameter need not be penalised as it has no role in the variable selection. The estimate of the intercept term can be obtained by
\begin{align*}
\beta^{(r)}_{0} = 
\frac{1}{\sum_{j=1}^{m} u_{j}}
\sum_{j=1}^{m} u_{j}\left(y^{*}_{j}-\mathbf{z}(s_{j}) \boldsymbol\beta'^{(r-1)}\right)
, \quad r = 1, 2, \ldots.
\end{align*}
The parameter estimates are updated until the algorithm converges. 
With regard to the tuning parameter $\lambda\in \left[\lambda_{max}, \lambda_{min}\right]$, we start with the smallest value $\lambda_{max}$ of the tuning parameter for which the entire vector is zero. That is, we begin with an $\lambda_{max}$ for which $\widehat{\boldsymbol\beta} = 0$ to obtain solutions for a decreasing sequence of $\lambda$ values. Using a prediction performance measure, e.g.~cross-validation, for the estimated model, the user can select a particular value of $\lambda$ from the candidate sequence of $\lambda$ values. Since the parameter estimation updating equation \eqref{addafacaane5} is obtained for elastic-net penalisation, we may set $\alpha = 0$ to implement ridge regression and $\alpha = 1$ to use the lasso approach; other elastic-net regularisation can be implemented by picking $\alpha\in(0, 1)$. Recall that elastic-net is useful when there are many correlated covariates in the statistical model and the data are high-dimensional, i.e.,~data with the property that $K\gg m$.  Cyclical coordinate descent methods are a natural approach to solving convex problems and each coordinate-descent step of the algorithm is fast with an explicit formula for each coordinate-wise optimisation. It also exploits the sparsity of the model and it has better computational speed both for high dimensional data and big data \citep{friedman2010regularization}. 
\begin{algorithm}[H]
\caption{\small Parameter estimation algorithm for regularized log-likelihood function of inhomogeneous Poisson point process.}
\label{alg:urgooftuuko} 
\begin{algorithmic}[1]	
\State Identify the spatial domain $W$,  
\State Generate a set of dummy points $\textbf{v} = \lbrace v_{1}, v_{2}, \ldots, v_{q} \rbrace$ in  $W$,
\State Combine the dummy points $\textbf{v} = \lbrace v_{1}, v_{2}, \ldots, v_{q}\rbrace$ with the data points $\mathbf{x} = \lbrace x_{1}, x_{2}, \ldots,  x_{n}\rbrace $ to form a set of quadrature points $\textbf{s} = \lbrace s_{j}\mid j = 1, 2, \ldots, m \rbrace$,
\State Compute the quadrature weights $w_{j}$,
\State Following equation \eqref{indicator}, determine the indicator $a_{j}$ and compute the variable $y_{j} = a_{j}/w_{j} $,
\State Obtain  the vector of spatial covariates $\mathbf{z}(s_{j}) = (z_{1}(s_{j}),  \ldots, z_{K}(s_{j}))$  at each quadrature point $s_{j}$,
\State Use existing model-fitting software such as \texttt{glmnet} \citep{JSSv033i01}, specifying that the model is a log-linear Poisson regression model, $\log\rho_{\boldsymbol\theta}(s_{j}) = \beta_{0} + \mathbf{z}(s_{j})\boldsymbol\beta'$, 
in order to fit the responses $y_{j}$ and vector of covariate values $\mathbf{z}(s_{j})$ with weights $w_{j}$,
\State  The coefficient estimates returned by the software give the approximate maximum log-likelihood estimate of  $\boldsymbol\theta$,
\end{algorithmic}
\end{algorithm}
The optimisation problem in equation \eqref{addafacaane4} can be implemented using Algorithm \ref{alg:urgooftuuko}.  The approximated log-likelihood function in equation \eqref{werqqq} and the log-likelihood function of the weighted generalised linear model (Poisson distribution) have the same deviance function $\mathcal{D}\left(\boldsymbol\theta\right)  = 2\lbrace \log \mathcal{L}(\mathbf{y}\mid \mathbf{y})-\log \mathcal{L}(\boldsymbol\theta\mid \mathbf{y})\rbrace$. Hereby, the deviance $\mathcal{D}$ of the regularised weighted generalised linear model (Poisson distribution) obtained by the model-fitting software, e.g.~\texttt{glmnet}, can be exploited to select an optimal tuning parameter in the optimisation of the regularised quadratic approximation of the log-likelihood function of the inhomogeneous Poisson process in equation \eqref{addafacaane3}. Choosing an optimal tuning parameter value can be done using K-fold cross-validation where, in short, we set a sequence of tuning parameter candidate values $\lambda_{1}>\lambda_{2}>\cdots>\lambda_{T}$ and split the data into K-folds.  Then, for each $\lambda$-candidate value, we leave out a data fold/piece and perform parameter estimation on all the remaining $K-1$ data folds, thus obtaining a deviance for the left-out data fold. We repeat the parameter estimation and deviance computation for the remaining $K-1$ possible folds to be left out. This means that we obtain $K$ out-of-sample deviances for each $\lambda$ value. Among the sequenced $\lambda$ values, the one giving the smallest mean deviance can be picked as an optimal estimate of the tuning parameter $\lambda$ of the regularised quadratic approximation of the log-likelihood function. We then use the selected optimal $\lambda$ to again carry out the regularized fitting, this time using the full dataset, in order to obtain a final estimate of the model parameter $\boldsymbol\theta$. Finally, the stopping criterion for the cyclic coordinate descent algorithm is generally based on the change of the fitted quadratic approximation of the log-likelihood function value.

\subsection{Semi-parametric intensity function modelling}\label{Semiparametric}

We propose the approach outlined below for the setting where i) the main goal is a prediction/predictive model, i.e.,~one wants to predict a collection of further/future events as precisely as possible, ii) one believes that the observed covariates can only describe a part of the spatial intensity variation, and iii) added spatial flexibility is warranted in the modelling. In the case of our ambulance data, as our end goal is to build optimal dispatching strategies, we mainly want a predictive model.  We want to mention again that the demographic covariates we have at hand only reflect where different demographic groups live but not how they move about.  Explicitly, we do not have access to explicit mobility covariates such as aggregated movement patterns in the population, and, as one may guess, people do not only need access to ambulances when they are at home.\\

Our solution is quite pragmatic and simple. We simply add a further spatial covariate to the existing collection of covariates, which is given by a non-parametric spatial intensity estimate  $\widetilde\rho(x)$, $x\in W$ and will be referred to as the {\em benchmark spatial intensity}. Hence, we include the spatial intensity estimate $\widetilde\rho$ as a covariate in the approximated log-likelihood expression in \eqref{werqqq} and the inclusion has the effect that the modelling steps away from a purely parametric setting to a semi-parametric approach.  This added covariate should pick up on regions where there is an increased intensity due to human mobility, which cannot be explained by the existing list of covariates. Note that this is similar in nature to the semi-parametric (spatio-temporal) log-Gaussian Cox process modelling approach advocated for in e.g.~\citet{diggle2013statistical}.  To be able to discern whether this added covariate is in fact necessary/useful in the presence of the other covariates, we carry out elastic-net regularisation-based variable selection (see Section \ref{s:Regularisation}) to indicate whether the benchmark spatial intensity has any added value in terms of describing the true intensity function.\\
 
A natural question here is what kind of non-parametric intensity estimator one should use to generate the benchmark spatial intensity. There are different candidates for this, and the main distinction one usually makes is between global and adaptive/local smoothers. Adaptive smoothing techniques include adaptive kernel intensity estimation \citep{davies2018tutorial} and (resample-smoothed) Voronoi intensity estimation \citep{ogata2003modelling, ogata2004space,moradi2019resample}. These have some clear benefits (in particular the latter, \citet{moradi2019resample}), but here we do not want to put too much weight on the local features since we may run the risk of overfitting. Instead, we here consider (global) kernel intensity estimation \citep{diggle1985kernel, baddeley2015spatial}, which is arguably the most prominent approach to global smoothing and is defined as 
$$
\widetilde\rho(x)=\sum_{y\in \x}\kappa_h(x-y)/w_h(x,y), \quad x\in W, 
$$
where $\kappa_h(\cdot)=h^{-1}\kappa(\cdot/h)$, $\kappa$ is a symmetric density function and the smoothing parameter $h>0$ is the bandwidth. 
The function $w_h(x,y)$ is a suitable edge correction factor which adjusts the effect of unobserved events outside $W$ on the intensity of the observed events \citep{baddeley2015spatial}; we here use the local corrector $w_h(x,y)=\int_W\kappa_h(u-y) du$ which ensures mass preservation, i.e.~that  $\int_W\widetilde\rho(x)dx=n$, the number of observed events. Practically, to carry out kernel intensity estimation we make use of the function \texttt{density.ppp} in the \textsf{R} package \textsf{spatstat} \citep{baddeley2015spatial}. \\

Although the choice of kernel may play a certain role, the choice of bandwidth is absolutely the main determinant for the quality of the intensity estimate; recall that the bandwidth governs how much we smooth the data. However, optimal bandwidth selection is a well-studied and challenging problem. Concerning the state of the art, the bandwidth selection criterion of \citet{cronie2018non} is generally the most stable with respect to accounting for spatial interaction; observed clusters of points in a point pattern may be the effect of aggregation/clustering (dependence) or intensity peaks, or a combination of the two.  However, there is one scenario where it tends to not perform too well, and that is when there are large regions in $W$ where there are no points present, which is the case for our ambulance dataset.  Other standard methods for bandwidth selection include the state estimation approach of \citet{diggle1985kernel} (called \texttt{bw.diggle} in \textsf{spatstat}), the Poisson process likelihood leave-one-out cross-validation approach in \citet{baddeley2015spatial} and \citet{loader1999local} (called  \texttt{bw.ppl} in \textsf{spatstat}), and the recent machine learning-based approach of \citet{bayisa2020large} (see Algorithm \ref{alg:Oroq}). 

\subsubsection{New heuristic algorithm for bandwidth selection}
In K-means clustering, the dataset is partitioned into a number of clusters, and each cluster consists of data points whose intra-point distances, i.e.,~ distances between points with in a cluster, are smaller than their inter-point distances, i.e.,~ their distances to points outside of the cluster.  In a recent study, \cite{bayisa2020large} proposed a K-means clustering-based bandwidth selection approach for kernel intensity estimation, where the average of the standard deviations of the clusters is used as an optimally selected bandwidth. Although the approach performed well in terms of non-parametrically describing the current ambulance call dataset, it has some limitations/issues.  Firstly, the number of clusters used in the K-means algorithm has been selected visually, and thereby subjectively. Secondly, clusters with high point densities and clusters with widely dispersed data points tend to uniformly determine the resulting bandwidth. Evidently, a cluster with widely dispersed data points has a larger standard deviation, which can be an outlier,  and hence, it distances the estimated bandwidth away from the standard deviations of clusters with highly clustered data points.  As a result, the selected bandwidth leads to oversmoothing of the spatial intensity. 
\begin{algorithm}[!htpb]
\caption{$K$--means clustering-based heuristic algorithm for bandwidth selection}
\label{alg:Oroq}
\begin{algorithmic}[1]
\State Consider the observed spatial location data: $\x = \left\lbrace x_{1}, x_{2}, \cdots, x_{n} \right\rbrace \subseteq W\subset\R^{2}$,
\State Candidates for the number of clusters: $K = 2, 3, \cdots, K_{max}$, 
\State Let $\tr\left(\cdot\right)$ denote the trace of a square matrix,
\State Let $d$ denote the number of variables,
\State Set while condition determining parameter: $\Delta = 1$,
\State Set the maximum number of iterations for the while loop: $max.iter$,
\State Let  $\varpi_{q}$, $q = 1, 2, \cdots, P$  represent the centroid of a cluster $q$,
\State Let $C_{q}$ denote the collection of observations in cluster $q$,
\State Let $n_{q}$ is the number of data points in  $C_{q}$,
\State Set a parameter determining the convergence of the algorithm: $\varepsilon$,
\State Let $D_{P} = \displaystyle\sum_{q = 1}^{P}\sum_{x\in C_{q}}  \left(x - \varpi_{q}\right)' \left(x -\varpi_{q}\right)$ denote within-cluster dispersion matrix for $P$ clusters, \label{withincluster}
\State Let $KL_{P}  = \displaystyle\left\lvert\frac{\left(P-1\right)^{2/d}\tr\left(D_{P-1}\right) - P^{2/d}\tr\left(D_{P}\right)}{P^{2/d}\tr\left(D_{P}\right) - \left(P+1\right)^{2/d}\tr\left(D_{P+1}\right)}\right\rvert$ represent $KL$ index, \label{KLIndex}
\Statex \For {$K \gets 2, 3, \cdots, K_{max}$} \label{DD1}
\For {$P \gets K+1$} \label{DD2}
\Statex \hspace{0.35in} Initialize the centroids: $\varpi^{\left(v\right) }_{q}$, $v = 0$ and $q = 1, 2, \ldots, P$,
\Statex \hspace{0.35in} Call algorithm \ref{alg:Subalgorithm},
\EndFor \label{For1}
\If {$K = 2$,}
\For {$P \gets   K$} 
\Statex \hspace{0.35in} Initialize the centroids: $\varpi^{\left(v\right) }_{q}$, $v = 0$ and $q = 1, 2, \ldots, P$,
\Statex \hspace{0.35in} Call algorithm \ref{alg:Subalgorithm},
\EndFor \label{For}

\For {$P \gets  K-1 = 1$} 
\Statex \hspace{0.35in}  $\varpi_{1} = \left( \displaystyle\displaystyle\sum_{i = 1}^{n}x_{i}\Biggm/n\right)$, 
\EndFor
\EndIf
\If {$K > 2$,} 
\For {$P \gets K-1, K$} 
\Statex \hspace{0.3in} Initialize the centroids: $\varpi^{\left(v\right) }_{q}$, $v = 0$ and $q = 1, 2, \ldots, P$,
\Statex \hspace{0.3in} Call algorithm \ref{alg:Subalgorithm},
\EndFor 
\EndIf\label{DD3}
\State \hspace{0.1in} Obtain the optimal centroids $\left\lbrace \varpi_{q}\right\rbrace_{q = 1}^{P} $ for $P = K-1$, $K$, and $K+1$ from step \ref{DD2} to \ref{DD3}, \label{ObtainCentroids2}
\State \hspace{0.1in} Using the result from step \ref{ObtainCentroids2}, compute the expression $D_{P}$ in step \ref{withincluster} for $P = K-1$, $K$, and $K+1$, \label{ObtainCentroidskl2}
\State \hspace{0.1in} Based on the result from step \ref{ObtainCentroidskl2}, compute $KL_{P}$ in step \ref{KLIndex} for $P = K$.
\EndFor  \label{DD4}
\State From step \ref{DD1} to \ref{DD4}, obtain the optimal number of clusters: $P_{0} = \displaystyle\argmax_{P\in\left\lbrace 2, 3,  \ldots, K_{max}\right\rbrace}\Big\{ KL_{P}\Big\}$, 
\State Initialize the centroids for the optimal number of clusters $P_{0}$: $\varpi^{\left(v\right) }_{q}$, $v = 0$ and $q = 1, 2, \ldots, P = P_{0}$, \label{InitializeD}
\State Based on step \ref{InitializeD}, call algorithm \ref{alg:Subalgorithm} and obtain the optimal centroids $\left\lbrace \varpi_{q}\right\rbrace_{q = 1}^{P} $ and classes of data points $\left\lbrace k_{i}\right\rbrace_{i = 1}^{n}$,

\State  Compute cluster dispersion measure: $\sigma_{q}^{2} = \displaystyle\frac{1}{2n_{q}}\sum_{x_{i}\in C_{q}} \left\|x_{i} - \varpi_{q} \right\|^{2}$, $q = 1, 2, \ldots, P$, 
\State Compute the weight: $w_{q} = \displaystyle\frac{1}{g_{q}\left(x, \varpi_{q}\right)}$, $q = 1, 2, \ldots, P$ and  $g_{q}\left(x, \varpi_{q}\right) =  \displaystyle\frac{1}{n}\sum_{i=1}^{n}\left\|x_{i} - \varpi_{q} \right\|^{2}$, 

\State Obtain an optimal bandwidth estimate:  $h = \sqrt{\displaystyle\sum_{q=1}^{P}w_{q} \Biggm/ \displaystyle\sum_{q=1}^{P}\frac{w_{q}}{\sigma_{q}^{2}}}$.
\end{algorithmic}
\end{algorithm}

\begin{algorithm}[H]
\caption{Subalgorithm of the main algorithm}
\label{alg:Subalgorithm}
\begin{algorithmic}[1]
\While{$\Delta > \varepsilon$ and $v\le max.iter$} 
\State Obtain optimal classes $k_{i} $ and  $1$-of-$P$ class indicator variables $\iota_{iq}$ for data points $ x_{i}$: \label{CLUSTER1}
\begin{align*}
k^{(v)}_{i} = \argmin_{q\in\left\lbrace 1, 2, \ldots, P\right\rbrace } \big \| x_{i} - \varpi^{\left(v\right)}_{q} \big \|^{2}, \quad \iota^{(v)}_{iq} =
\1_{\left\{q = k^{(v)}_{i}\right\}}, \quad q = 1, 2, \ldots, P; i = 1, 2, \ldots, n, 
\end{align*} 
\State Update the centroids of the clusters (the mean locations of the clusters) $\varpi_{q}$: \label{CLUSTER2}
\begin{align*}
\varpi^{\left(v+1\right)}_{q} = \left\lbrace \sum_{i = 1}^{n} \iota^{\left(v\right)}_{iq}x_{i}\right\rbrace 
\Bigg/
\sum_{i = 1}^{n}\iota^{\left(v\right)}_{iq}, 
\quad q = 1, 2,  \ldots,P,
\end{align*}
\State $\Delta = \displaystyle\sum_{q=1}^{P}\left\| \varpi^{(v+1)}_{q} - \varpi^{(v)}_{q} \right\|^{2}$. \label{CLUSTER3}
\EndWhile
\end{algorithmic}
\end{algorithm}
To overcome these limitations, we propose a new heuristic algorithm, which is outlined in Algorithm \ref{alg:Oroq}, to establish the ideal number of clusters and thereby to obtain an optimal estimate of the bandwidth. The main algorithm, which is Algorithm \ref{alg:Oroq},  consists of two crucial steps. It continually invokes Algorithm \ref{alg:Subalgorithm}, which is a K-means algorithm, to establish the optimal number of clusters using the $KL$ index of \citet{krzanowski1988criterion}.  Once an optimal number of clusters has been obtained, the main algorithm determines an optimal bandwidth by calling the K-means algorithm and using a weighted harmonic mean of dispersion measures for the clusters, where the weight for each cluster is given by the inverse of the average of the squares of the distances from the centroid of the cluster to each observation. When establishing the bandwidth, the weights help in balancing the contributions of clusters with closely spaced spatial points and clusters with widely spaced spatial points. In Algorithm \ref{alg:Subalgorithm}, the spatial data points are re-assigned to clusters (see step \ref{CLUSTER1} in Algorithm \ref{alg:Subalgorithm}), the cluster means are re-computed (see step \ref{CLUSTER2} in Algorithm \ref{alg:Subalgorithm}), and these steps are repeated until the sum composed of the squared Euclidean distances between all successive centroids is smaller than a user-specified value (see step \ref {CLUSTER3} in Algorithm \ref{alg:Subalgorithm}). Alternatively, one may  control the convergence of Algorithm \ref{alg:Subalgorithm} by repeating steps \ref{CLUSTER1} and \ref{CLUSTER2} until there is either no further change in the assignments of data points to clusters or until some maximum number of iterations has been reached. In our case, we have used $\varepsilon = 10^{-5}$ and $max.iter = 100$.
\section{Created covariates}\label{s:CreatedCovariates}
\begin{figure}[!htpb]
\centering
\includegraphics[height=22cm, width=18cm]{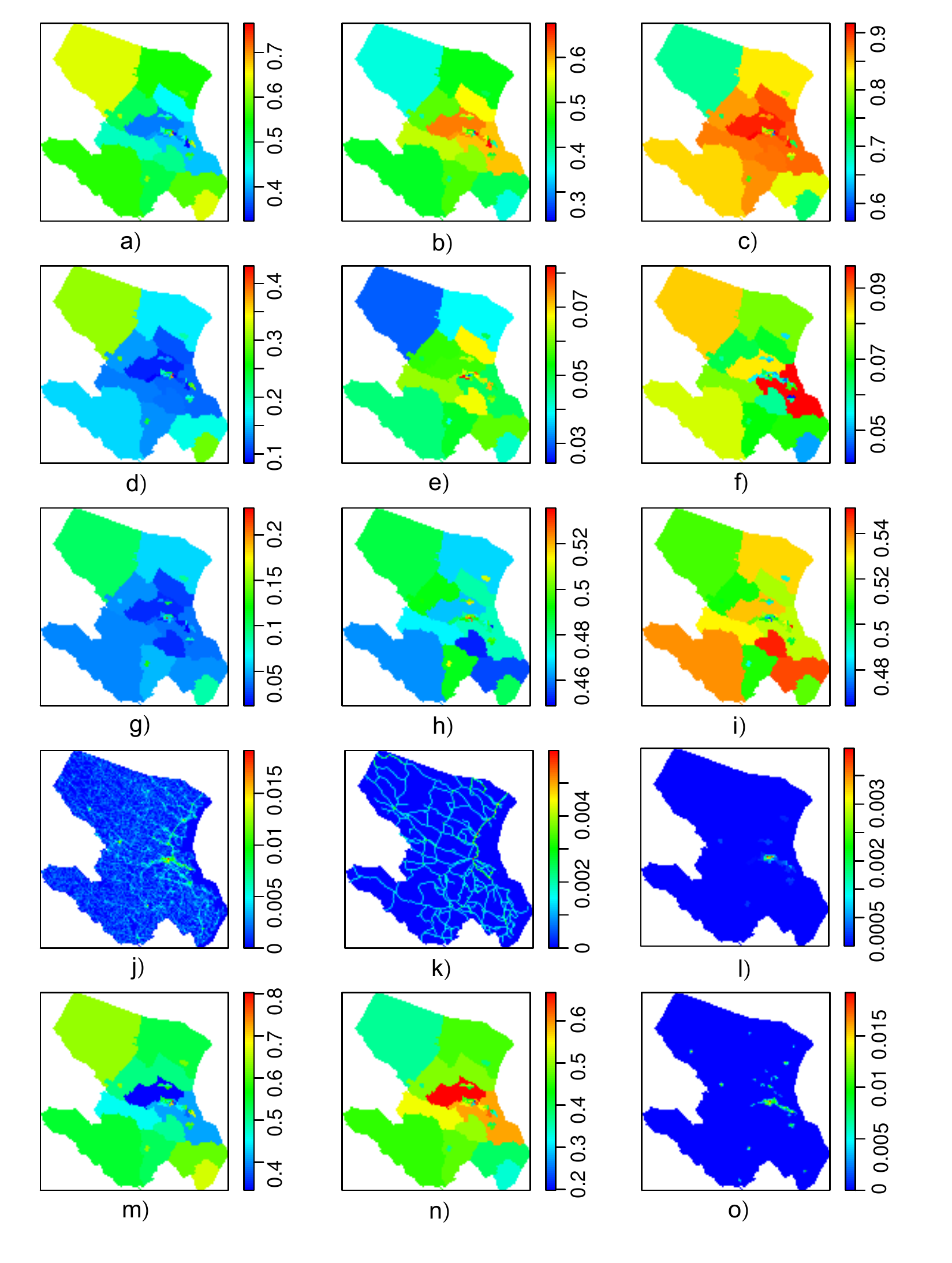}
\caption{Spatial demographic and road network-related covariates. Population 20+ years old income status: a) below and b) above median income. Population 20-64 years old employment status: c) employed and d) unemployed. Population by age: e) 35-39 years old, f) 55-59 years old, and g) 80+ years old. Population by sex: h) male and i) female. j) complete road network line density. k) main road network line density. l) population density. Households 20+ years old economic status: m) below and n) above median income. o) Densely populated line density. }
\label{Covariates}
\end{figure}

Recall the notions of 'original' and 'created' covariates from Section \ref{data}.  We illustrate some of the selected  'original' and 'created' covariates considered in this study in Figure \ref{Covariates}. Below, we  provide a description of the construction of all created covariates. 

\subsection{Benchmark intensity covariate for the ambulance data}
\begin{figure}[!htpb]
\centering
\includegraphics[width= 0.8\textwidth, height=0.6\linewidth]{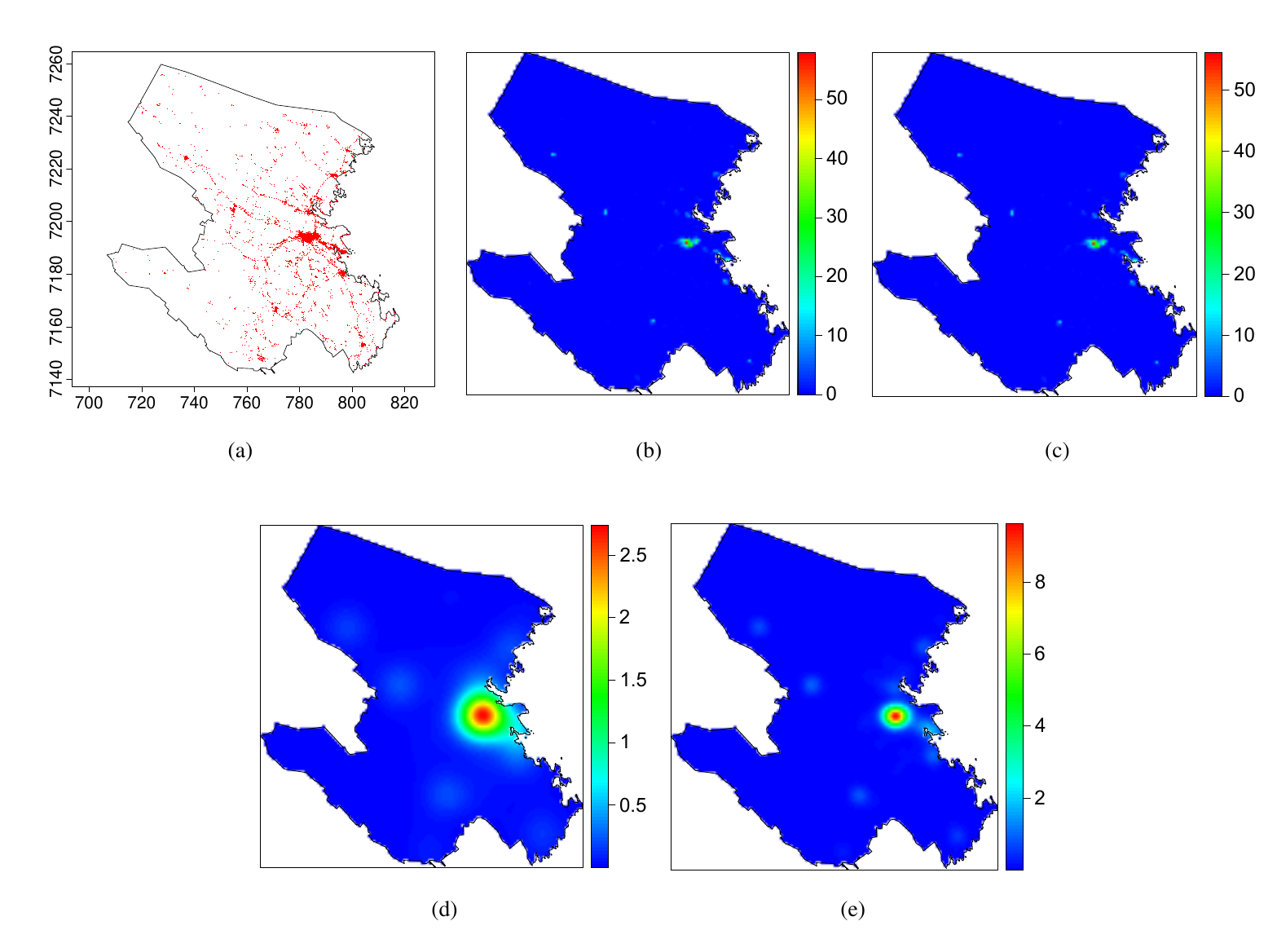}
\caption{The roles of different bandwidth selection methods using Gaussian kernel density estimation for the unmarked ambulance call data in panel (a). b) State estimation \citep{diggle1985kernel}. c) Poisson likelihood cross-validation \citep{loader1999local}. d) K-means clustering \citep{bayisa2020large}. e)  The new heuristic algorithm developed in this work. Note that what interests us here are the relative scales rather than the raw scales; the values in the plots above have been multiplied by 1000 for ease of visualisation.}
\label{BandWidth} 
\end{figure}

Recall that our semi-parametric approach is based on the idea of including a non-parametric intensity estimate, the so-called benchmark intensity, as a covariate in our log-linear intensity form.  To generate the benchmark intensity for the ambulance data, we employ our new algorithm to obtain the intensity estimate found in panel (e) of Figure \ref{BandWidth}. We further compare the result with the aforementioned approaches, namely the state estimation approach, the Poisson process likelihood leave-one-out cross-validation approach, and the approach of \citet{bayisa2020large}, which is based on 5 clusters (this number has been obtained through visual inspection). The resulting intensity estimates for the ambulance data can be found in panels (b)-(d) in Figure \ref{BandWidth}. Note that throughout we have used a Gaussian kernel in combination with the aforementioned local edge correction factor. We argue that the state estimation approach and the Poisson process likelihood leave-one-out cross-validation approach tend to under-smooth the data and thereby do not reflect the general overall variations of the data, whereas the K-means clustering based bandwidth selection of \citet{bayisa2020large} instead tends to over-smooth the ambulance data.  Recall that the number of clusters, which is a necessary input in the K-means clustering based bandwidth selection of \citet{bayisa2020large}, has to be selected through visual inspection.  Looking at panel (e) of Figure \ref{BandWidth}, we see that by employing our proposed $KL$ index to automatically select the number of clusters, we obtain a total of 16 clusters. We argue that, compared to the different panels in Figure \ref{BandWidth}, the new heuristic algorithm  performs the best in terms of balancing over- and under-smoothing of the  ambulance call events. 

\subsection{Creation of (road) network-related covariates}\label{s:LineDensity}
\begin{figure}[H]
\centering
\includegraphics[width= 1\textwidth, height=0.9\linewidth]{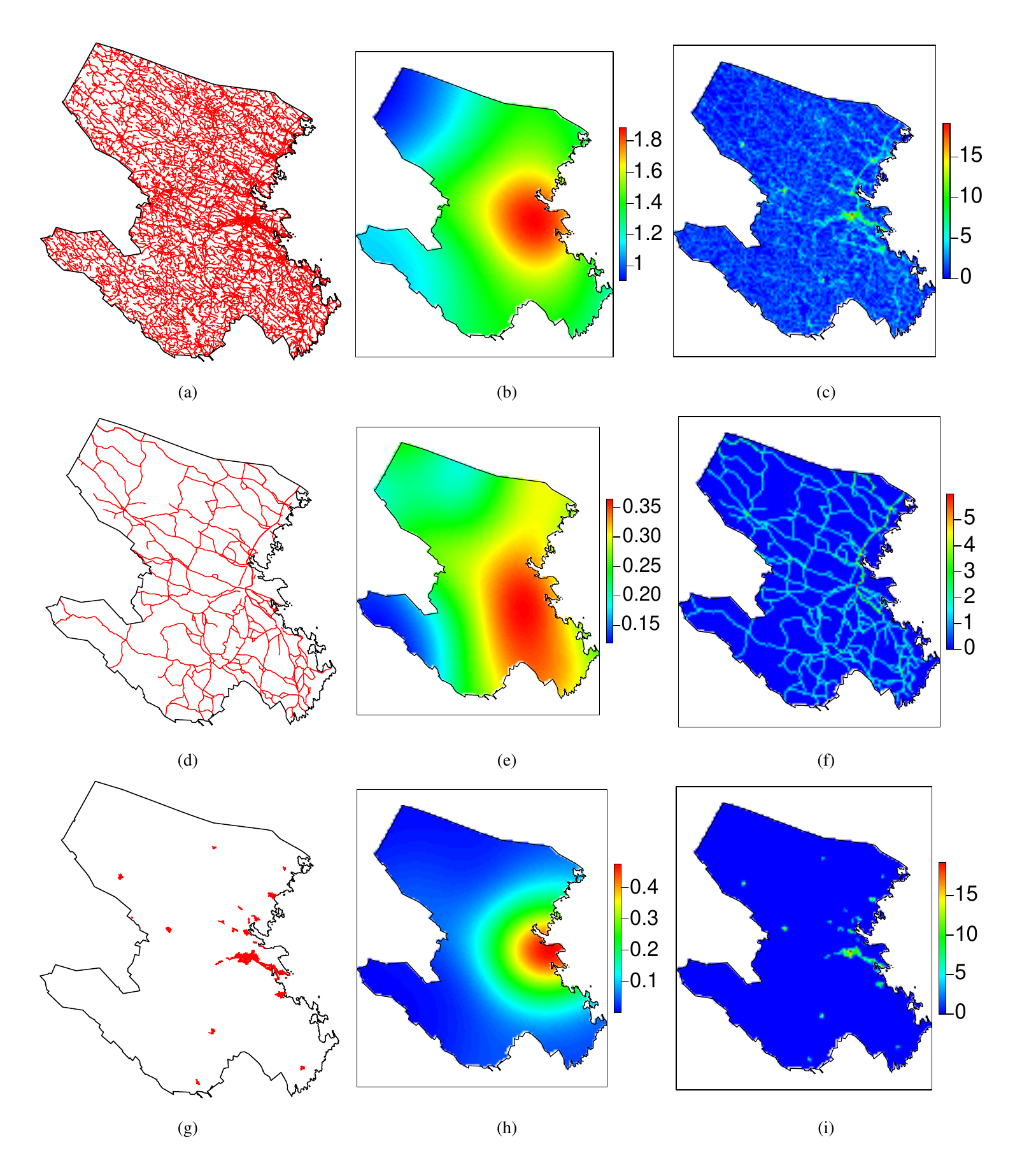}
\caption{Creation of covariates from line segment patterns. The first column represents a) the complete road network,  d) the main road network and g) the densely populated area road network, respectively.  Created road network covariates using the \textsf{spatstat} function \texttt{density.psp}:  (b), (e) and (h) have been obtained using the default bandwidth of \texttt{density.psp}, whereas (c), (f) and (i) have been obtained using the new heuristic bandwidth selection algorithm.  The values in the plots have been multiplied by 1000 for ease of visualisation.}
\label{BandWidthLineDensity}
\end{figure}

The role of the road network-related covariates is to control that the fitted model generates events close to the underlying road network. But, as previously indicated, the original road network-related covariates are not of the form $z_i(s)$, $s\in W$. 
\begin{itemize}
\item Complete road networks line density. It is a spatial pattern of line segments, which is converted to a pixel image. The value of each pixel in the image is measured as the total length of intersection between the pixel and the line segments [1].
\item Main road networks line density. It is a spatial pattern of line segments, which is converted to a pixel image. The value of each pixel in the image is measured as the total length of intersection between the pixel and the line segments [1].
\item Densely populated line density. It is a spatial pattern of line segments, which  is converted to a pixel image. The value of each pixel in the image is measured as the total length of intersection between the pixel and the line segments [1].
\item Bus stops density. It is a spatial point pattern, which is converted to a pixel image. The value of each pixel is an intensity, which is measured as "points per unit area" [1]. 
\item Shortest distance to bus stops. It is the shortest distance in meters from the ambulance location data to the bus stops [1].
\item Shortest distance to densely populated areas. It is the shortest distance in meters from the ambulance location data to the densely populated areas [1].
\item Shortest distance to  main road networks. It is the shortest distance in meters from the ambulance location data to  the main road networks [1].
\item Shortest distance to complete road networks. It is the shortest distance in meters from the ambulance location data to  the complete road networks [1].
\end{itemize} 
We here propose to treat the road networks under consideration as line segment patterns \citep{baddeley2015spatial}, which essentially means that each road network considered is a realisation of a point process in the space of line segments in $\R^2$. 
The spatial covariate corresponding to a given line segment pattern is then given by the estimated line segment intensity, which is obtained as the convolution of an isotropic Gaussian kernel with the line segments of the pattern in question. Practically, such an estimate may be obtained through the function \texttt{density.psp} in the \textsf{R} package \textsf{spatstat}, and the standard deviation of the Gaussian kernel, the bandwidth, determines the degree of smoothing. The default bandwidth choice in \texttt{density.psp} is given by the diameter of the observation window multiplied by 0.1. As an alternative, we propose to use our new heuristic algorithm for bandwidth selection, which is achieved by letting $W = [0,\infty)$, and letting the observations $\x = \left\lbrace x_{1}, x_{2}, \cdots, x_{n} \right\rbrace \subseteq W$ considered in Algorithm \ref{alg:Oroq} represent the lengths of the line segments. Figure \ref{BandWidthLineDensity} compares the default bandwidth choice of \texttt{density.psp} to our heuristic algorithm,  and it clearly suggests that the line segment intensities generated using the heuristic algorithm bandwidth selection have captured the spatial pattern of the line segments in the observed data better than the default bandwidth choice.  Note that what interests us here are the relative scales rather than the raw scales; the values in the plots have been multiplied by 1000 for ease of visualisation.

\section{Data analysis}\label{Results} 
Having presented the semi-parametric regularised fitting procedure to be employed, we next turn to its actual modelling of the ambulance call data; recall the structure of the data (including the missing sex/gender label issue) and that we may model the intensity of each marginal process $Y_j$ separately.\\

As carefully emphasised and laid out above, we have chosen to follow \citet{yue2015variable} and thus chosen to employ elastic-net regularisation of a Poisson process log-likelihood function, where one of the covariates in fact is a non-parametric intensity estimate of the data (the so-called benchmark intensity). Our motivation for elastic-net regularisation is mainly that we have a large number of spatial covariates, which may be highly correlated. The choice of $\alpha$ is subjective, and we usually pick an $\alpha$ between 0 and 1, and then proceed to the estimation of the proposed model \citep{friedman2010regularization}. Recall further that elastic-net performs like a lasso but avoids unstable behaviour due to extreme correlation if the elastic-net parameter $\alpha$ is large enough but less than one \citep{yue2015variable}; $\alpha=1$ yields the lasso penalty and $\alpha=0$ yields the ridge penalty.  Since we here are interested in carrying out lasso-like elastic-net regularization, we let $\alpha = 0.95$, following a recommendation of \cite{friedman2010regularization} and \citet{yue2015variable}.  Such lasso-like elastic-net regularisation results in variable selection (coefficients of less determining covariates are set to zero), but the penalty also forces highly correlated features to have similar coefficients.

\subsection{Modelling the ambulance call intensity function} \label{SpatialCovariates}
\begin{figure}[!htpb]
\centering
\includegraphics[width=0.8\textwidth, height=0.4\linewidth]{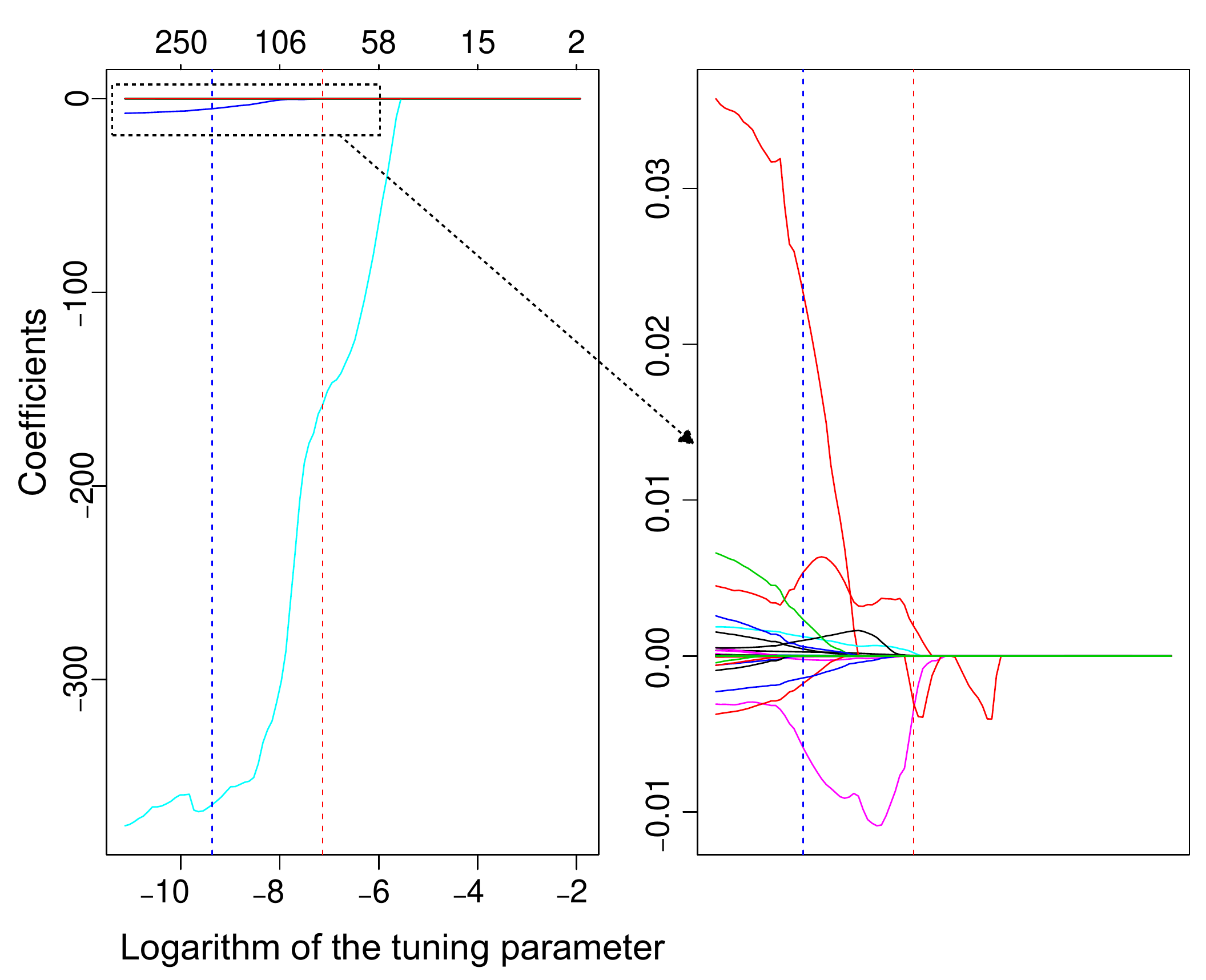}\\
\includegraphics[width=0.8\textwidth, height=0.4\linewidth]{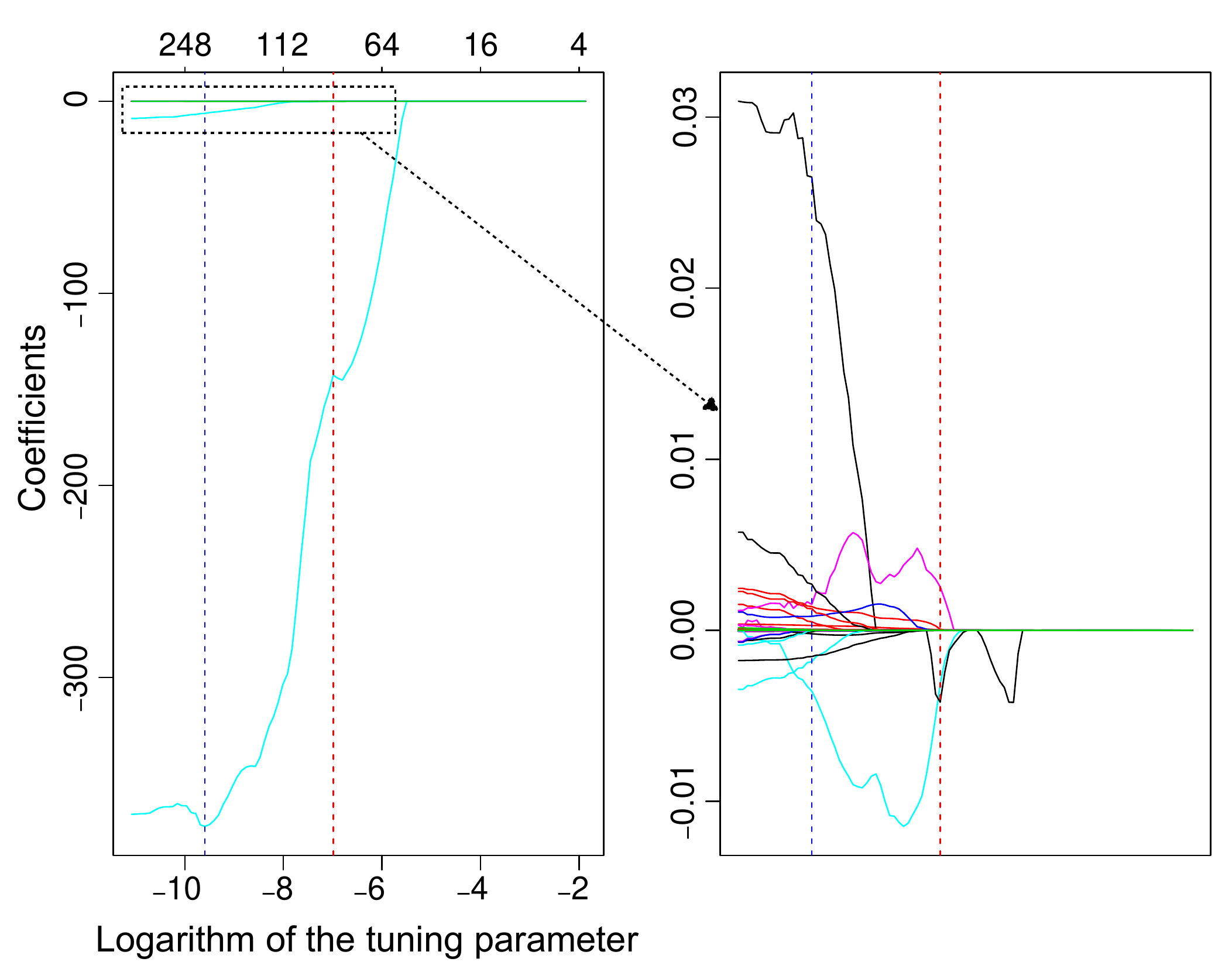}
\caption{A trace plot of the estimated coefficients for lasso (the first row) and lasso-like elastic-net (the second row) regularisations. The numbers at the top of each panel indicate the number of spatial covariates with non-zero estimated coefficients.  The plots in the  right panel of the left plots represent the zoomed in portion of the region marked by the rectangular region. For the purpose of visualization, the estimated coefficients are scaled (divided) by $10^{10}$. }
\label{Orgr2LassoElastic}
\end{figure}

 Figure \ref{Orgr2LassoElastic} shows the estimated coefficients of the spatial covariates, using both lasso regularisation and  lasso-like elastic-net regularisation, for the spatial point pattern constituting the events with priority label 1. Note that the numbers at the top of each panel in Figure \ref{Orgr2LassoElastic} indicate the number of spatial covariates with non-zero coefficients, i.e.~the number of covariates that are associated with the fitted spatial intensity functions for the indicated value of $\lambda$. We clearly see how a large number of covariates are quickly excluded as we increase $\log\lambda$.\\

Using the aforementioned covariates and their first-order interaction terms, the trace plots of the estimated parameters for lasso regularization  and lasso-like elastic-net regularisation with  $\alpha = 0.95$ are shown in Figure \ref{Orgr2LassoElastic}. At their corresponding optimal tuning parameters, which are shown by blue-coloured vertical lines in Figure \ref{Orgr2LassoElastic}, lasso and lasso-like elastic-net regularisations have selected 185 (18.69\%)  and  207 (20.91\%) of the covariates that are associated with the intensity of the events. As it is expected, even though the lasso-like elastic-net regularisation selects more variables  than the lasso regularization, it forces the highly correlated covariates to have similar coefficients \citep{friedman2010regularization, yue2015variable}. As a result, the correlated covariates have similar roles on the intensity of the events, and their presence in the fitted model can be advantageous for interpretation. Therefore, hereafter, we use lasso-like elastic-net regularisation with $\alpha = 0.95$.

\begin{figure}[!htpb]
\centering
\includegraphics[width=0.7\textwidth, height=0.5\linewidth]{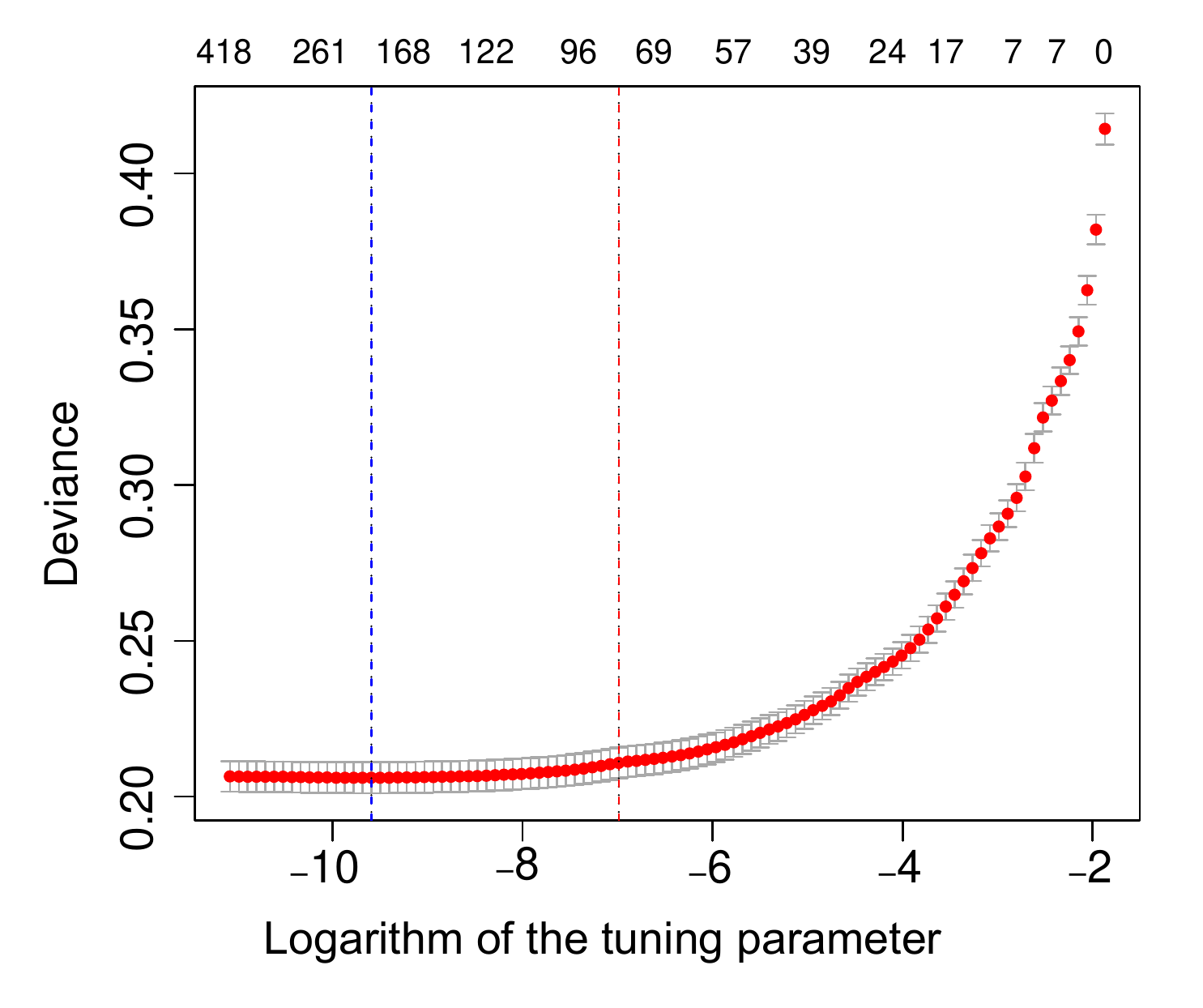}
\caption{An optimal tuning parameter selection via ten-fold cross-validation: location of the logarithm of the optimal estimate of the tuning parameter (blue)  and  the location of the logarithm of the estimate of the tuning parameter that is one standard error away from the optimal estimate (red).}
\label{Orgr2}
\end{figure}

A grid of  $\lambda$ values has been exploited to train the proposed model, and among the candidate $\lambda$ values, the one which gives the smaller deviance, $\mathcal{D}$, has been selected as an optimal estimate of $\lambda$.  The optimal elastic-net regularisation parameter $\lambda$ has been selected using ten-fold cross-validation as shown in Figure \ref{Orgr2}. 
\noindent The two vertical lines in the figure have been drawn to show the location of the logarithm of the optimal estimate of the tuning parameter (blue) and the location of the logarithm of the estimate of the tuning parameter that is one standard error away from the optimal estimate (red); the one-standard-error-rule \citep{hastie2017elements} says that one should go with the simplest model, which is no more than one standard error worse than the best model.\\
 
\begin{figure}[H]
\centering
\includegraphics[width=0.25\textwidth, height=0.2\linewidth]{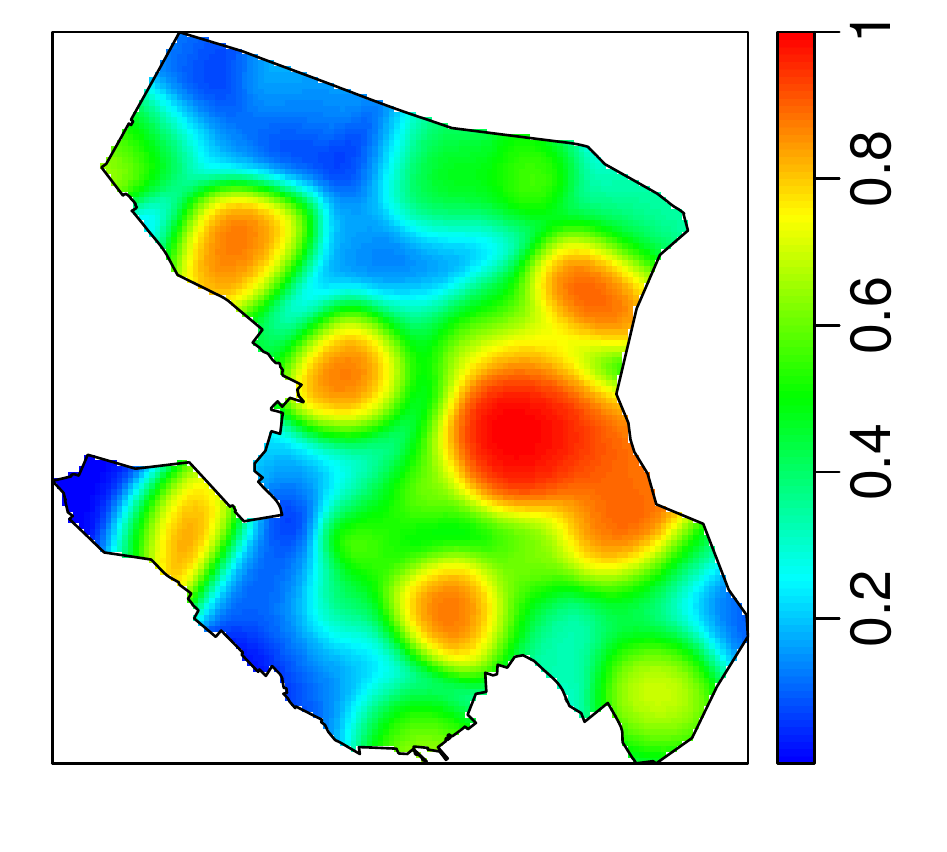}
\includegraphics[width=0.2\textwidth, height=0.2\linewidth]{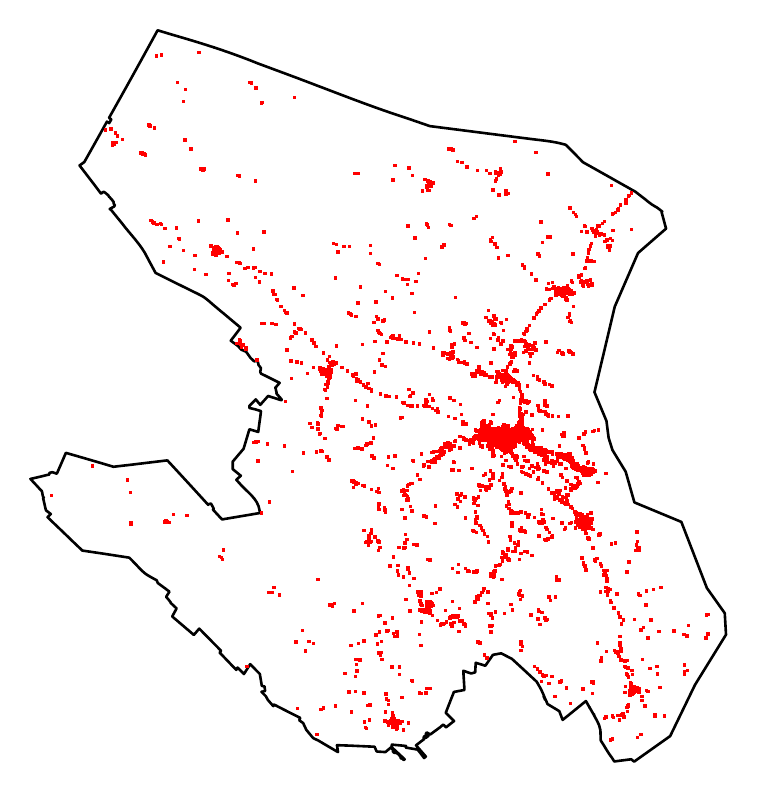}
\includegraphics[width=0.25\textwidth, height=0.2\linewidth]{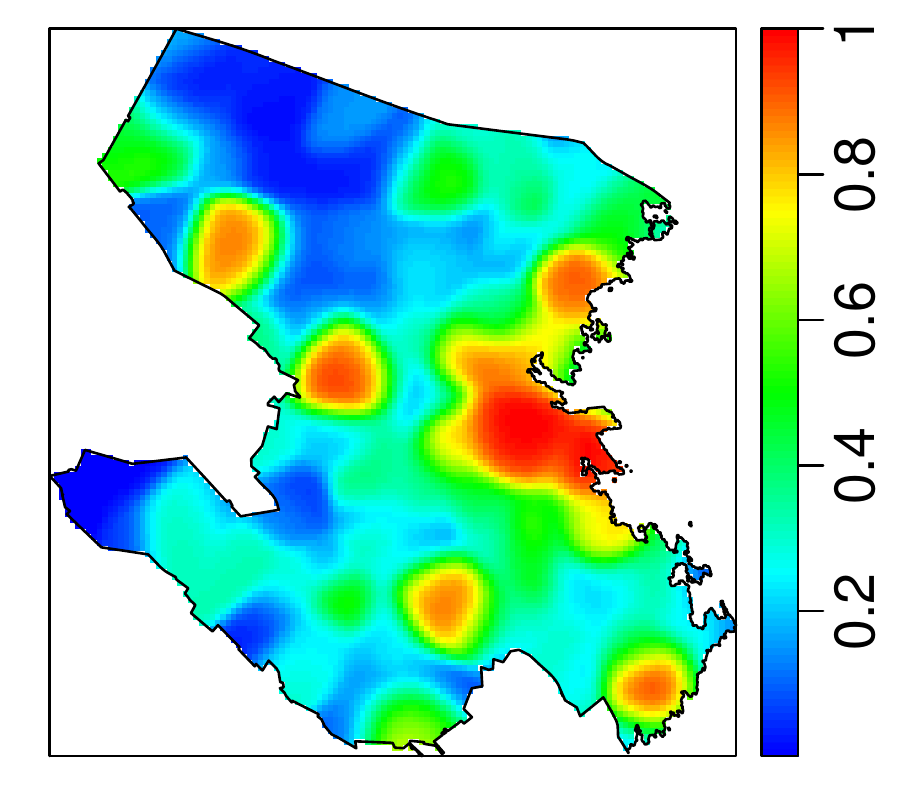}
\includegraphics[width=0.2\textwidth, height=0.2\linewidth]{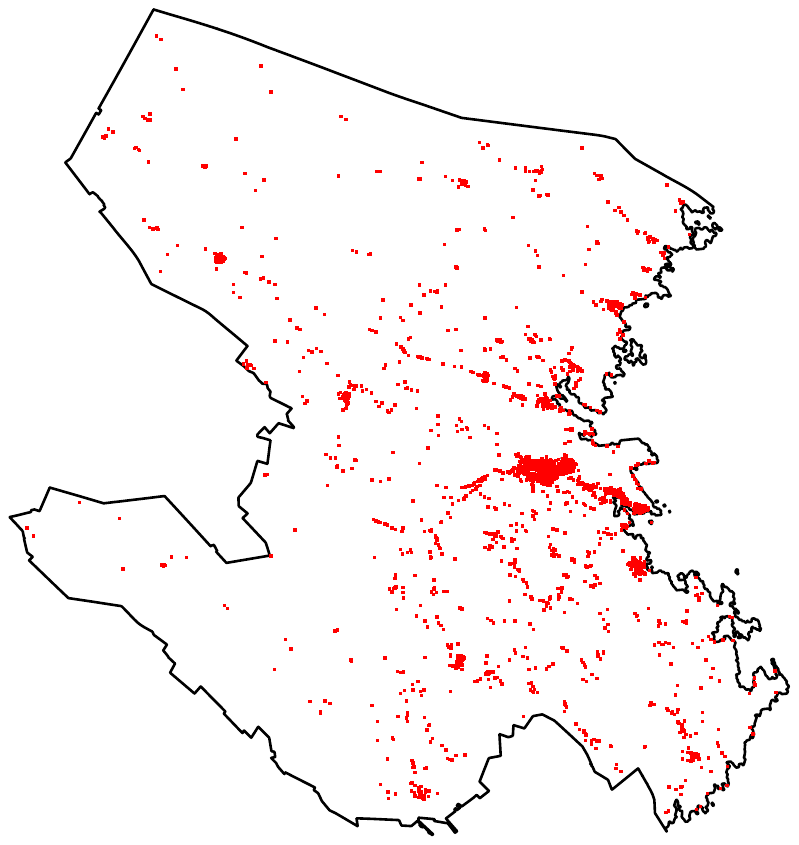}\\
\includegraphics[width=0.25\textwidth, height=0.2\linewidth]{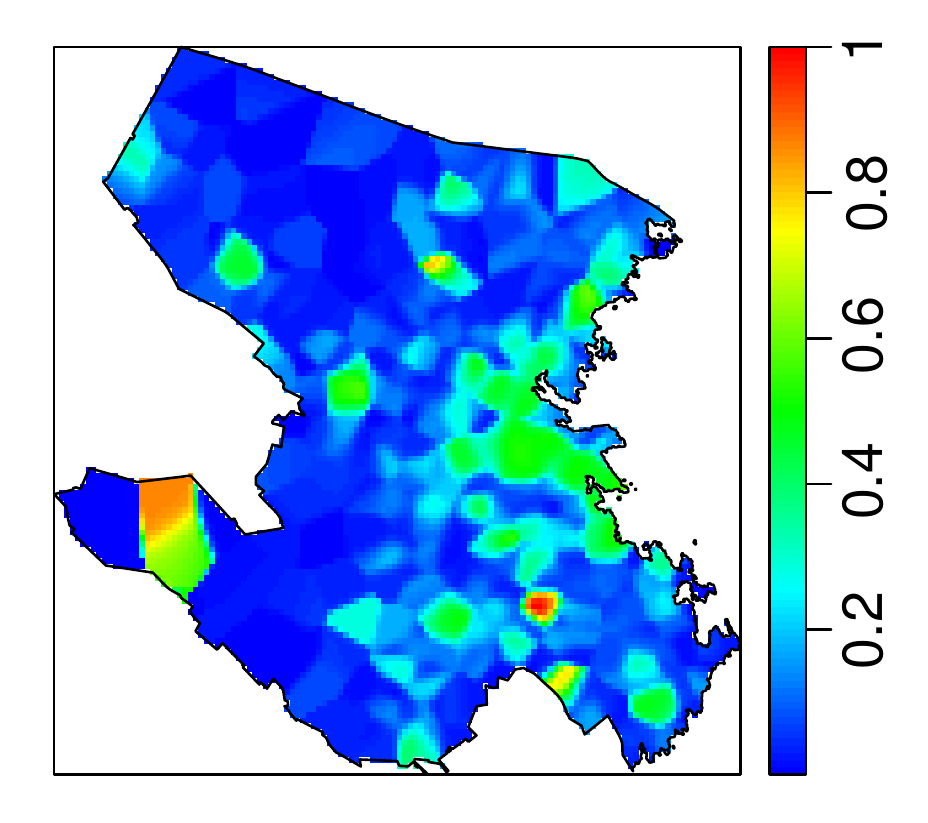}
\includegraphics[width=0.2\textwidth, height=0.2\linewidth]{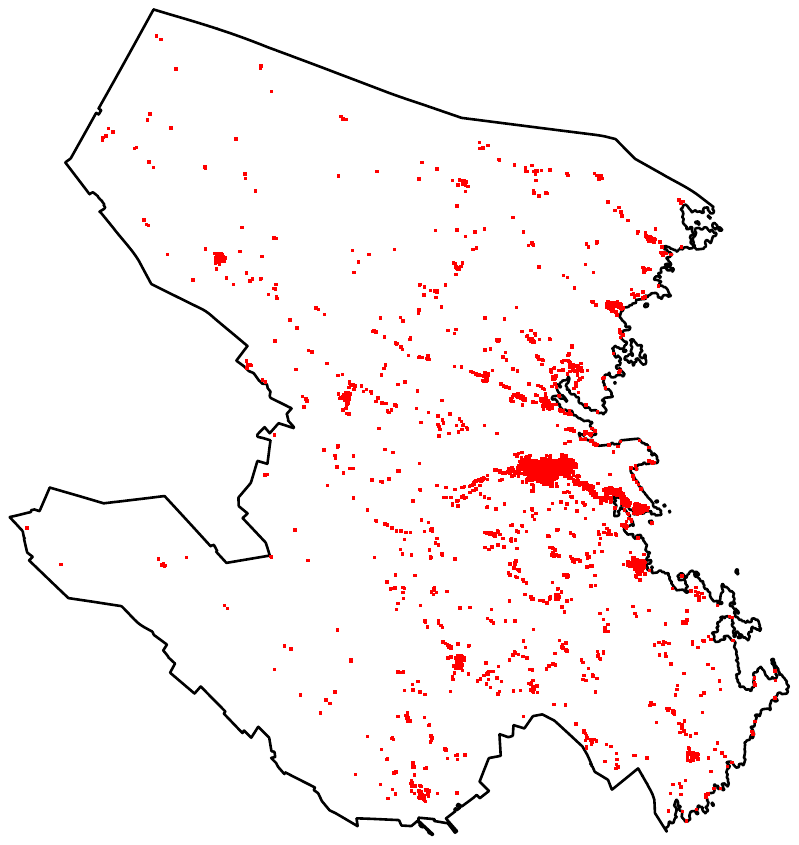}
\includegraphics[width=0.25\textwidth, height=0.2\linewidth]{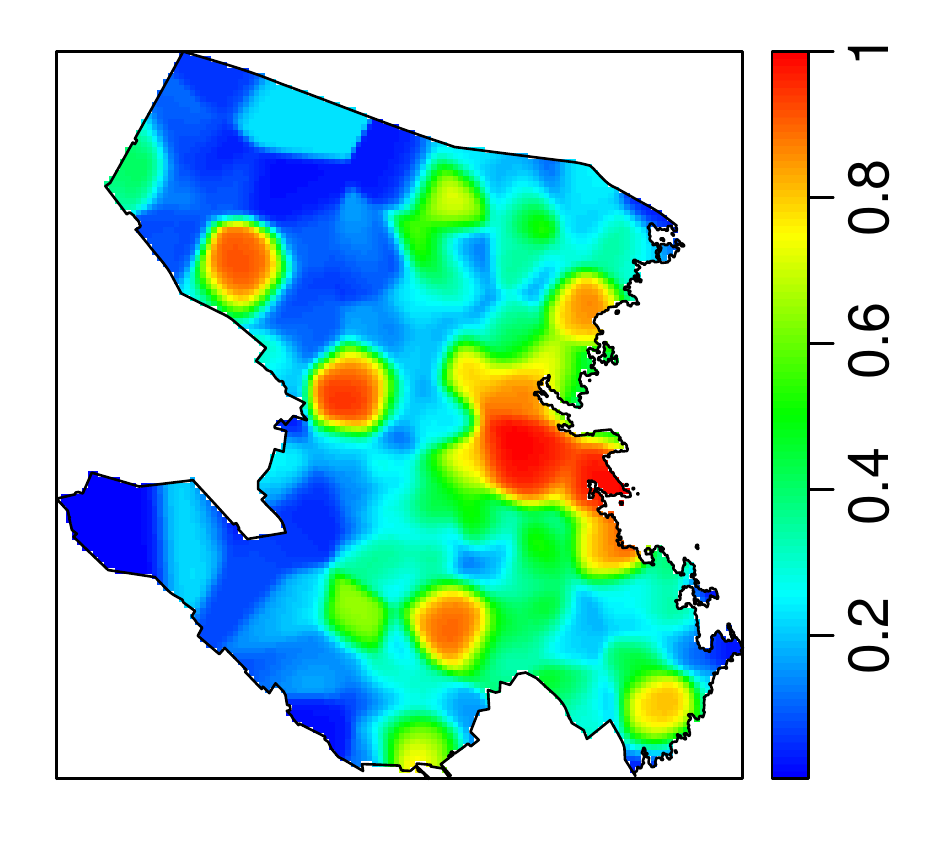}
\includegraphics[width=0.2\textwidth, height=0.2\linewidth]{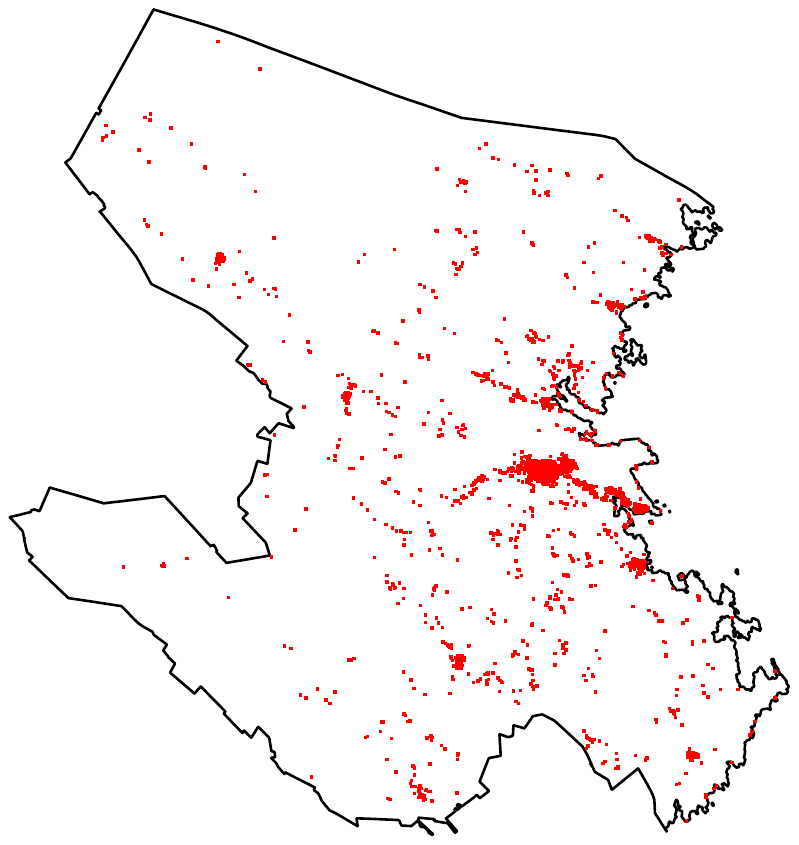}
\caption{The estimated spatial intensities of the emergency alarm calls and their corresponding observed marginal spatial point patterns. The first column presents estimated intensities for priority levels 1 and 2, respectively, while the second column demonstrates their corresponding observed marginal spatial point patterns. In a similar manner, the third column presents the estimated intensities for genders, male and female, while the fourth column shows their corresponding observed marginal spatial point patterns. Note that we have scaled the intensity estimates to range between 0 and 1 so that we may compare them more easily.}
\label{Orgr3}
\end{figure}

The spatial intensity function of the events has been obtained at each observed spatial location using the estimated model, and it has been obtained at the desired spatial locations in the study area using kernel-smoothed spatial interpolation of the estimated intensity function, which is used as the mark of the observed point pattern. The estimated spatial intensities of the marginal spatial point patterns are shown in Figure  \ref{Orgr3}; note that we have scaled the intensity estimates to range between 0 and 1 so that we may compare them more easily. For the priority level 1 and 2 events, about 20.91\% and 36.06\% of the spatial covariates have been associated with/included in the final intensity function estimates, respectively. The lasso-like elastic-net has discerned about  17.68\% and 35.76\% of the spatial covariates which determine the spatial intensities of the point patterns corresponding to male and female, respectively. 
The aforementioned analyses have utilised the spatial covariates  $z_i(s)$, $s\in W$, and the interaction terms of the form $z_i(s)z_j(s)$, which holds for the cases $i = j$ and $i \ne j$. However, with this approach, it is not easy to interpret the results and present the estimated models. Therefore, we are interested in further exploring the estimation of the intensity of the emergency alarm call events using only the original spatial covariates $z_i(s)$, $s\in W$, i.e.,~no first-order  and higher-order interactions of spatial covariates in the modelling of the spatial intensity. Modelling the spatial intensity of the emergency alarm call events using only the original spatial covariates can help to identify the spatial covariates that play a key role in determining the spatial distribution of the emergency alarm call events, and it also ease the interpretation and presentation of the results. Moreover, we can also compare the estimated intensities that are obtained by the two approaches (i.e.,~modelling with only the original spatial covariates and modelling  with the original spatial covariates and  the first-order  interaction terms) to discern the approach that can be practicable.  Using only the original spatial covariates,  a lasso-like elastic-net ($\alpha = 0.95$) with an optimal tuning parameter estimate does not provide sparse solutions, i.e.,~most of the coefficients of the spatial covariates are non-zero. That is to say, the lasso-like elastic-net provides dense solutions, i.e.,~it has a low rate of variable exclusion, see Table \ref{tab:EstimatedModels1}. 

\begin{center}
\begin{longtable}{|l|l|ccccc}
\caption{\normalsize The estimated dense models for the marginal and unmarked emergency alarm call events/patterns. The dots in the table represent small coefficients of covariates that are shrunk to zero.}
\label{tab:EstimatedModels1}\\

\hline \multicolumn{1}{|c|}{\textbf{No.}} & \multicolumn{1}{c|}{\textbf{Covariates}} & \multicolumn{1}{c}{\textbf{Priority 1}} & \multicolumn{1}{c}{\textbf{Priority 2}}  & \multicolumn{1}{c}{\textbf{Male}} & \multicolumn{1}{c}{\textbf{Female}} & \multicolumn{1}{c}{\textbf{Unmarked}}\\ \hline 
\endfirsthead

\multicolumn{3}{c}%
{{\bfseries \tablename\ \thetable{} -- continued from previous page}} \\
\hline \multicolumn{1}{|c|}{\textbf{No.}} & \multicolumn{1}{c|}{\textbf{Covariates}} & \multicolumn{1}{c}{\textbf{Priority 1}} & \multicolumn{1}{c}{\textbf{Priority 2}}  & \multicolumn{1}{c}{\textbf{Male}} & \multicolumn{1}{c}{\textbf{Female}} & \multicolumn{1}{c}{\textbf{Unmarked}} \\ \hline 
\endhead

\hline \multicolumn{3}{l}{{Continued on next page}} \\ \hline
\endfoot

\hline \hline
\endlastfoot
&\textbf{Ambulance location data}: \\\cline{1-2}
1\footnote[1]{\label{1sttablefoot}The parameter estimates are divided by $10^{-5}$ for better visualisation of the estimates}&\hspace{2cm}x & 0.823 &0.826 &0.951  &0.522&0.725\\
2\footref{1sttablefoot}&\hspace{2cm}y &-0.144 &-0.225&-0.175  &-0.463&-0.243\\\hline
&\textbf{Shortest distance to}: \\\cline{1-2}
3\footref{1sttablefoot}& Bus stops&-42.810 &-52.120&-53.820&-49.810& -44.776\\
4\footref{1sttablefoot}&Densely populated area &-3.685 &-4.380&-2.344& -4.246&-3.674\\
5\footref{1sttablefoot}&Main road networks &-32.030 &-16.740&-16.890&-20.800&-24.907\\
6\footref{1sttablefoot}&Complete road networks &-206.300&-238.600&-203.100&-259.200&-229.287\\\hline
7&Population density &18.991 &-141.524&68.728&$\cdot$& -31.185\\
8&Benchmark intensity & 949.209 &3802.066&-15119.642 &$\cdot$& 276.511\\\hline
&\textbf{Line densities}:\\\cline{1-2}
9&Complete road networks & 243.727&280.890 &308.540 &304.280& 275.264\\
10&Main road networks &42.073  &-54.408 &-78.472 & -57.196&25.836\\
11&Densely populated & -158.792 &-155.344  &-175.181  &-157.708& -161.800\\\hline
12&Bus stops density &548072.243&-163337.873 &604558.360 & 81888.976&494114.346\\\hline
&\textbf{Population by}: \\\cline{1-2}
&a) \textbf{employment status:} \\
13&Employed & $\cdot$ & $\cdot$ & $\cdot$  &$\cdot$&  $\cdot$\\
14&Unemployed& $\cdot$ & $\cdot$ & $\cdot$ &$\cdot$& $\cdot$\\\hline
&b) \textbf{income status:} \\
15&Below national median income & 4.747 &0.387  &0.365 &0.866&4.446\\
16&Above national median income& -0.116 &-0.001 &-0.001  & -0.010&-0.079\\\hline
&c) \textbf{educational level}: \\
17&Pre-high school & -0.355 &-2.112 & -1.996  & -2.260& -2.150 \\
18&High school & 1.808 & 0.962 & 2.001 & 1.365& 1.190 \\
19&Post-secondary -- less than 3 years & 3.105 & 1.196&2.534  &$\cdot$& 1.034\\ 
20&Post-secondary -- 3 years or longer& $\cdot$ & $\cdot$ &$\cdot$  &$\cdot$&  $\cdot$\\\hline
&d) \textbf{age:} \\
21&0 -- 5 & 7.999 & 5.926  &4.683 & 6.189&6.000\\
22&6 -- 9& -1.115 &-12.052   &-3.768  &-12.189&-4.575\\
23&10 -- 15 &4.505  &1.310   &5.901  & 3.828&7.542\\
24&16 -- 19 & $\cdot$ &4.433  &2.148  &0.247&-3.813\\
25&20 -- 24 & $\cdot$ &-3.429 &$\cdot$  &$\cdot$& 0.063\\
26&25 -- 29 &  -0.179 &-0.764 &-1.902  & -4.533&-0.163\\ 
27&30 -- 34 & -3.306 & -3.821  &$\cdot$  & $\cdot$&-0.599\\ 
28&35 -- 39 &1.486 & 12.824  &4.757 &  15.668&5.583\\
29&40 -- 44 &-3.386  &0.112 &-8.503  &  -3.040& 0.737\\
30&45 -- 49 & -1.064 &-3.946  &-4.763 & -4.098& -4.794\\ 
31&50 -- 54 & -10.938 &-8.441  &-7.539 &  -5.996&-10.981\\ 
32&55 -- 59 & 3.340 & 4.501 &6.950  & 7.621&6.450\\ 
33&60 -- 64 &-1.003  &0  & 1.370 & -0.171&-0.422\\
34&65 -- 69 & -13.011 &-5.694 &-6.573  & -8.789& -7.743\\
35&70 -- 74 & 0.131 &0.025  &$\cdot$  & 0.098&-4.211\\
36&75 -- 79 & 3.020 & $\cdot$ &$\cdot$  & 3.359& 2.524\\ 
37&80 + &  $\cdot$ &0.294  &$\cdot$  & $\cdot$&2.507\\\hline
&e) \textbf{gender:}\\
38&Female &-0.816  &2.586  & 1.431  & 2.145&-1.302\\
39&Male &0.058  & -0.017 &-0.004  & -0.043& 0.106\\\hline
&f) \textbf{background:}\\
40&Swedish background & -0.177  & -1.460  &-1.715  & $\cdot$&-0.671\\
41&Swedish foreign background&0.046 &  0.028 &0.062  & $\cdot$&0.044\\\hline
&g) \textbf{household economic standard:} \\
42&Below the national median & -3.741  & $\cdot$ &$\cdot$  & $\cdot$ &-3.195\\
43&Above the national median & 0.131 & $\cdot$& $\cdot$ & $\cdot$&0.079\\\hline
&\textbf{Intercept parameter estimate}: & 2.410 &8.409 & 3.646 &25.929&10.886\\
\end{longtable}
\end{center}

\noindent On the other hand, exploiting the one-standard-error rule in \cite{JSSv033i01}, where we pick the most parsimonious estimated model within one standard error of the minimum, in the context of regularisation, we present the result of the estimation for each of the intensity functions of the marginal point processes/patterns in Table \ref{tab:EstimatedModels2}. 

\begin{center}
\begin{longtable}{|l|l|ccccc}
\caption{\normalsize The estimated parsimonious/sparse models for the marginal and unmarked emergency alarm call events/patterns. The dots in the table represent small coefficients of covariates that are shrunk to zero.}
\label{tab:EstimatedModels2}\\

\hline \multicolumn{1}{|c|}{\textbf{No.}} & \multicolumn{1}{c|}{\textbf{Covariates}} & \multicolumn{1}{c}{\textbf{Priority 1}} & \multicolumn{1}{c}{\textbf{Priority 2}}  & \multicolumn{1}{c}{\textbf{Male}} & \multicolumn{1}{c}{\textbf{Female}} & \multicolumn{1}{c}{\textbf{Unmarked}}\\ \hline 
\endfirsthead

\multicolumn{3}{c}
{{\bfseries \tablename\ \thetable{} -- continued from previous page}} \\
\hline \multicolumn{1}{|c|}{\textbf{No.}} & \multicolumn{1}{c|}{\textbf{Covariates}} & \multicolumn{1}{c}{\textbf{Priority 1}} & \multicolumn{1}{c}{\textbf{Priority 2}}  & \multicolumn{1}{c}{\textbf{Male}} & \multicolumn{1}{c}{\textbf{Female}} & \multicolumn{1}{c}{\textbf{Unmarked}} \\ \hline 
\endhead

\hline \multicolumn{3}{l}{{Continued on next page}} \\ \hline
\endfoot

\hline \hline
\endlastfoot
&\textbf{Ambulance location data}: \\\cline{1-2}
1\footnote[1]{\label{1sttablefoot}The parameter estimates are divided by $10^{-5}$ for better visualisation of the estimates}&\hspace{1.75cm}x & 0.098 &0.324&0.390&0.119& 0.128\\
2\footref{1sttablefoot}&\hspace{1.75cm}y &-0.077 &$\cdot$  &-0.002  &-0.015& -0.133\\\hline
&\textbf{Shortest distance to}: \\\cline{1-2}
3\footref{1sttablefoot}&Bus stops&-38.910  &-45.250 &-48.560&-44.100 &-41.486\\
4\footref{1sttablefoot}&Densely populated area &-2.633 &-2.882 &-1.567 & -2.899&-2.898\\
5\footref{1sttablefoot}&Main road networks &-22.740 &-8.121&-8.359  & -11.310& -19.045\\
6\footref{1sttablefoot}&Complete road networks &-65.710&-54.770  &-57.790  &  -64.720&-101.270\\\hline
7&Population density &$\cdot$  &$\cdot$ &$\cdot$  &$\cdot$& $\cdot$\\
8&Benchmark intensity & $\cdot$ &$\cdot$  &$\cdot$  &$\cdot$ & $\cdot$\\\hline
&\textbf{Line densities}: \\\cline{1-2}
9&Complete road networks & 273.826&315.707  &319.707  & 335.114&309.844\\
10&Main road networks &74.692 &$\cdot$  &$\cdot$  & $\cdot$ &  34.869\\
11&Densely populated&-141.438 &-129.585  &-152.325 & -141.335& -159.128\\\hline
12&Bus stops density &601370.222&52605.907  &76587.453 & $\cdot$& 604564.297\\\hline
&\textbf{Population by}:\\\cline{1-2}
&a) \textbf{employment status:}\\
13&Employed & $\cdot$ &$\cdot$ &$\cdot$  & $\cdot$ &$\cdot$\\
14&Unemployed& $\cdot$ &$\cdot$ &$\cdot$  & $\cdot$&$\cdot$\\
&b) \textbf{income status:}\\
15&Below national median income &$\cdot$ &$\cdot$  &$\cdot$  &$\cdot$  &$\cdot$\\
16&Above national median income&$\cdot$ & $\cdot$ &$\cdot$  & $\cdot$ &$\cdot$\\\hline
&c) \textbf{educational level:} \\
17&Pre-high school &$\cdot$ &$\cdot$  &$\cdot$  &$\cdot$  &$\cdot$\\
18&High school &$\cdot$ &$\cdot$  &$\cdot$  &$\cdot$ &$\cdot$\\
19&Post-secondary -- less than 3 years &$\cdot$ &$\cdot$ &$\cdot$  & $\cdot$ &$\cdot$\\ 
20&Post-secondary -- 3 years or longer& $\cdot$ & $\cdot$ &$\cdot$  & $\cdot$ &$\cdot$\\\hline
&d) \textbf{age:}\\
21&0 -- 5 &1.917 &2.120  & 2.780  & 3.782&3.786\\
22&6 -- 9& 0.045 & $\cdot$  &1.197 & $\cdot$&$\cdot$\\
23&10 -- 15 &2.123 &2.552  &3.126 & 0.937&0.924\\
24&16 -- 19 &$\cdot$ &$\cdot$   &$\cdot$  &$\cdot$ &$\cdot$\\
25&20 -- 24 &$\cdot$ &$\cdot$  &$\cdot$  & $\cdot$ &$\cdot$\\
26&25 -- 29 &$\cdot$ &$\cdot$  &$\cdot$  & $\cdot$ &$\cdot$\\ 
27&30 -- 34 &$\cdot$ &$\cdot$  &$\cdot$  & $\cdot$&$\cdot$\\ 
28&35 -- 39 &$\cdot$ &4.098  &$\cdot$  & 9.585&1.289\\
29&40 -- 44 &$\cdot$  &$\cdot$  &$\cdot$  & $\cdot$ &$\cdot$\\
30&45 -- 49 &$\cdot$ &$\cdot$  & $\cdot$ & $\cdot$ &$\cdot$\\ 
31&50 -- 54 &$\cdot$ &$\cdot$  &$\cdot$  & $\cdot$&-1.541\\ 
32&55 -- 59 &$\cdot$ &-0.901  & $\cdot$ & $\cdot$&$\cdot$\\  
33&60 -- 64 &-6.267  &-2.451  &-3.532 & $\cdot$ &-3.204\\
34&65 -- 69 &-0.466 &$\cdot$  &$\cdot$  & $\cdot$ &-0.144\\
35&70 -- 74 &$\cdot$ &$\cdot$  &$\cdot$  & $\cdot$&$\cdot$\\
36&75 -- 79 &$\cdot$ &$\cdot$  &$\cdot$  & $\cdot$ &$\cdot$\\ 
37&80 + &  $\cdot$ &$\cdot$  &$\cdot$  &$\cdot$&$\cdot$\\\hline
&e) \textbf{gender:} \\
38&Female &$\cdot$  &$\cdot$   &$\cdot$  & $\cdot$&$\cdot$\\
39&Male &$\cdot$ &$\cdot$ &$\cdot$  & $\cdot$&$\cdot$\\\hline
&f) \textbf{background}: \\
40&Swedish background &$\cdot$  &$\cdot$   &$\cdot$  & $\cdot$&$\cdot$\\
41&Swedish foreign background&$\cdot$ &$\cdot$  &$\cdot$  & $\cdot$&$\cdot$\\\hline
&g) \textbf{household economic standard}: \\
42&Below the national median &$\cdot$  &$\cdot$  &$\cdot$  & $\cdot$ &$\cdot$\\
43&Above the national median &$\cdot$ &$\cdot$ &$\cdot$  & $\cdot$ &$\cdot$\\\hline
&\textbf{Intercept parameter estimate}: &3.255 &-4.821 &-4.872 & -2.757&6.651\\
\end{longtable}
\end{center}

\begin{figure}[H]
\centering
\includegraphics[width=0.23\textwidth, height=0.2\linewidth]{SpatialIntensityplotp2}
\includegraphics[width=0.23\textwidth, height=0.2\linewidth]{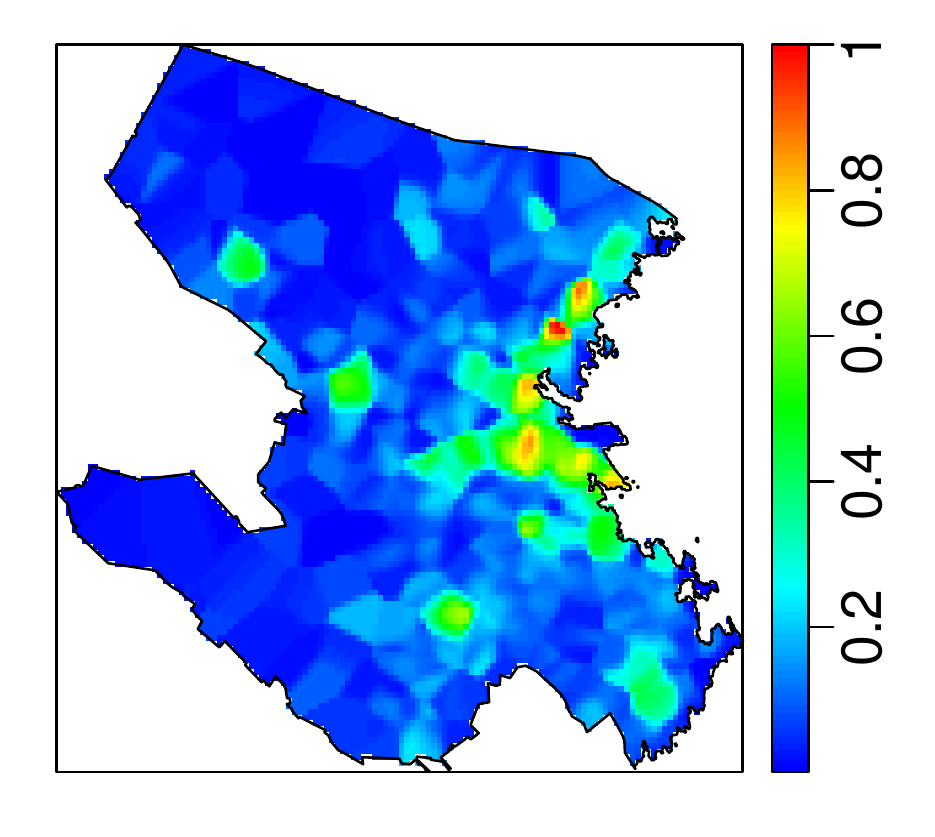}
\includegraphics[width=0.23\textwidth, height=0.2\linewidth]{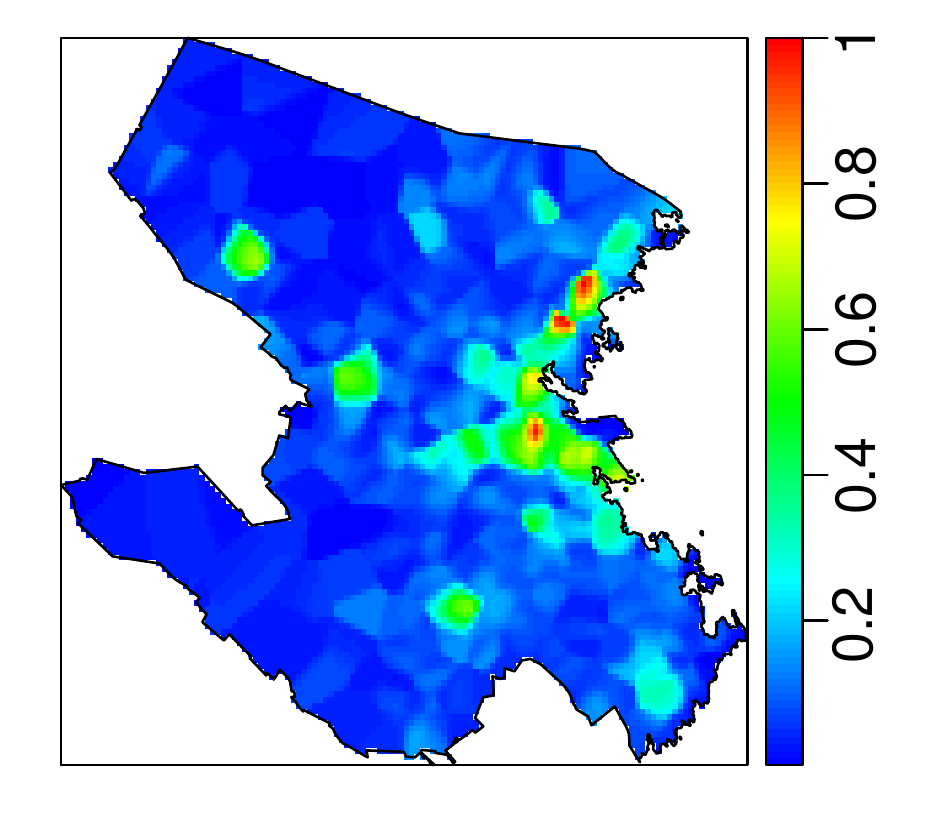}
\includegraphics[width=0.18\textwidth, height=0.2\linewidth]{Dataplotp2}
\caption{The estimated spatial intensities of the emergency alarm call events with priority level 2 using the estimated model based on the original spatial covariates and the first-order interaction terms, the estimated dense and sparse models, which are based on only the original covariates,  respectively. The performance of the estimated models can be evaluated based on the pattern of the events in the observed marginal spatial point pattern. Note that we have scaled the intensity estimates to range between 0 and 1 so that we may compare them more easily.}
\label{OrgrgGd3Compare}
\end{figure}

The three approaches, which are the estimated spatial intensity based on the original spatial covariates and the first-order interaction terms, and the estimated spatial intensities that are obtained using the estimated dense model  and parsimonious/sparse model based on only the original spatial covariates, can all be compared for how well they capture the spatial variation of the events.  Here, we need to remark that when we say dense model and parsimonious/sparse model, we are referring to the estimated models obtained based on only the original spatial covariates, i.e.,~no interaction term in the model setting. Figure \ref{Orgr3} indicates that the spatial distributions of the events are well captured by their corresponding estimated spatial intensities, with the exception that the distribution of the events with priority  level 2 is not well captured by its corresponding estimated spatial intensity.  This can be viewed as the shortcoming of employing the original covariates and the first-order interaction terms in modelling the ambulance call events. Following this, we are interested in comparing the performance of the estimated model based on the original spatial covariates and the first-order interaction terms with that of the estimated dense and sparse models based on only the original covariates on the marginal point pattern corresponding to priority level 2. In a broad sense, the three approaches have similar performance for the marginal point patterns corresponding to the marks priority level 1, male, and female. Here we present the results of the three approaches, i.e.,~the estimated spatial intensity using the estimated model based on the original spatial covariates and the first-order interaction terms; the estimated spatial intensities using the estimated dense and the parsimonious/sparse models based on only the original spatial covariates. Figure \ref{OrgrgGd3Compare} presents the estimated spatial intensities  using the three approaches for the marginal point pattern with priority level 2.
\noindent The figure clearly demonstrates that the dense and sparse models capture the spatial distribution of the events more accurately than the estimated model based on the original spatial covariates and the first-order interaction terms. Those spatial covariates in the dense model with non-zero estimated coefficients that continue to exist in the sparse model, i.e.,~those spatial covariates with non-zero estimated coefficients, have a strong association with the spatial intensity of the events. As the figure shows, the estimated dense and sparse models perform well in capturing the spatial variation of the events, and, relatively speaking,  we can loosely and strongly interpret the results from the dense and sparse models, respectively. \\

\begin{figure}[H]
\centering
\includegraphics[width=0.23\textwidth, height=0.2\linewidth]{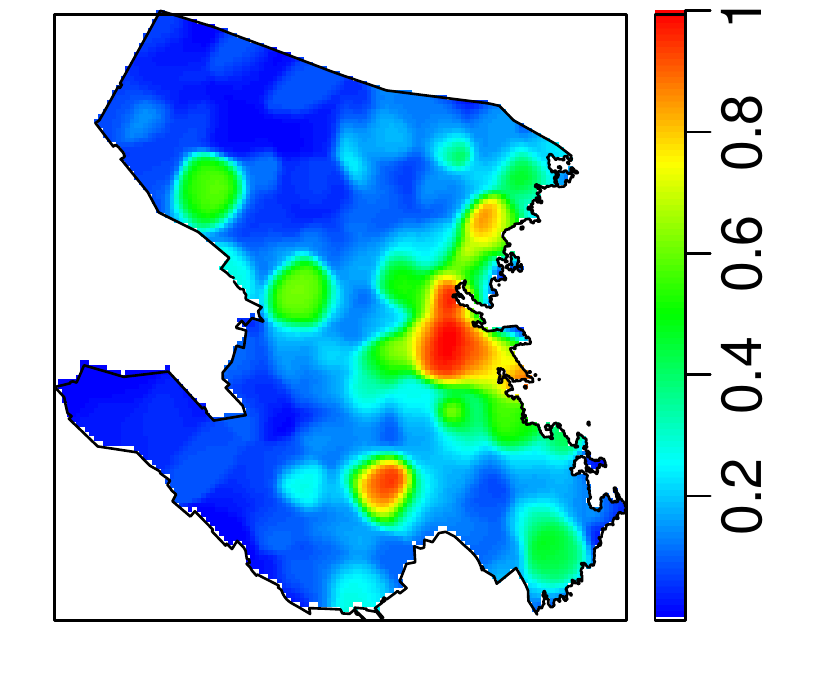}
\includegraphics[width=0.23\textwidth, height=0.2\linewidth]{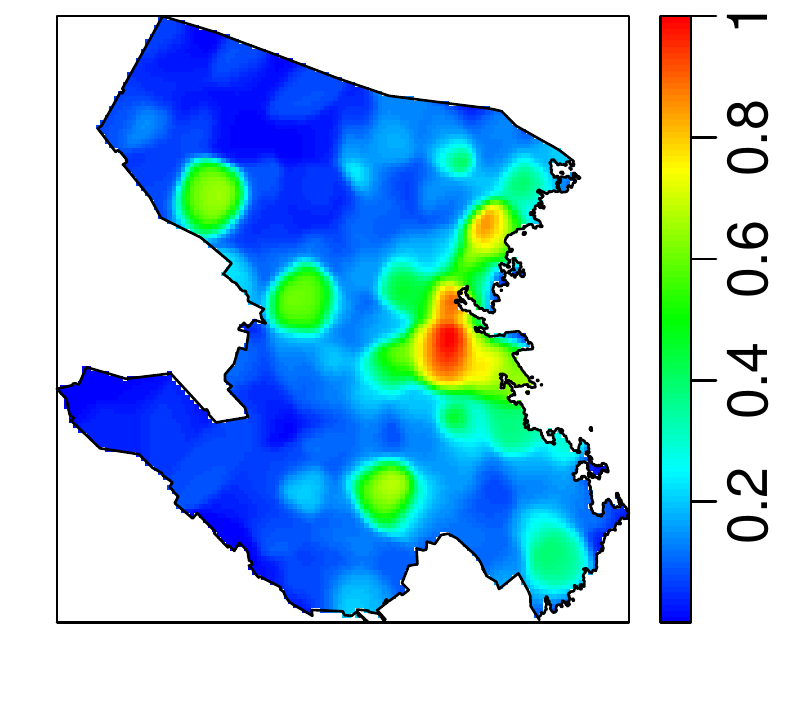}
\includegraphics[width=0.17\textwidth, height=0.2\linewidth]{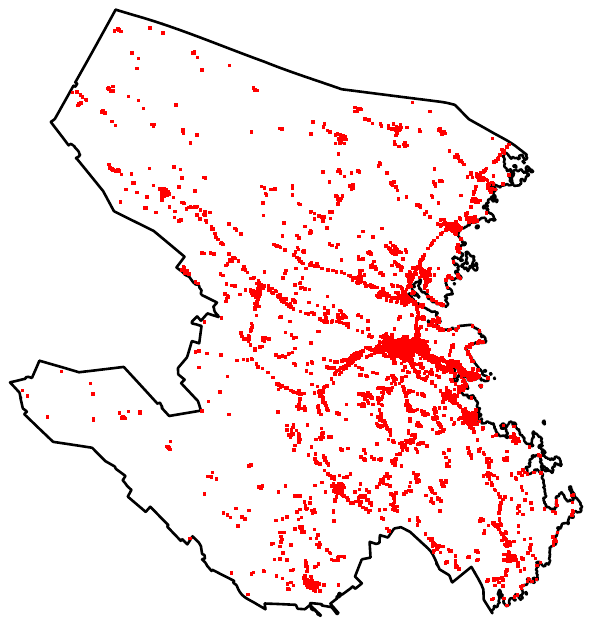}
\caption{The estimated spatial intensities of the unmarked ambulance call data using the estimated dense model (left) and sparse model (middle).  The pattern of the events in the observed unmarked spatial point pattern (right) can be used to assess the performance of the estimated models. To make it easier to compare the intensity estimates, we scaled them to have a range between 0 and 1.}
\label{UnmarkedOrgrg3}
\end{figure}

Next, we examine the modelling of the spatial intensity function for unmarked ambulance call data, i.e.,~we ignore the marks of the ambulance call data,  using only the original spatial covariates.  Here, we are interested in recognising how the marks influence the inclusion of different covariates. As in the modelling of each marked spatial point pattern, the spatial intensity function modelling of the unmarked spatial point pattern based solely on the original spatial covariates is expected to better capture the spatial distribution of the emergency alarm call events than the corresponding model setting including both the original spatial covariates and the first-order interaction terms. Then, using only the original spatial covariates, we focus on modelling the spatial intensity function of the unmarked ambulance call data. The last column in Table \ref{tab:EstimatedModels1} presents the estimated dense model for the unmarked spatial point pattern, while the last column in Table  \ref{tab:EstimatedModels2} displays the corresponding  sparse model for the unmarked spatial point pattern that is obtained by utilising the one-standard-error rule. Figure \ref{UnmarkedOrgrg3} shows the estimated spatial intensities of the unmarked ambulance call data using the estimated dense and sparse models.  
\section{Evaluation of the fitted model}\label{Evaluation}
\begin{figure}[!htpb]
\centering
\includegraphics[width=0.18\textwidth, height=0.17\linewidth]{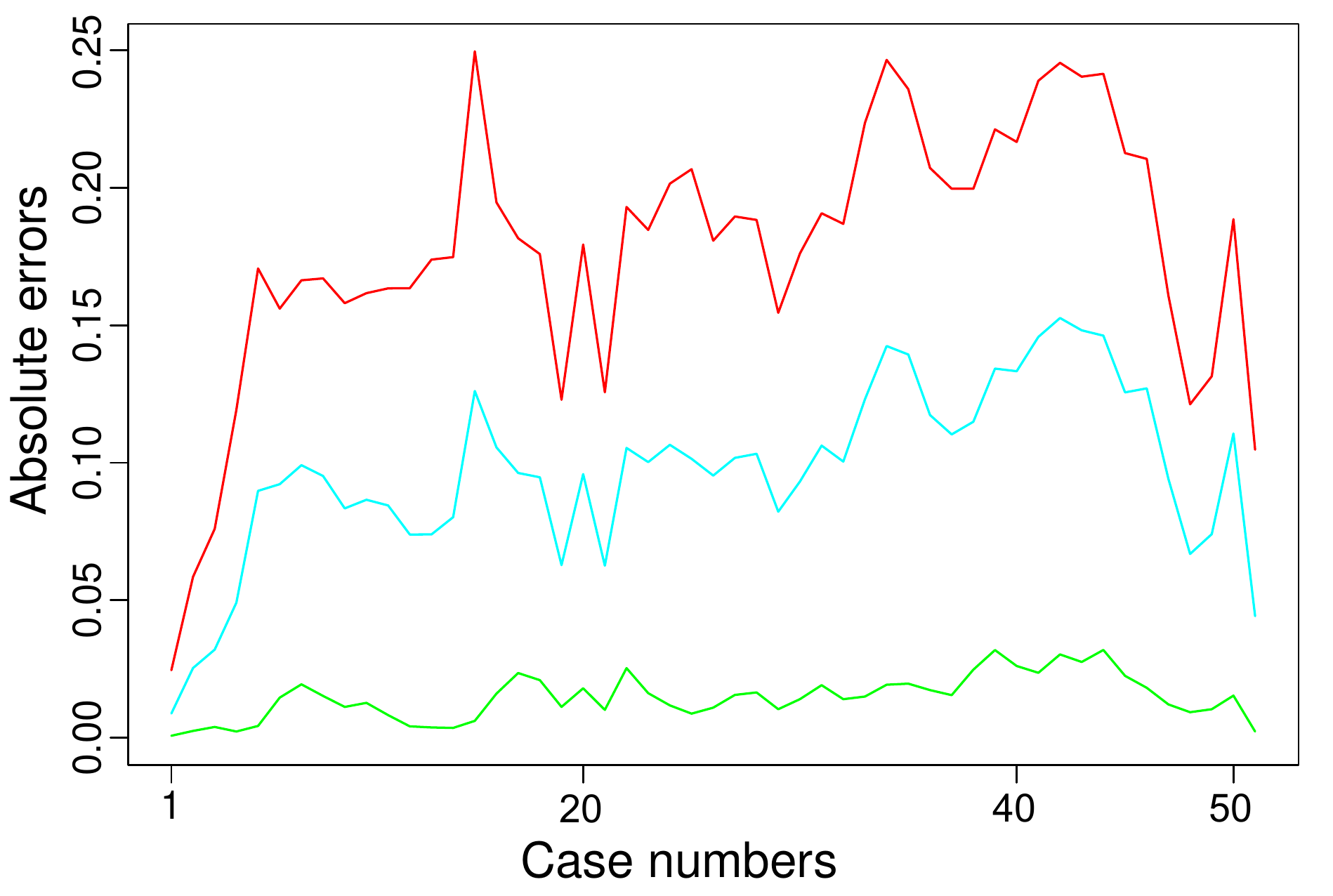}
\includegraphics[width=0.18\textwidth, height=0.17\linewidth]{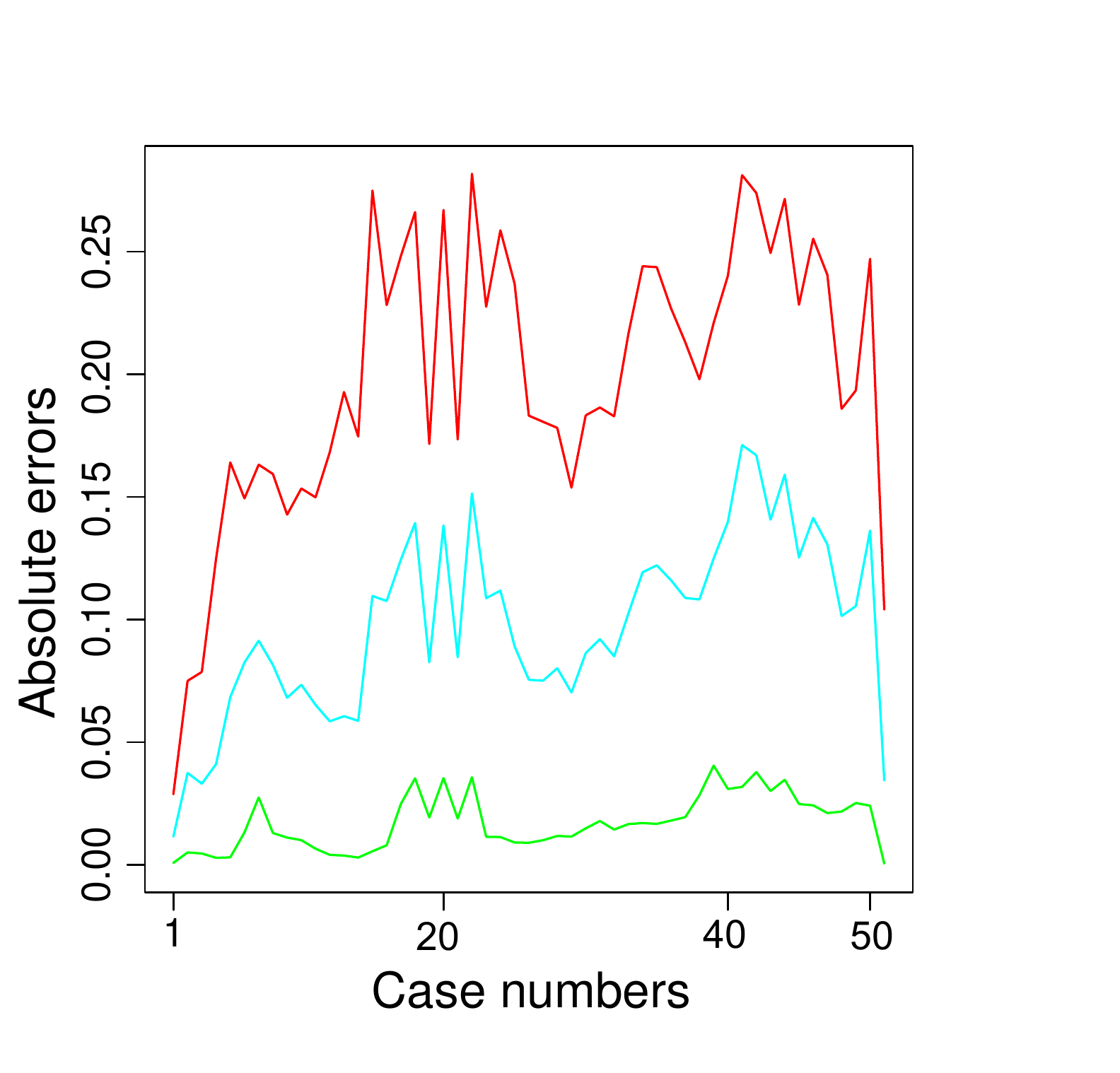}
\includegraphics[width=0.18\textwidth, height=0.17\linewidth]{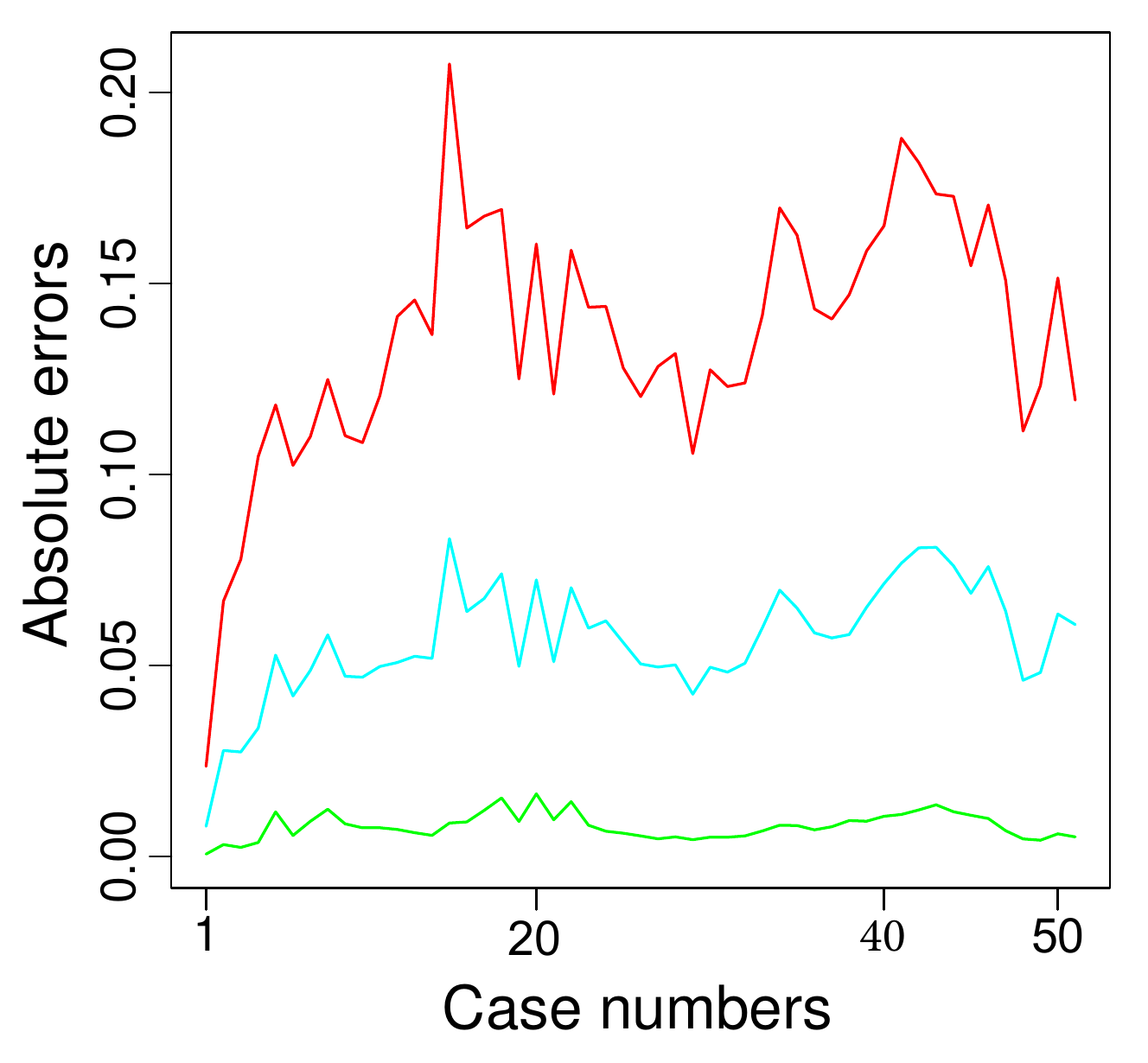}
\includegraphics[width=0.18\textwidth, height=0.17\linewidth]{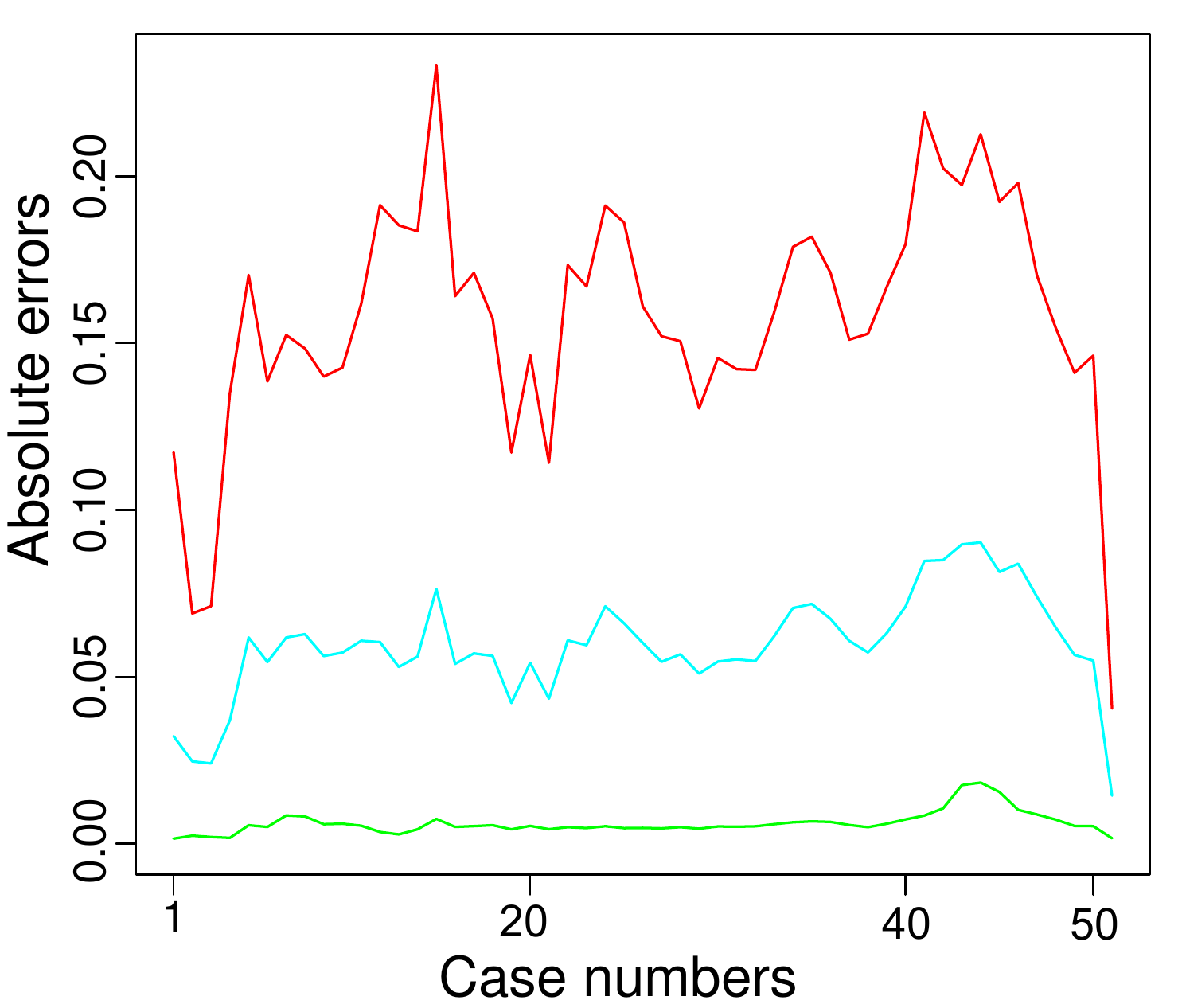}
\includegraphics[width=0.18\textwidth, height=0.17\linewidth]{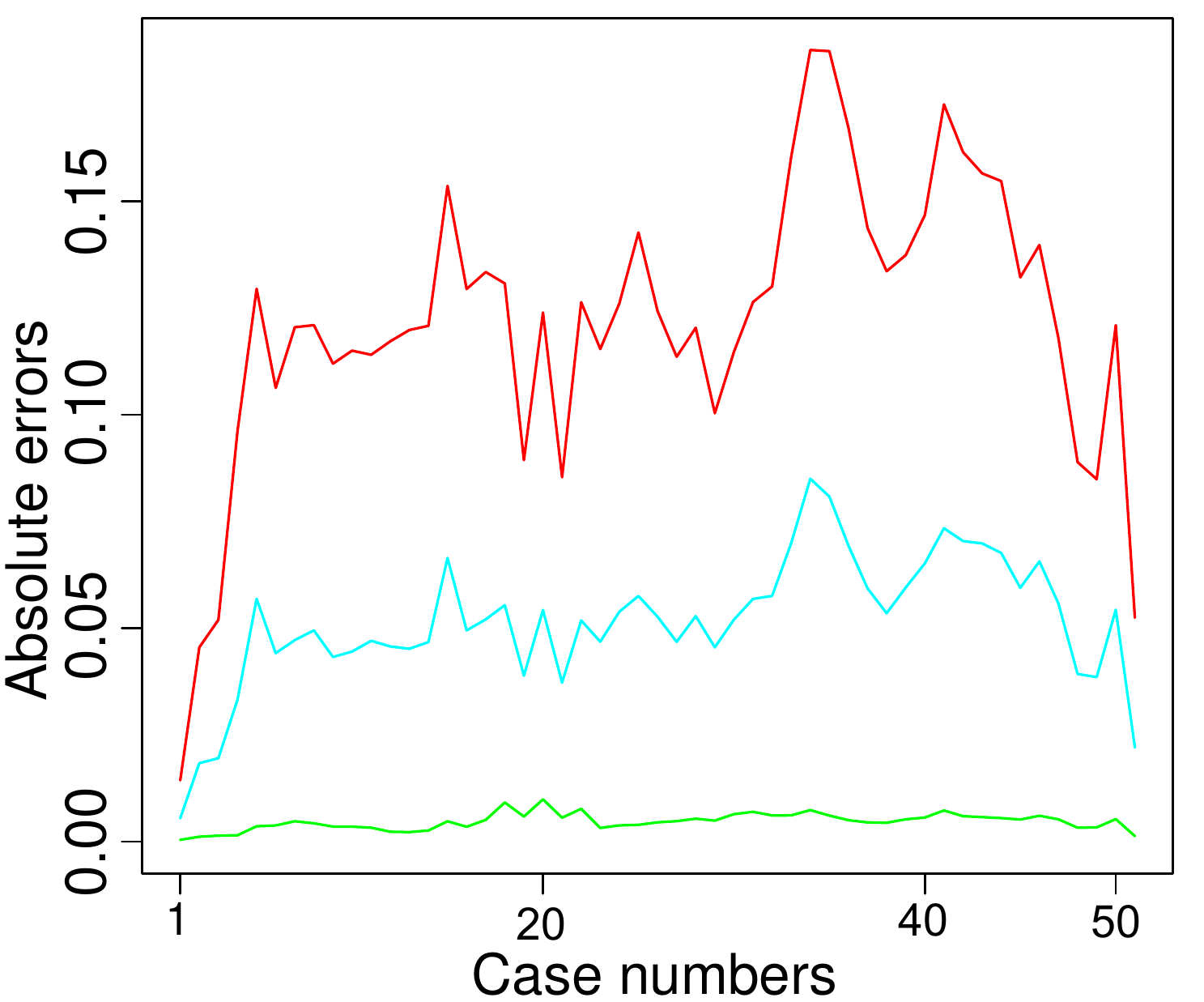}
\caption{The smoothed plots of pixel-wise mean absolute errors, the 5\% and 95\% quantiles of pixel-wise absolute errors of the true intensities and the estimated spatial intensities based on the undersampled point patterns. Row-wise, the plots show the evaluation of the estimated dense models for the point patterns with priority level 1, priority level 2, male, and female, as well as for the unmarked point pattern. The top, middle, and bottom curves in each plot represent the 95\% quantiles, the mean absolute errors, and the 5\% quantiles.}
\label{Densemae}
\end{figure}

\begin{figure}[!htpb]
\centering
\includegraphics[width=0.18\textwidth, height=0.17\linewidth]{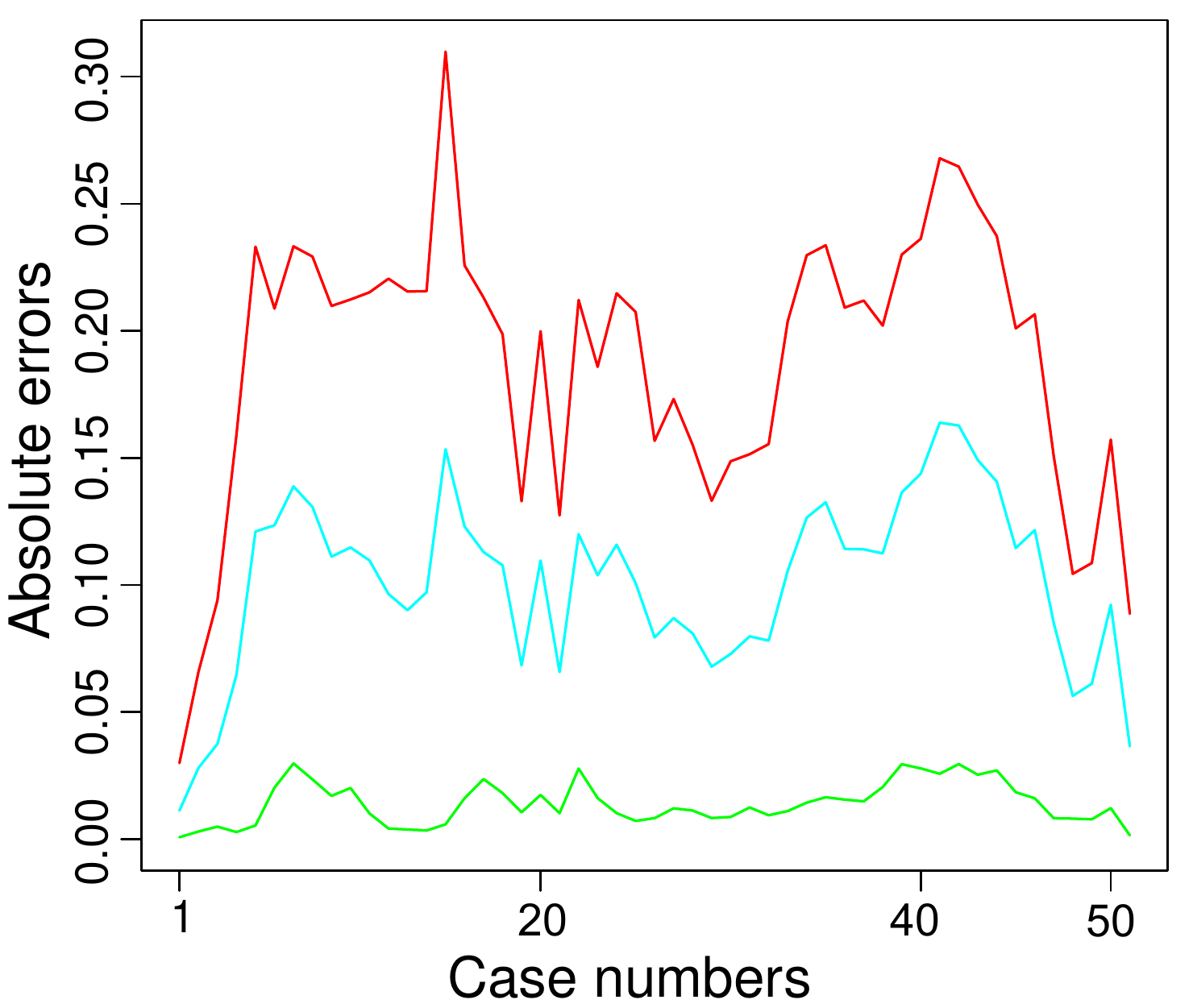}
\includegraphics[width=0.18\textwidth, height=0.17\linewidth]{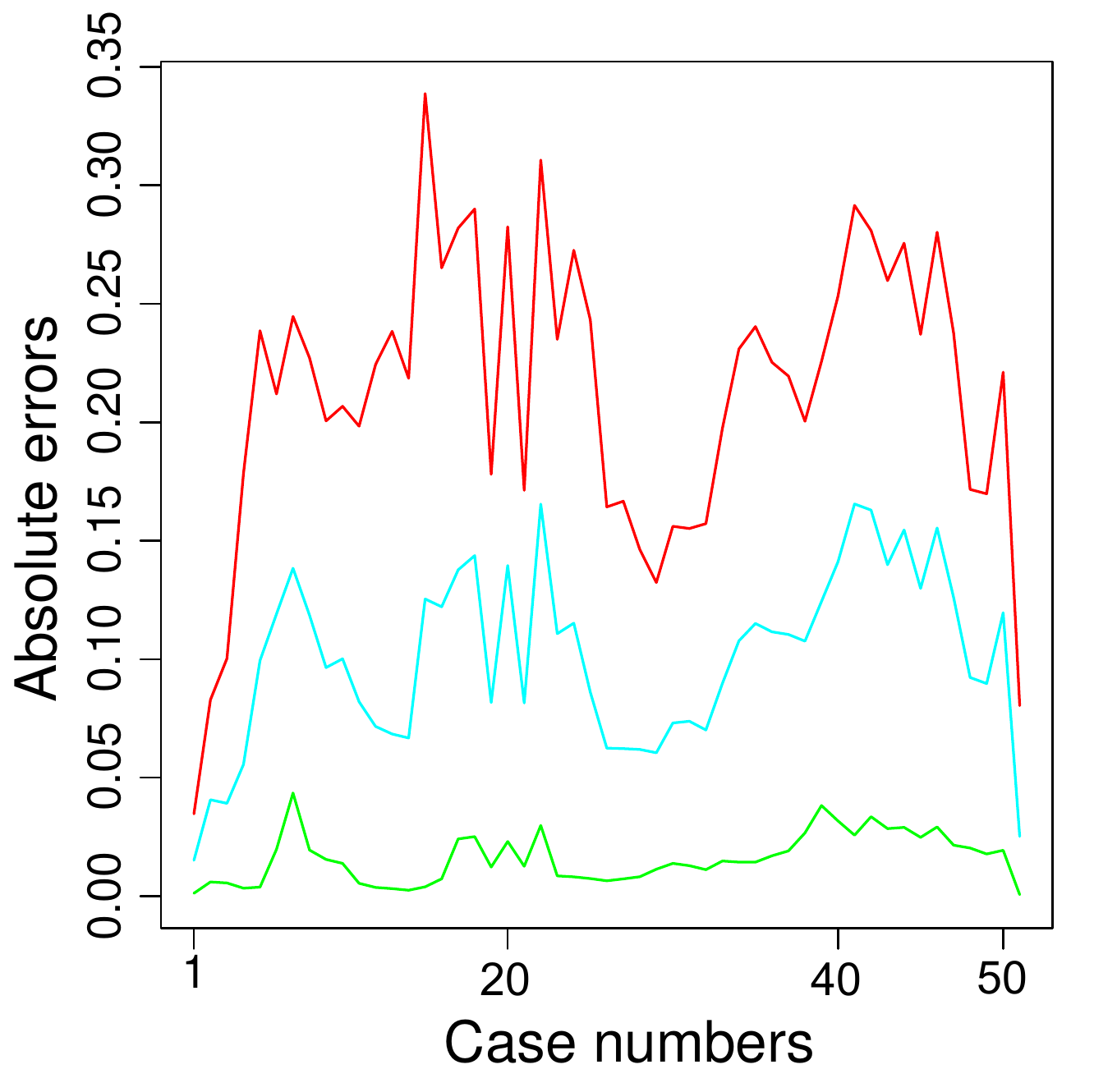}
\includegraphics[width=0.18\textwidth, height=0.17\linewidth]{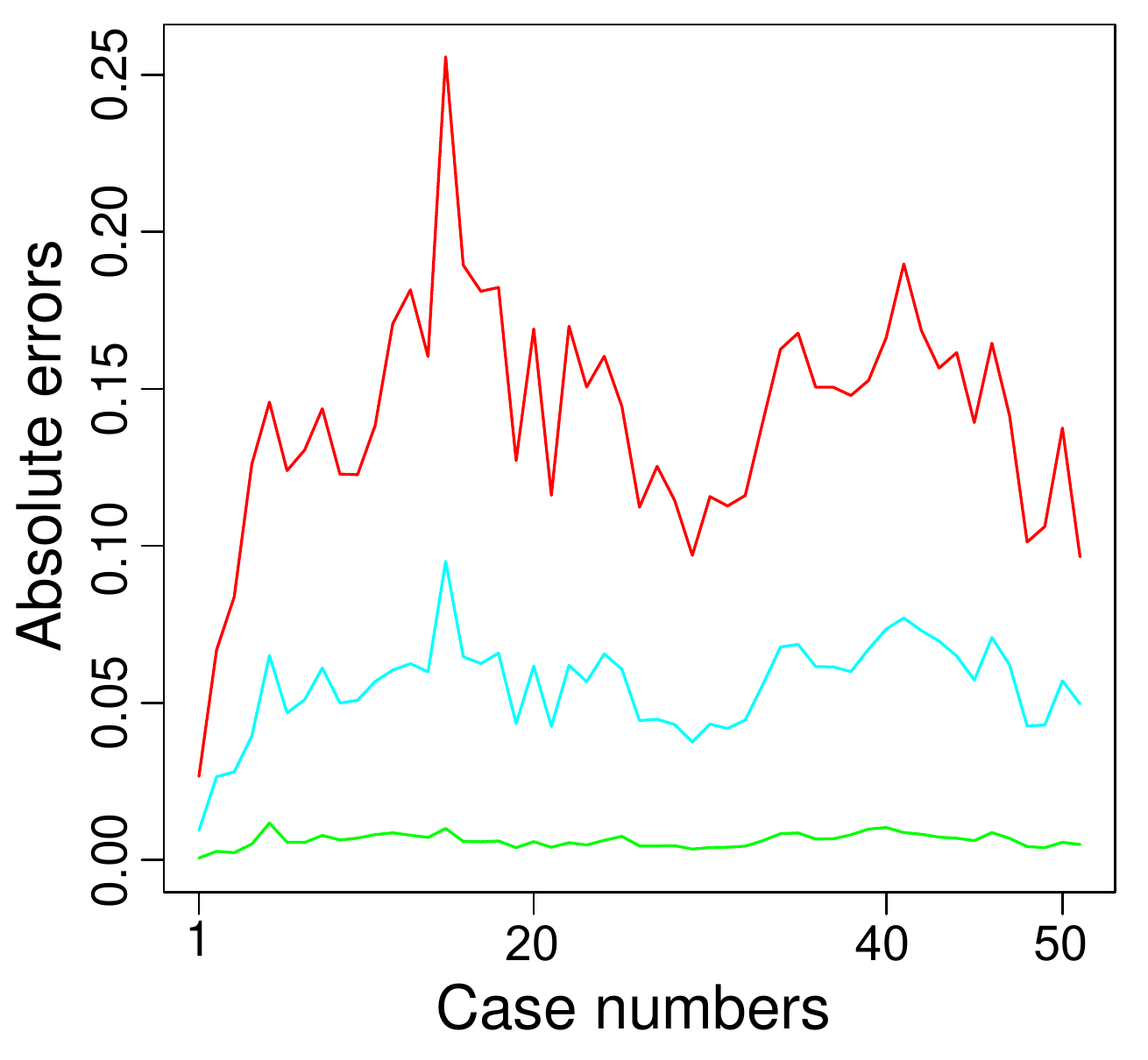}
\includegraphics[width=0.18\textwidth, height=0.17\linewidth]{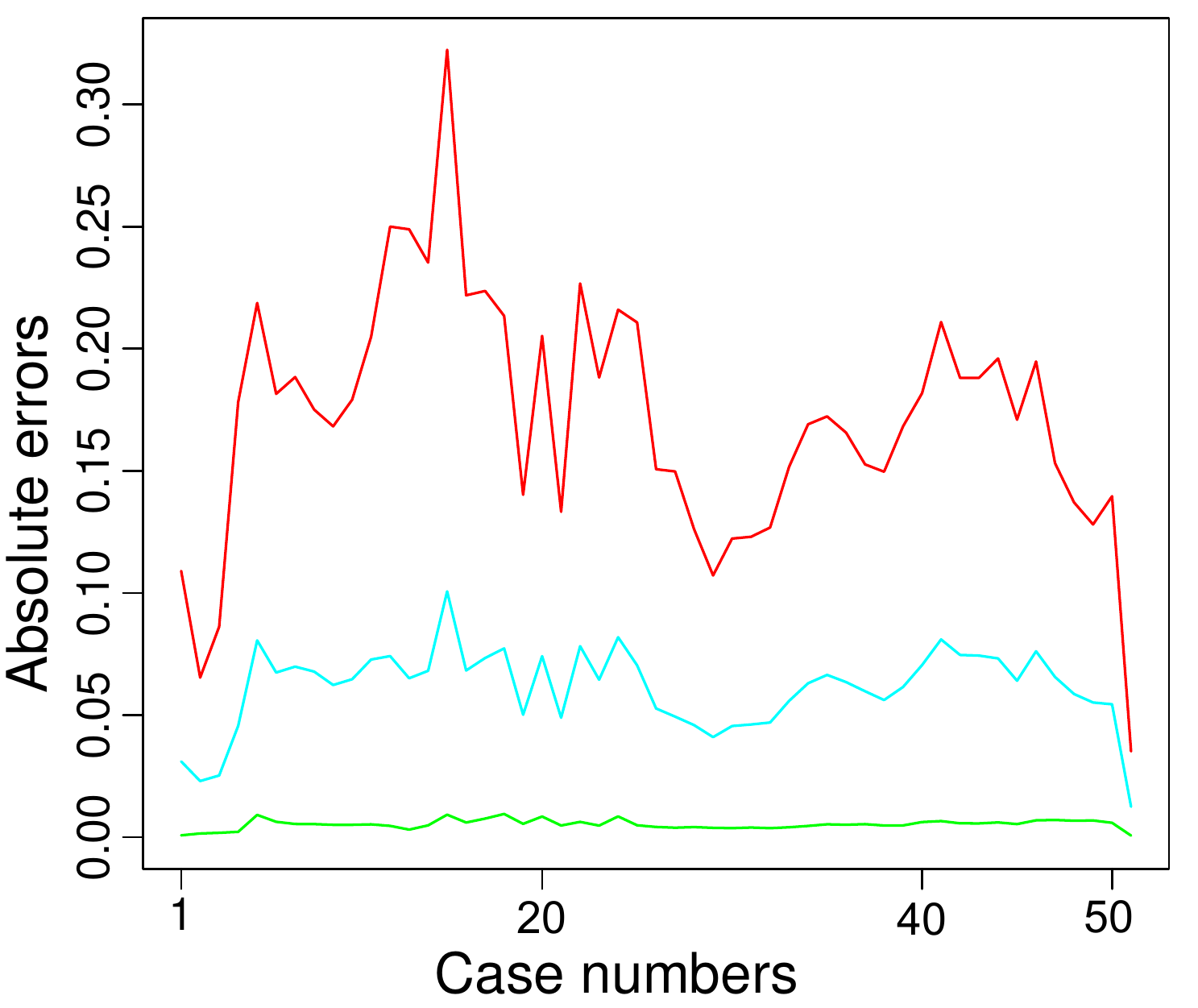}
\includegraphics[width=0.18\textwidth, height=0.17\linewidth]{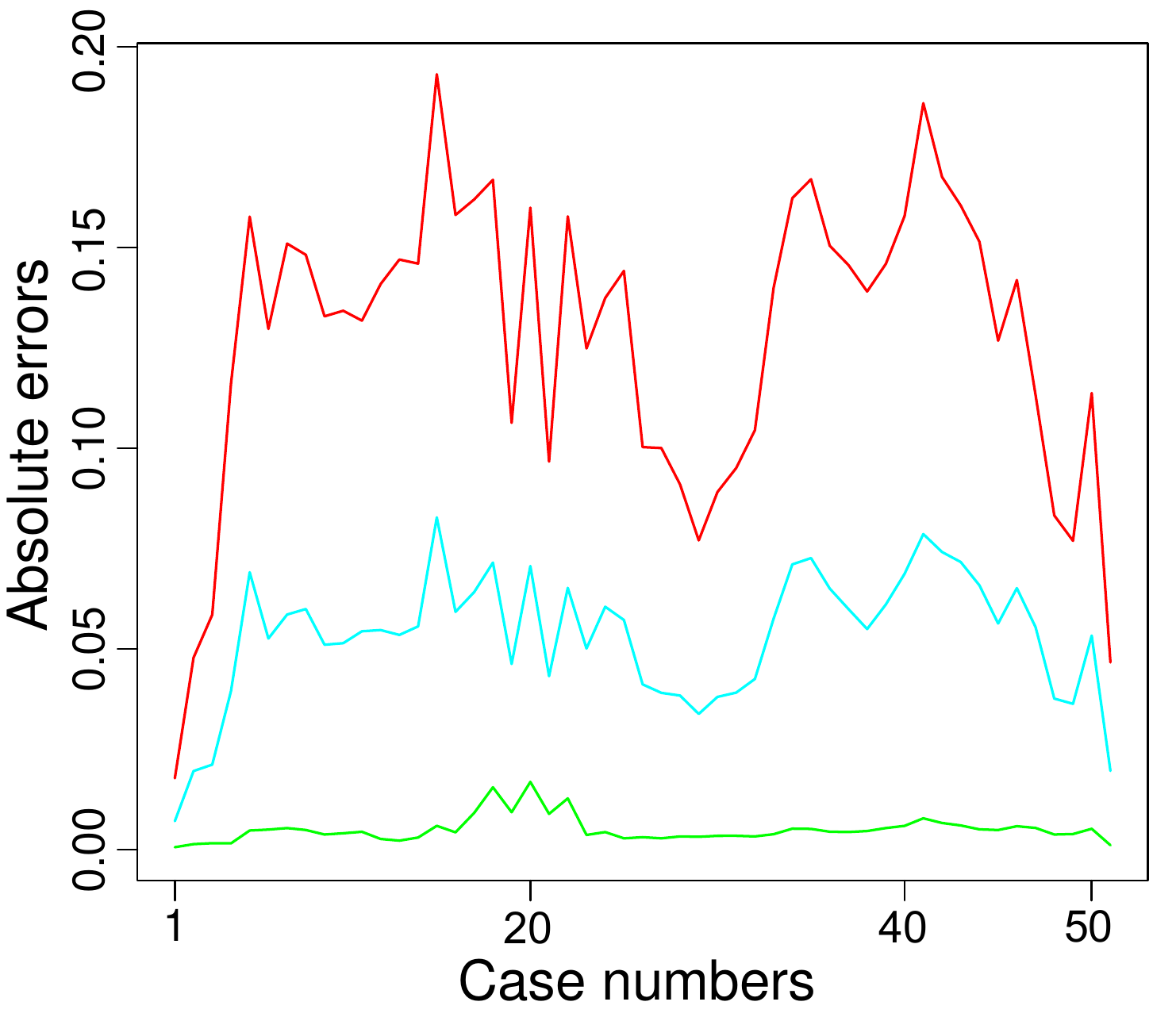}
\caption{The smoothed plots of pixel-wise mean absolute errors, the 5\% and 95\% quantiles of pixel-wise absolute errors of the true intensities and the estimated spatial intensities based on the undersampled point patterns. Row-wise, the plots show the evaluation of the estimated sparse models for the point patterns with priority level 1, priority level 2, male, and female, as well as for the unmarked point pattern. The top, middle, and bottom curves in each plot represent the 95\% quantiles, the mean absolute errors, and the 5\% quantiles.}
\label{Sparsemae}
\end{figure}

\begin{figure}[H]
\centering
\includegraphics[width=0.18\textwidth, height=0.17\linewidth]{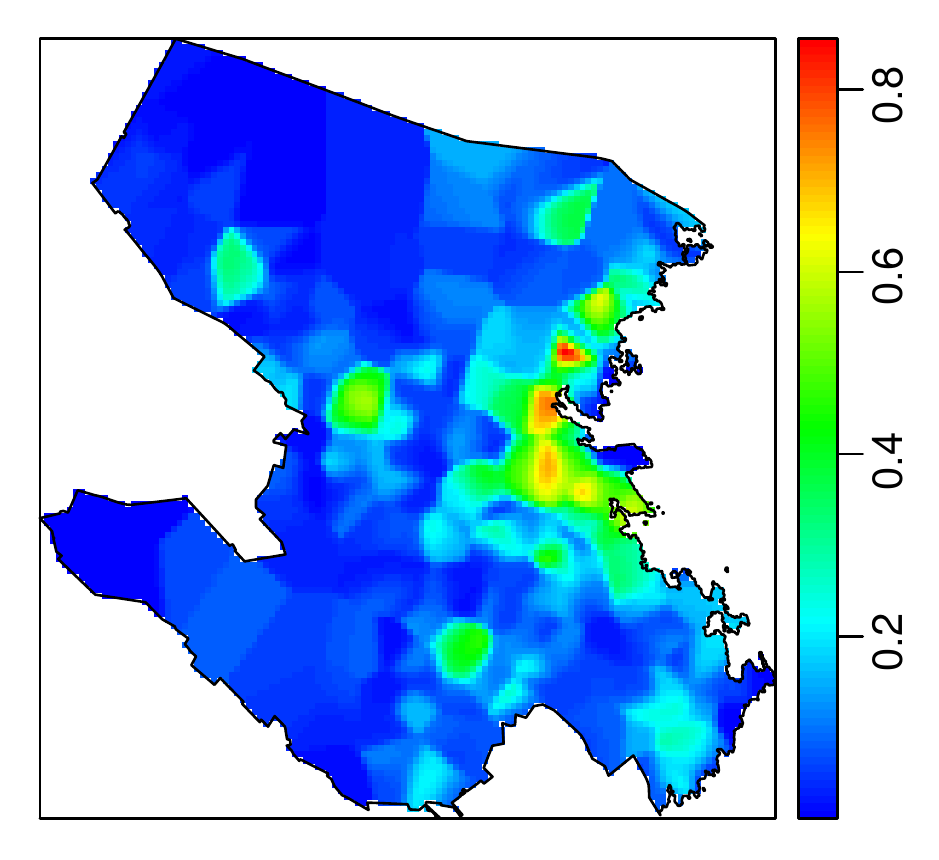}
\includegraphics[width=0.18\textwidth, height=0.17\linewidth]{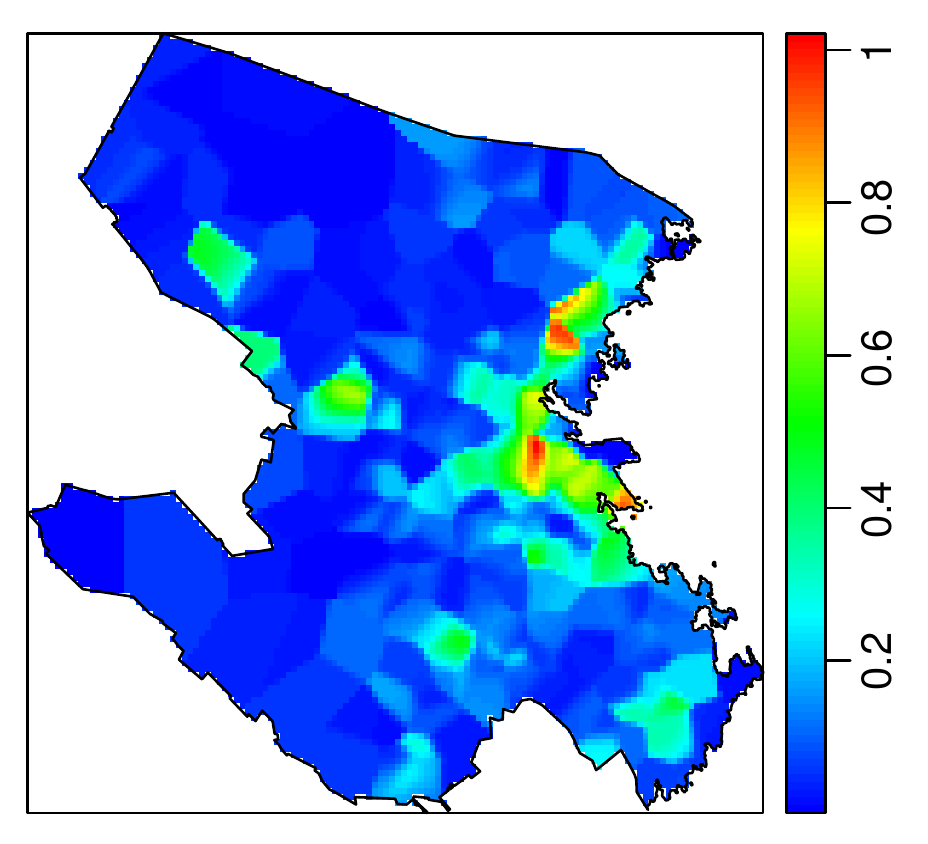}
\includegraphics[width=0.18\textwidth, height=0.17\linewidth]{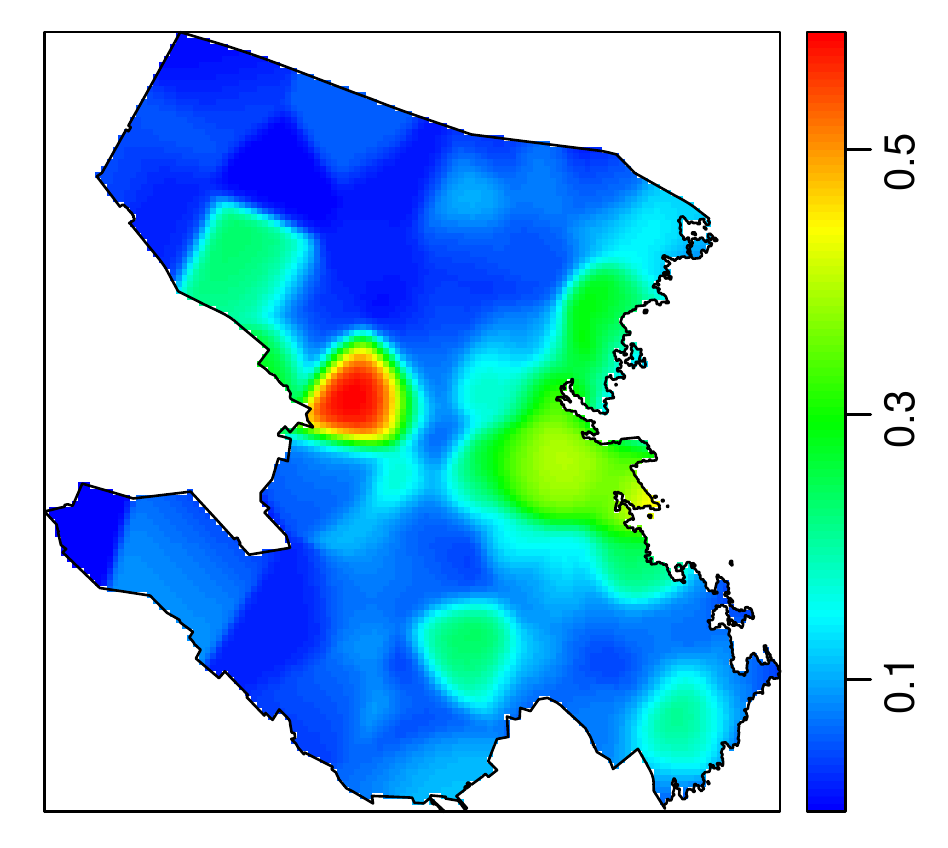}
\includegraphics[width=0.18\textwidth, height=0.17\linewidth]{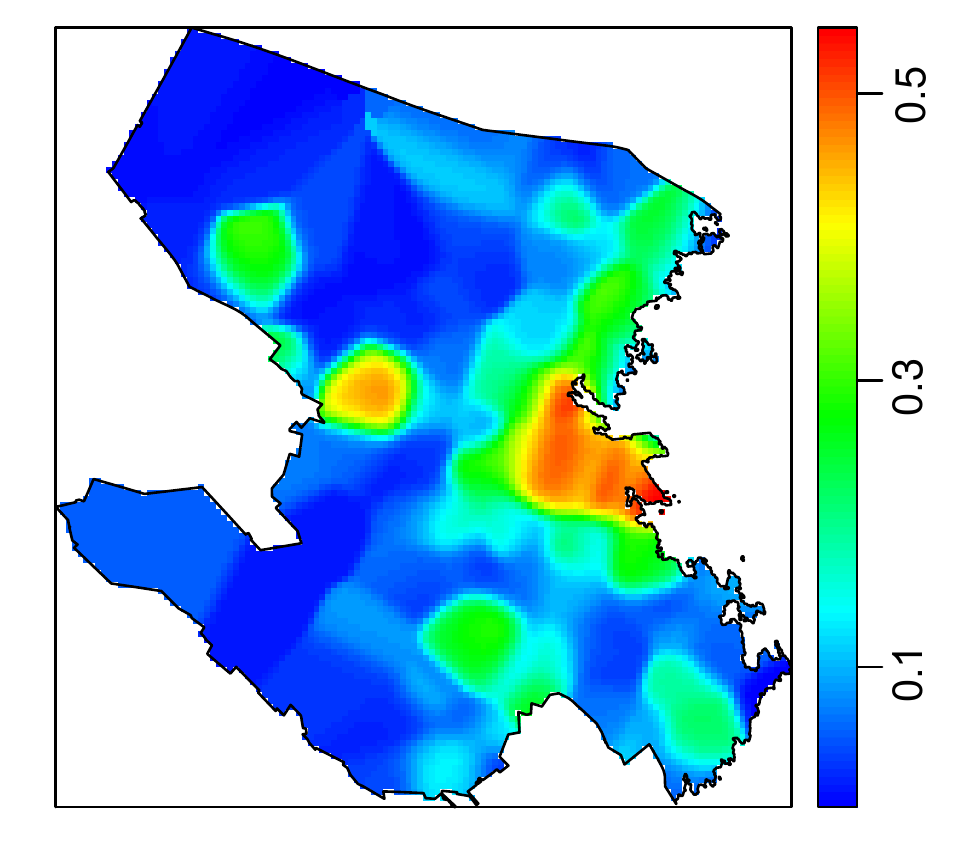}
\includegraphics[width=0.18\textwidth, height=0.17\linewidth]{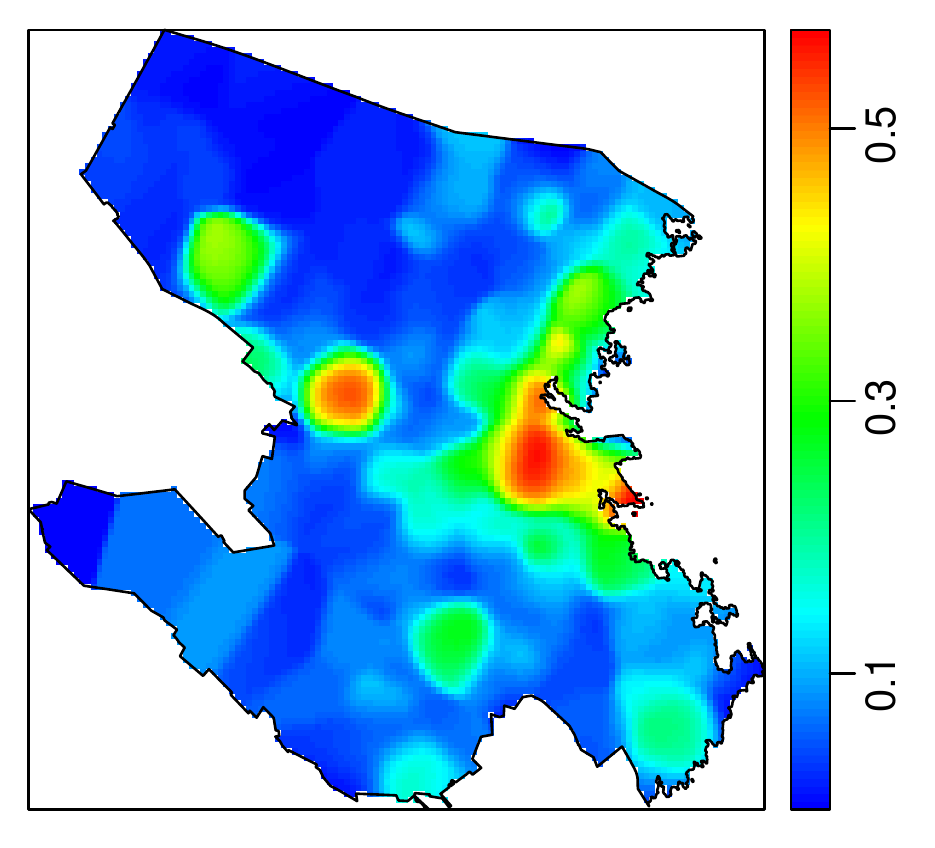}\\
\includegraphics[width=0.18\textwidth, height=0.17\linewidth]{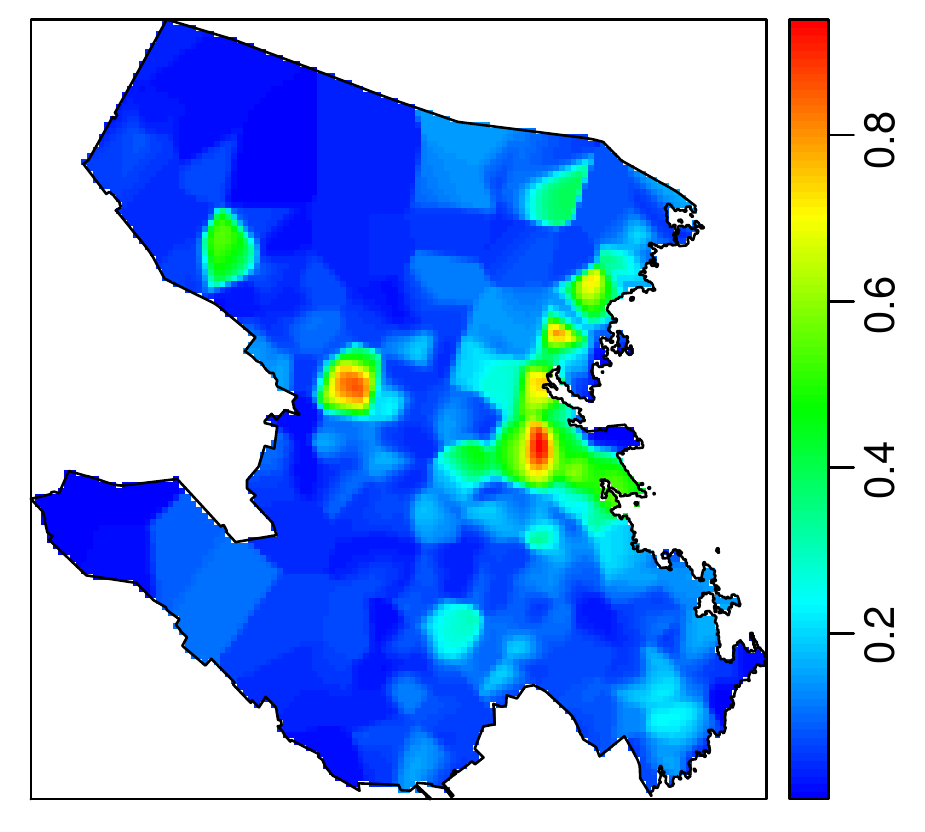}
\includegraphics[width=0.18\textwidth, height=0.17\linewidth]{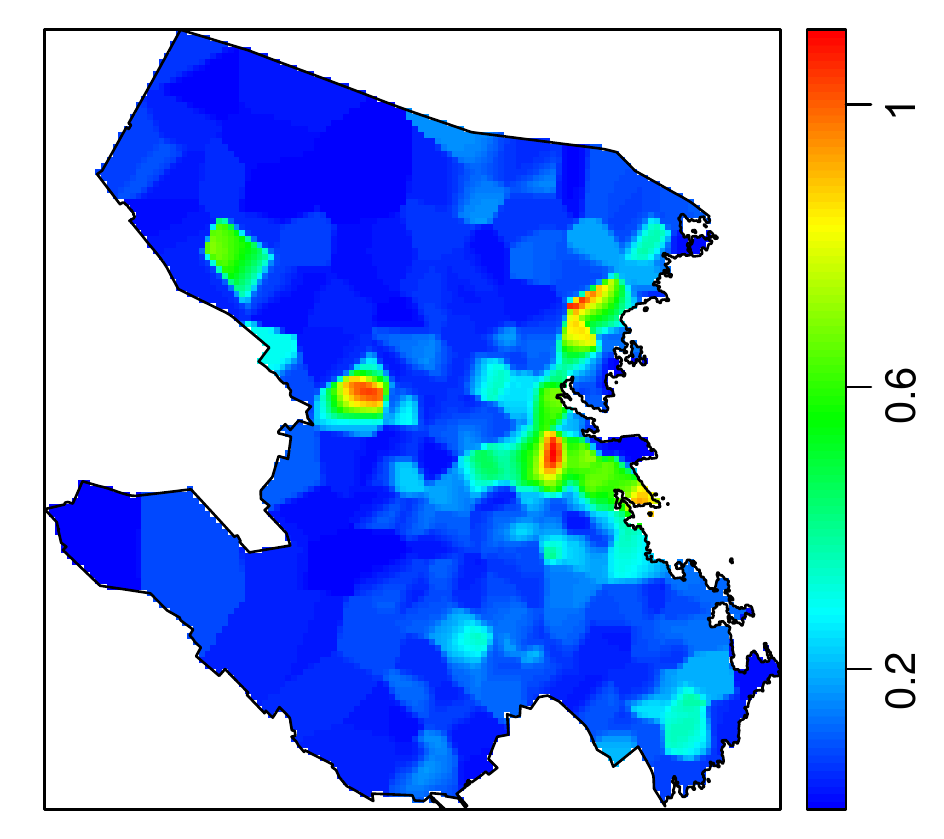}
\includegraphics[width=0.18\textwidth, height=0.17\linewidth]{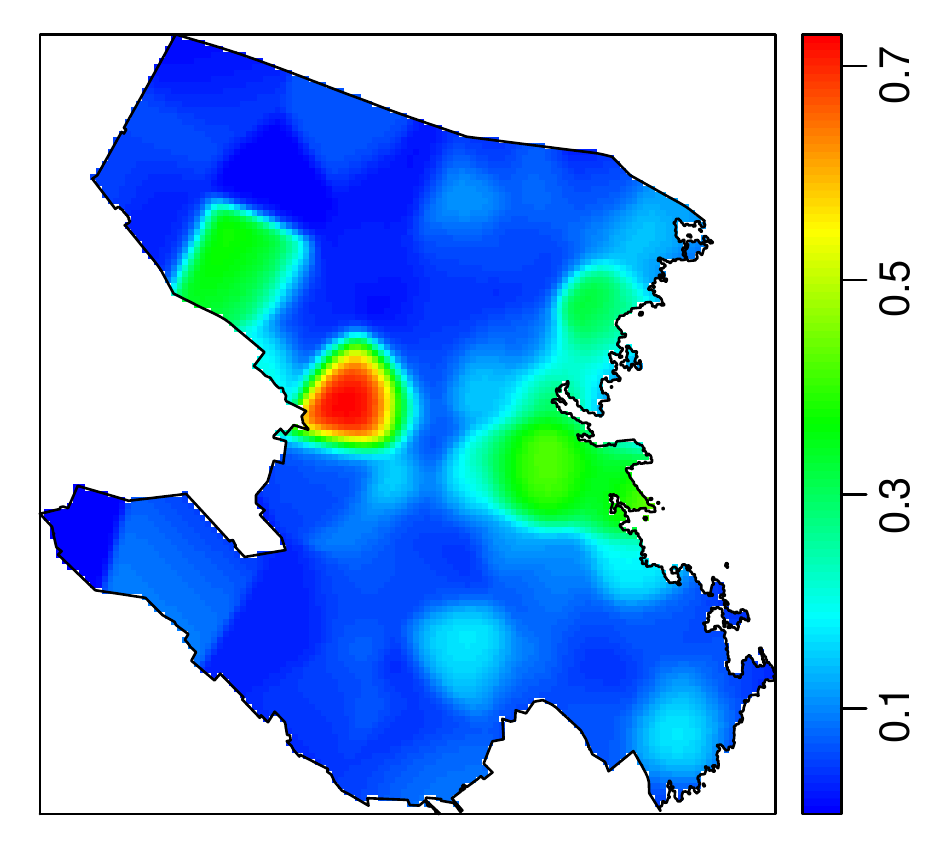}
\includegraphics[width=0.18\textwidth, height=0.17\linewidth]{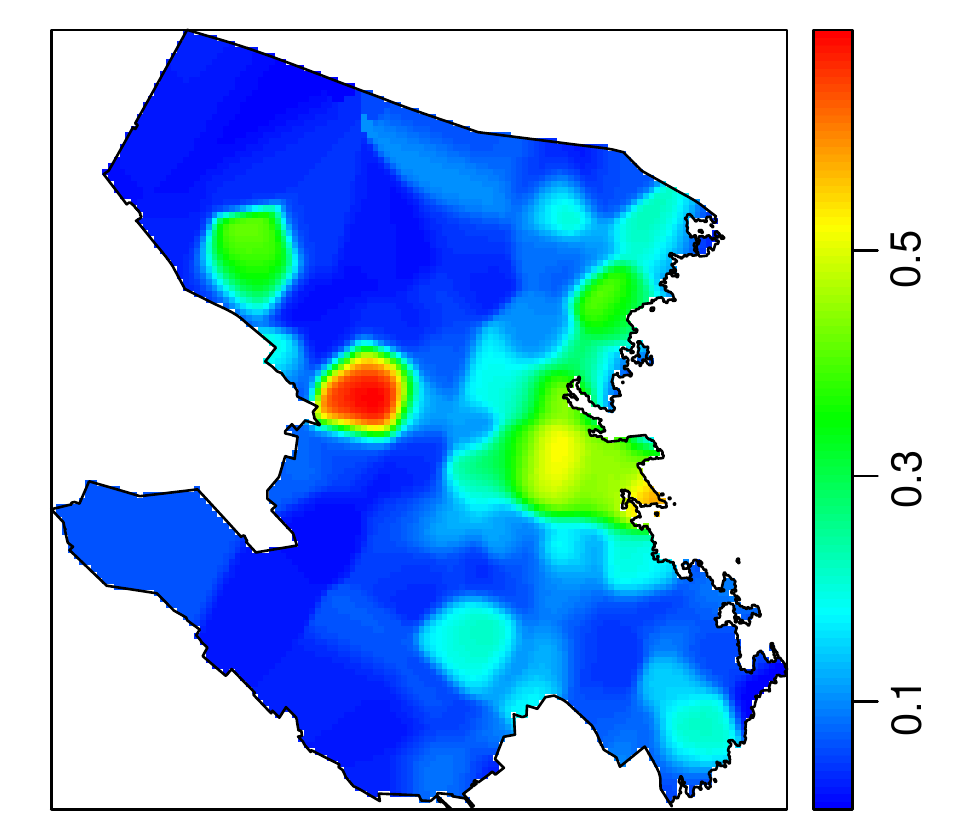}
\includegraphics[width=0.18\textwidth, height=0.17\linewidth]{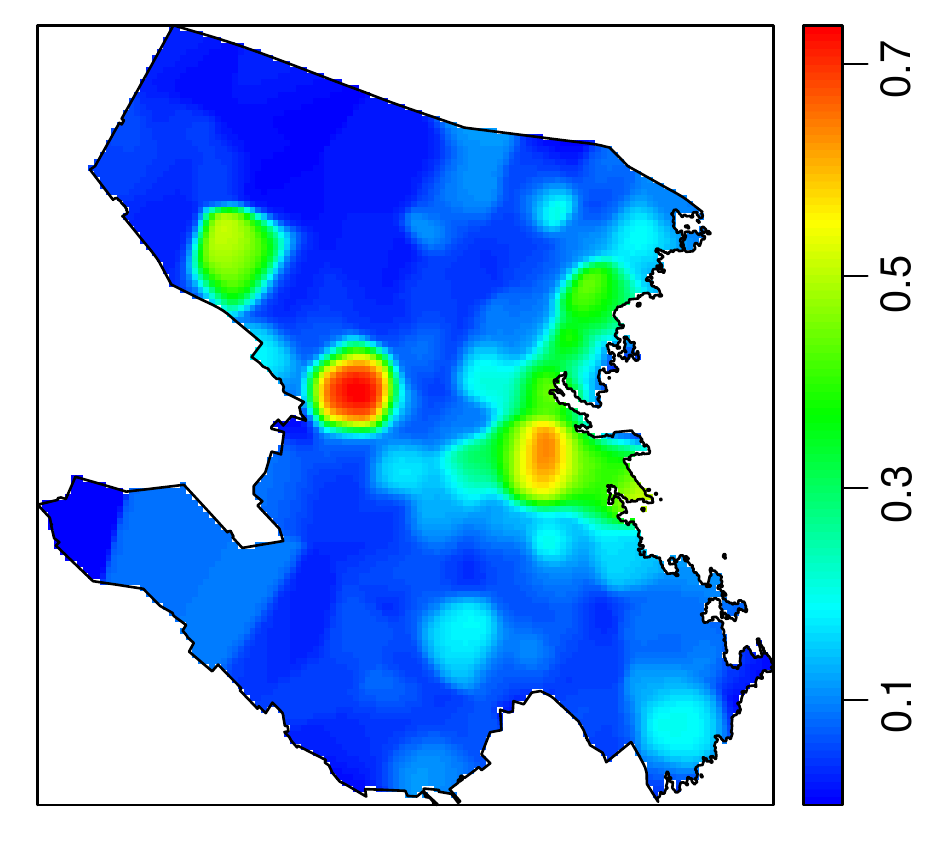}\\
\includegraphics[width=0.18\textwidth, height=0.17\linewidth]{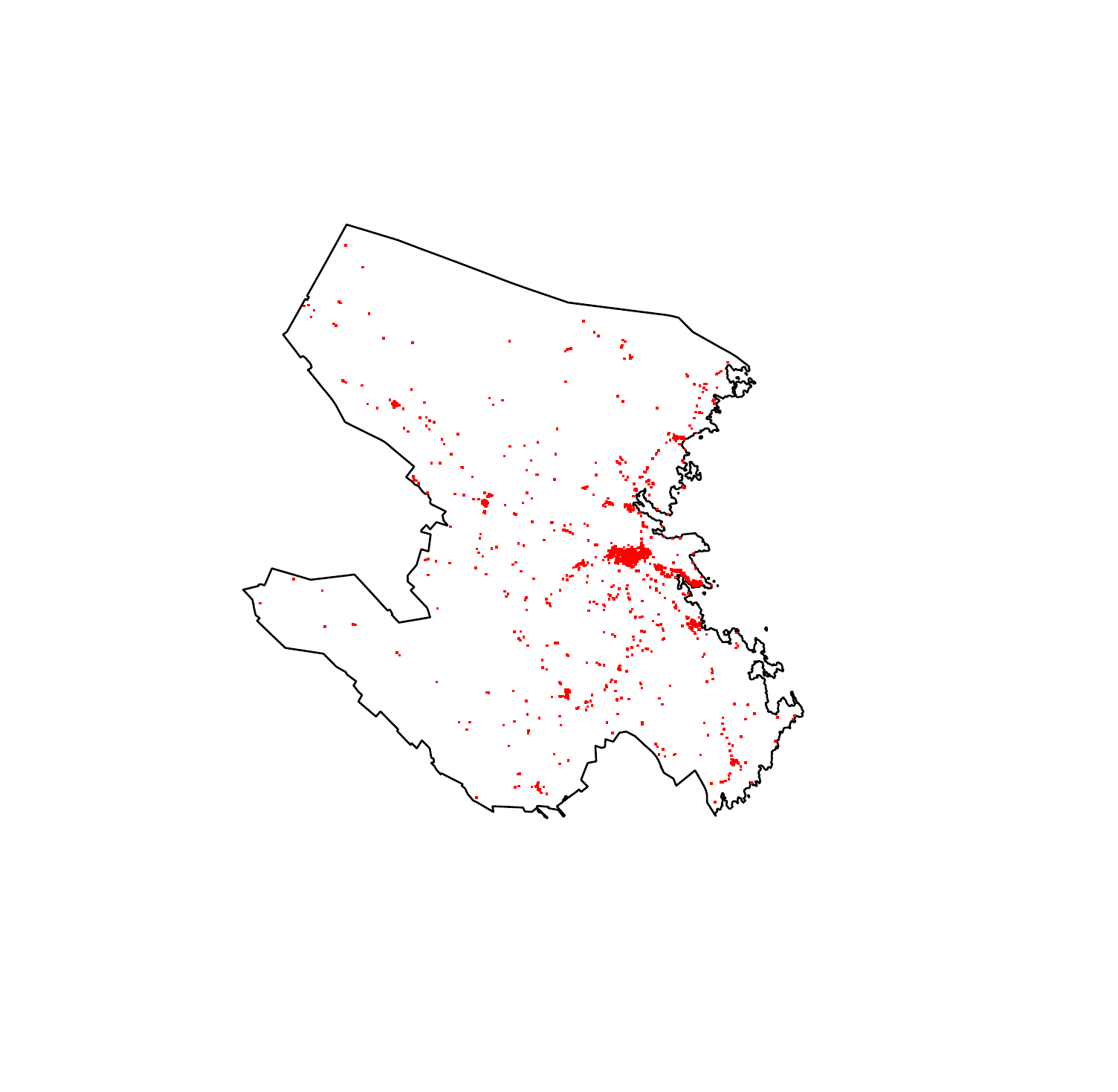}
\includegraphics[width=0.18\textwidth, height=0.17\linewidth]{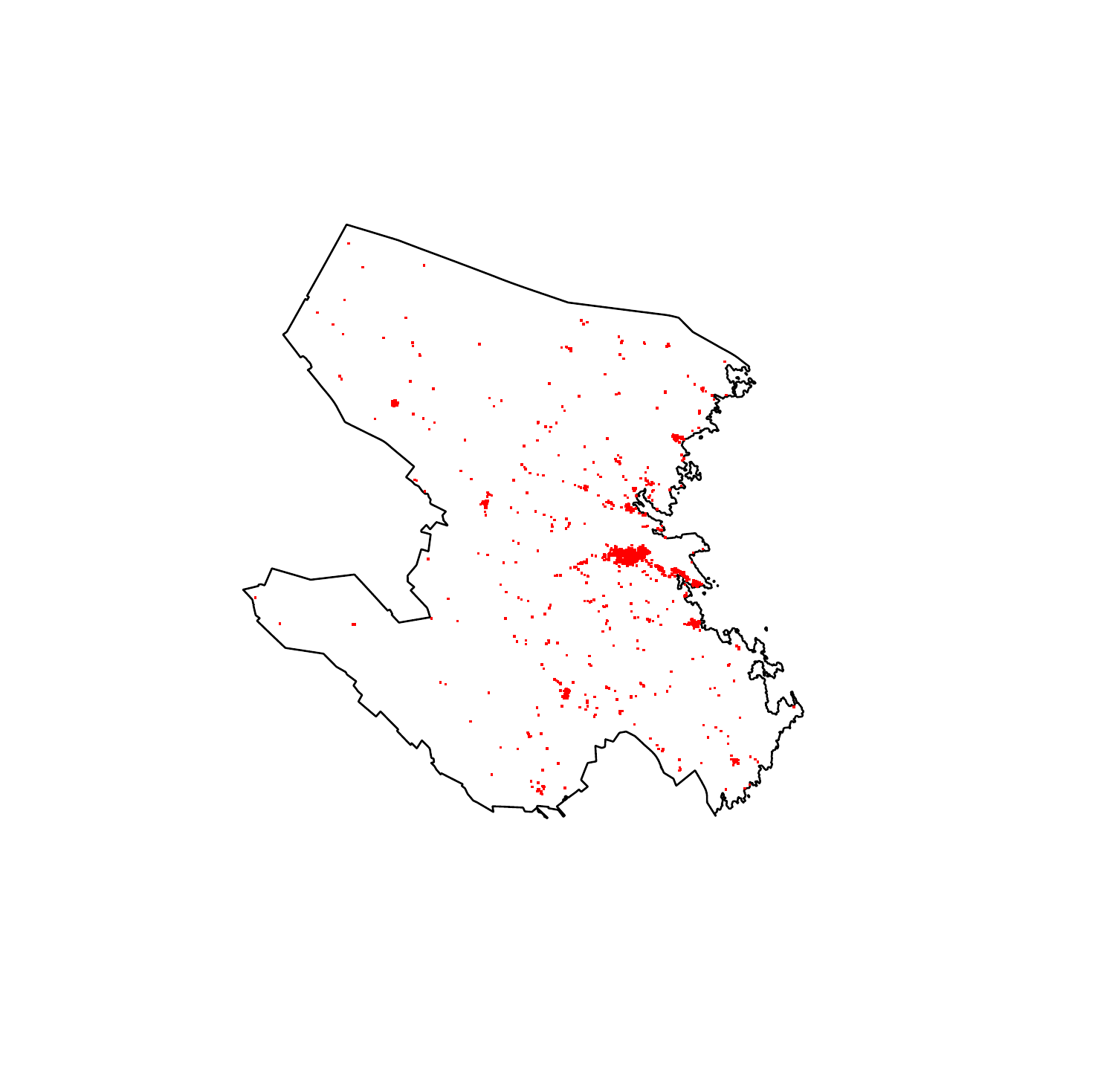}
\includegraphics[width=0.18\textwidth, height=0.17\linewidth]{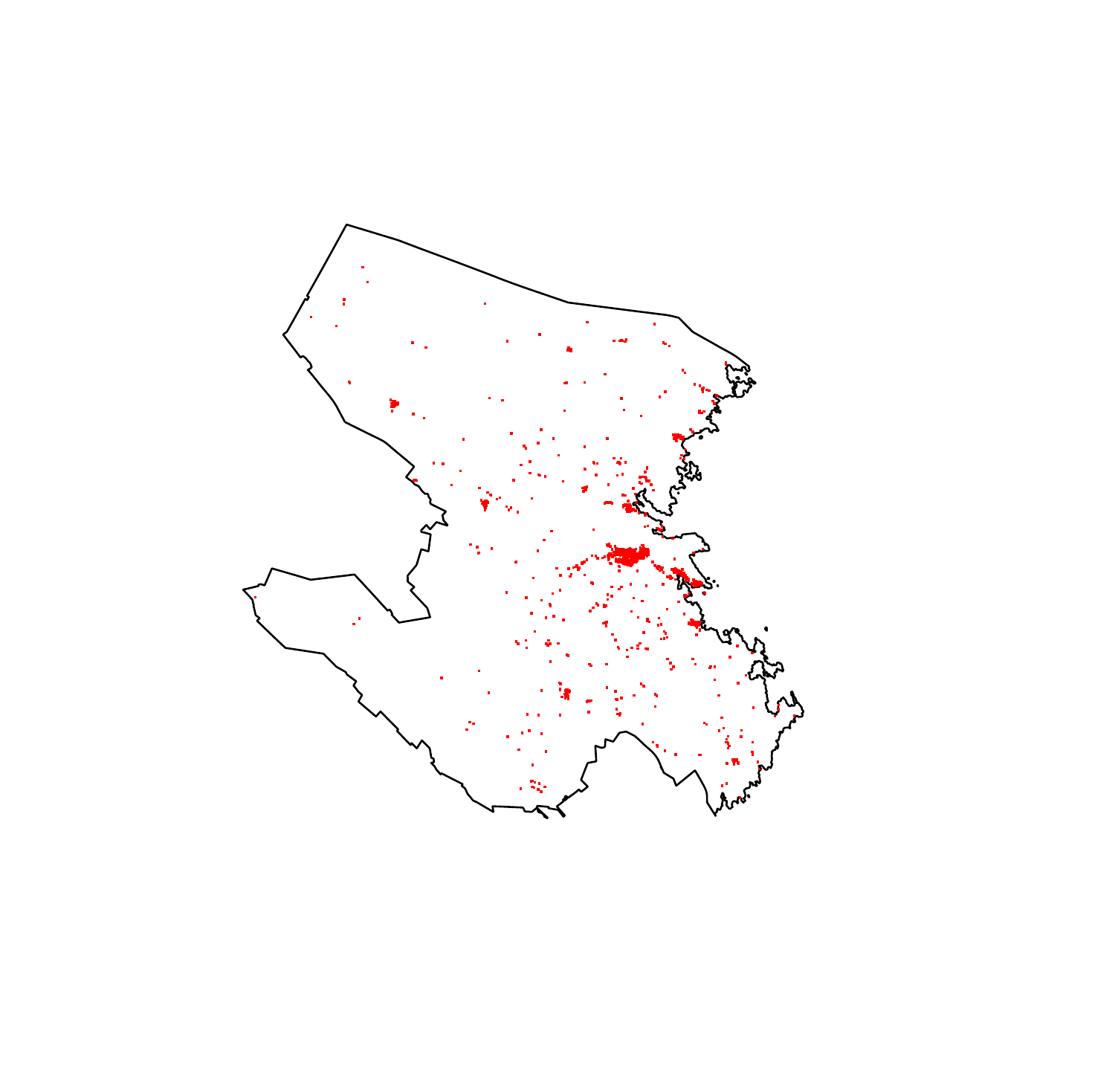}
\includegraphics[width=0.18\textwidth, height=0.17\linewidth]{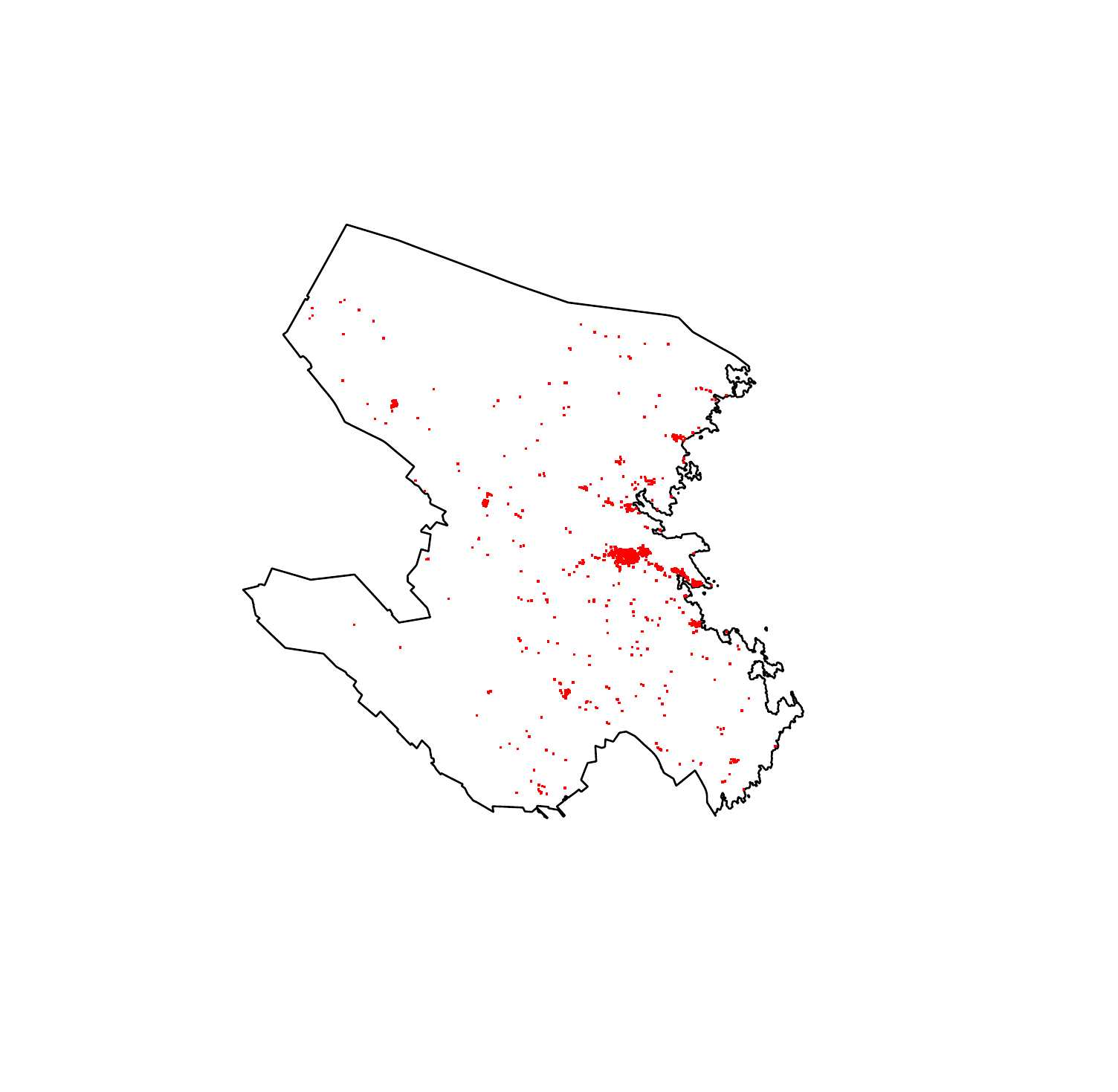}
\includegraphics[width=0.18\textwidth, height=0.17\linewidth]{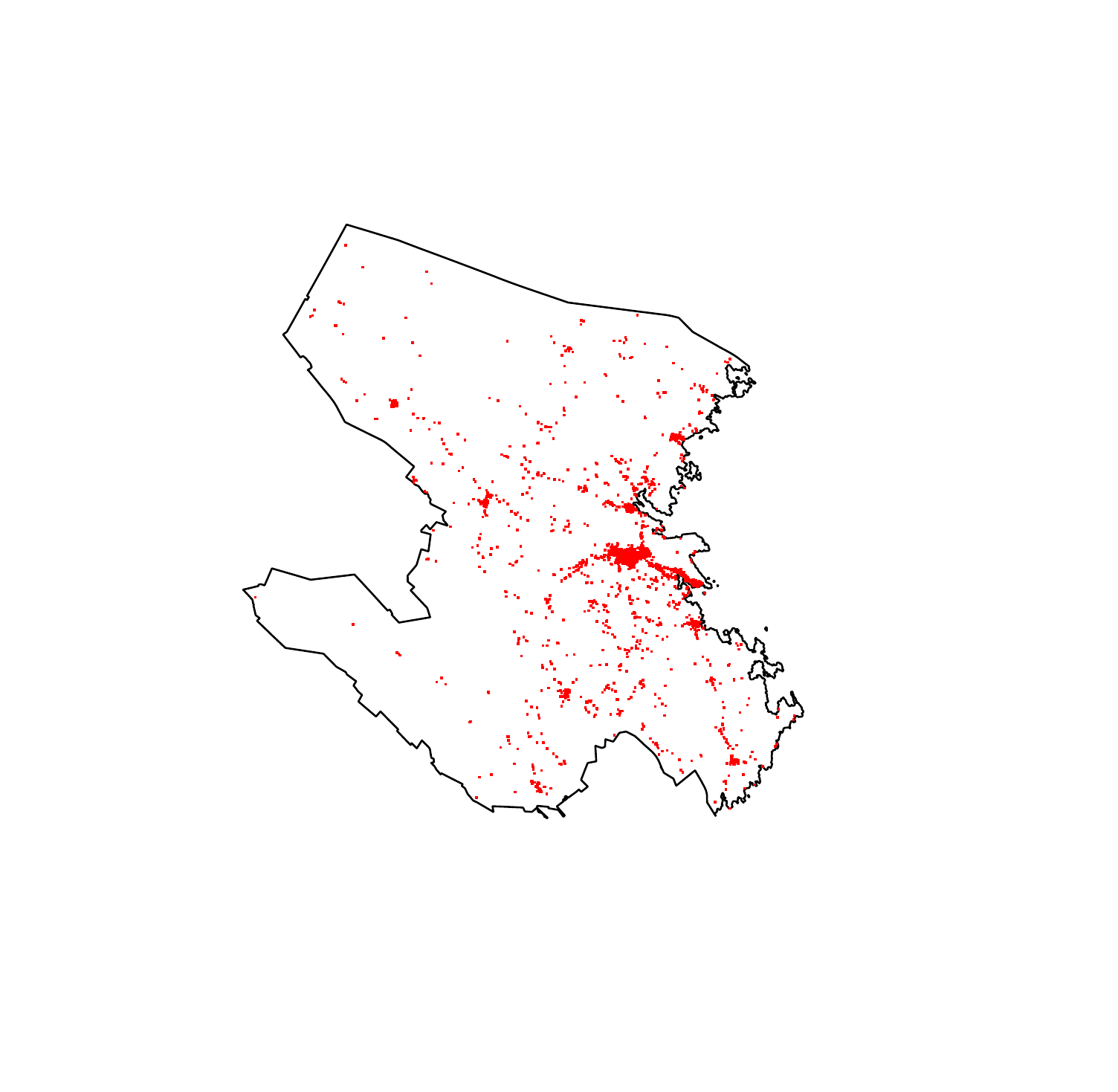}
\caption{A cross-validation based evaluation of the performance of the estimated dense and sparse models in estimating the spatial intensities of the emergency alarm call events. The estimated spatial intensities are shown in the first and second rows using the estimated dense and sparse models based on the corresponding test datasets in the third row. From left to right, the plots in the figure  present the estimated spatial intensities for the test point patterns with priority level 1, priority level 2, male, and female, as well as for the unmarked test point pattern. Note that we have scaled the intensity estimates to range between 0 and 1 so that we may compare them more easily. We also multiplied the normalized versions by 1000 for ease of visualisation.}
\label{CVDenseSparseData}
\end{figure}

The evaluation of the fitted spatial models is generally challenging, and here we use two approaches. Firstly, we evaluate the stability of the estimated models. We employ undersampling of the observed point patterns, i.e.,~70\% for the unmarked and marginal point patterns, to compare the stability of the fitted models. The estimates of the spatial intensities obtained for the unmarked and marginal point patterns using the fitted dense and sparse models are treated as the true spatial intensities of the corresponding point patterns. In this context, hereafter, we refer to the estimated spatial intensities based on the marginal and unmarked point patterns as the true spatial intensities.  We generated one hundred undersampled random realisations, i.e.,~70\% for each of the unmarked and marginal spatial point patterns, in order to produce one hundred estimated spatial intensities for each of the unmarked and marginal spatial point patterns.  Pixel-wise mean absolute errors, the 5\% and 95\% quantiles of pixel-wise absolute errors of the true intensities, and each of the one hundred estimated spatial intensities are used to assess the stabilities of the fitted dense and sparse models for each of the unmarked and marginal point patterns. Figure \ref{Densemae} shows the evaluation of the stabilities of the estimated dense models in estimating the spatial intensities of the emergency alarm call events. The overall means and standard deviations of the pixel-wise mean absolute errors for the dense models are 0.098 and 0.107, 0.099 and 0.123, 0.057 and 0.052, 0.060 and 0.052, and 0.052 and 0.051, respectively, for the point patterns with priority level 1, priority level 2, male, and female, as well as for the unmarked point pattern. Figure \ref{Sparsemae}  demonstrates the evaluation of the stabilities of the estimated sparse models in estimating the spatial intensities of the emergency alarm call events. The overall means and standard deviations of the pixel-wise mean absolute errors for the sparse models for the priority level 1, priority level 2, male, and female point patterns, as well as for the unmarked point pattern, are 0.103 and 0.128, 0.103 and 0.140, 0.056 and 0.058, 0.062 and 0.066, and 0.054 and 0.061, respectively. Based on a comparison of the plots in Figures \ref{Densemae} and \ref{Sparsemae}, the estimated dense models seem to be more stable than the sparse models. Even though the 95\% and 5\% quantiles of pixel-wise absolute errors for the estimated dense models do not entirely lie below the corresponding 95\% and 5\% quantiles of pixel-wise absolute errors for the estimated sparse models,  the maximum 95\% and 5\% quantiles can be used to compare the stabilities of the estimated dense and sparse models. An estimated model can be unstable if it has maximum 95\% and 5\% quantiles and has a broader band between the two quantiles. Despite the fact that the overall means and standard deviations of the pixel-wise mean absolute errors for the estimated dense and sparse models do not differ much as the aforementioned results suggest, we still believe that the estimated dense models are more stable than the estimated sparse models. Here, we want to emphasize that the spatial covariates that continue to exist in the sparse model and have non-zero estimated coefficients have a strong association with the emergency alarm call events, and as a result, we may draw a strong interpretation based on the estimated sparse model.
We can also visually assess how well the estimated models capture the spatial distribution of the emergency alarm call events. In order to do this, each of the datasets corresponding to the unmarked and marginal point patterns can be randomly divided into two parts. That is, we estimate the proposed model using 70\% of each dataset corresponding to the unmarked and marginal point patterns, and we use the remaining 30\% of each dataset to assess or validate the performance of the estimated models. We refer to the 30\% of each dataset that is retained for model validation in this context as a test point pattern. The estimated intensities for the test point patterns and their corresponding test point patterns can then be used to visually evaluate the performance of the estimated models. As can be seen in  Figure \ref{CVDenseSparseData}, the hotspot regions in the test point patterns are well captured by their respective estimated spatial intensities.  The plots in the figure also suggest that the estimated dense models seemingly perform better than their corresponding estimated sparse models, which are taken to be submodels of the estimated dense models. 

\section{Discussion}\label{Discn}
The purpose of this study is to model the spatial distribution of ambulance/medical emergency alarm call events in order to establish the framework for developing optimal ambulance dispatching strategies/algorithms. In order to develop the optimal dispatching strategies, it is essential to consider the response times and operational costs of prehospital resources such as ambulances. The optimal design can assist the concerned body in providing emergency medical assistance to life-threatening emergencies as quickly as possible. The locations of hotspot areas within the study region are crucial components of such designs. This means that identifying study area subregions with a high risk of events plays a crucial role in developing the optimal dispatching strategies. This work focuses on discerning such hotspot areas as well as selecting/exploring spatial covariates that are associated with the spatial distribution of the call events.  \\

The data at hand consists of the locations of the events, which are the medical emergency alarm calls from patients in Skellefte{\aa}, Sweden, during the years 2014 -- 2018. The events have marks associated with them in addition to their spatial locations. These marks take the form of priority levels (1 or 2, where 1 denotes the highest priority)  and the genders (male, female, or  missing) of the patients linked to the calls. We have 14,919 events in total, with 7,204 and 7,715 locations having priority level 1 and priority level 2 marks, respectively. Of the events, 5,236 and 5,238 have male and female marks, respectively. In this case, the marks of some events are missing. We want to emphasise  that we have chosen to treat the spatial locations associated with males, females, priority level 1s, and priority level 2s as individual point patterns, and the main reason for this is that we are interested in identifying spatial covariates associated with the intensity function of a spatial point process corresponding to each mark. This approach can also help to pinpoint subregions of the study area that have a high risk of events of a given mark. The other reason is that we have a large number of events with missing sex or gender labels. It should be emphasised that we could have, e.g.,~treated the missing entries as an independent thinning (a $p$-thinning) of the complete data set and thus scaled the intensity accordingly.\\

An exploratory analysis of the data shows that the observations tend to be along road networks, making the statistical modelling difficult (see Figure \ref{Or213}). Although a non-parametric estimate such as an (adaptive) kernel intensity estimate \citep{cronie2018non,davies2018tutorial,diggle1985kernel, baddeley2015spatial} or a (resample-smoothed) Voronoi intensity estimate \citep{ogata2003modelling, ogata2004space,moradi2019resample} does shed some light on the hotspots, we have observed that it does not do so to the fullest extent. More specifically, non-adaptive kernel estimators seem to over-smooth the data, whereas the other estimators, which are all adaptive estimators, tend to under-smooth the data (cf.~\citet{moradi2019resample}).  Kernel intensity estimation, where a bandwidth was selected using a machine learning technique, i.e., K-means clustering, attempts to balance the over-and under-smoothing roles of the aforementioned intensity estimation methods, see \cite{bayisa2020large}, which selected the number of clusters through visual evaluation. In this work, a new heuristic algorithm is developed to obtain an optimal number of clusters and, thus, optimal bandwidth. The spatial distribution of the events or hotspot regions is well captured using the kernel intensity estimation based on the new bandwidth selection algorithm, as shown in \ref{BandWidth}. In addition, and more importantly, since we have access to various spatial covariates, we have shed some light on the spatial covariates in Subsection \ref{SpatialCovariates}  that may influence the spatial variation of the intensity of the events. Here, it is important to emphasise that the spatial covariates we employ to model the spatial intensity function of the events are either based on demographics or the structure of the road networks.\\

Our modeling strategy/approach is to treat the data as a realization of a spatial point process and model its intensity function to quantify the spatially varying call risk. We assume that the intensity function is a log-linear function of the various spatial covariates under consideration, and we fit the intensity function by means of  a Poisson process log-likelihood function. To carry out variable selection and adjust for over-/under-fitting, a regularisation term is added to the (approximated) log-likelihood function. Following \citet{yue2015variable}, we have chosen to employ  an elastic-net penalty (a convex combination of the ridge and lasso penalties), which is useful when the number of covariates considered in the modeling exceeds the total point count or when the model contains several correlated spatial covariates. The elastic-net penalty is governed by a parameter $\alpha$, which controls how much weight we put on either the lasso or the ridge penalty, and by setting $\alpha = 1-\epsilon$ for some small $\epsilon > 0$, it performs much like the lasso, but avoids any unstable behaviour caused by extreme correlations.  We used $\alpha = 0.95$, i.e.,~a lasso-like elastic-net, and ten-fold cross-validation to select an optimal estimate of the tuning parameter $\lambda.$\\
 
 We considered two scenarios for the spatial covariates in this study. The first scenario is that the individual spatial covariates (we have also called them the original spatial covariates) and their products/pairwise interaction terms have been used in the modelling of the call events. In this case, we have  nine hundred eighty-nine spatial covariates, which have been used in a regularised modelling of medical emergency alarm call events. The regularisation can select the original and pairwise interaction variables, making the results difficult to interpret. The second scenario involves only using the original spatial covariates in the modelling of the emergency alarm call events. In this work, we have also considered the cases of marked/marginal and unmarked spatial point patterns. Furthermore, we have investigated the dense model (we may call it the full model) and the sparse model (we may call it the submodel) for each of the spatial point patterns.  A cyclical coordinate descent algorithm has been used for fast estimation of the parameters of the model. The intensity function of the events at the observed locations has been estimated using the estimated model parameters. To obtain the estimated intensities at the desired spatial locations in the study area, we have smoothed the estimated intensities at the observed locations, which are irregular spatial locations in the study area, using a Gaussian kernel.\\

Using the original spatial covariates and the first-order interaction terms, the estimated model parameters/coefficients for spatial point pattern with priority level 1 under lasso and lasso-like elastic-net regularisations have been shown in Figure \ref{Orgr2LassoElastic}.  According to the figure, lasso and lasso-like elastic-net have chosen approximately 185 (18.69\%) and 207 (20.91\%) of the spatial covariates as important covariates in determining the spatial distribution of emergency alarm call events, respectively. Following the suggestion by \cite{yue2015variable}, we have used elastic-net regularisation, in particular, lasso-like elastic-net with $\alpha = 0.95$.  Figure \ref{Orgr2} shows a ten-fold cross-validation method for selecting an optimal estimate of the tuning parameter for the spatial point pattern with priority level 1. The tuning parameter values corresponding to the blue and red vertical lines in the figure show an optimal and one standard error away from the optimal tuning parameter estimate. We may call the estimated model corresponding to the optimal estimate of the tuning parameter a dense model (or full model), while the estimated model corresponding to one standard error away from the optimal estimate of the tuning parameter may be referred to as a sparse model (or submodel). The optimal tuning parameter that was chosen has been used to obtain optimal parameter estimates, which have been used to generate the spatial intensities of the emergency alarm call events. The lasso-like elastic-net identified approximately 20.91\% and 36.06\% of the spatial covariates that are associated with the emergency alarm call events with priority levels 1 and 2, respectively. Furthermore, it has identified approximately 17.68\% and 35.76\% of the spatial covariates for the male and female marks of the alarm call events, respectively. The estimated intensity models based on the original spatial covariates and first-order interaction terms are depicted in Figure \ref{Orgr3}. The spatial variations of emergency alarm call events are not well captured using the estimated models based on the original spatial covariates and first-order interaction terms, as shown in the figure, particularly for emergency alarm call events with priority level 2. This finding prompted us to look into estimating the spatial intensities of emergency alarm call events using only the original covariates. \\

Using only the original spatial covariates, the lasso-like elastic-net provides dense solutions at the optimal tuning parameter estimates. That is, the majority of the spatial covariate coefficients are non-zero, see Table \ref{tab:EstimatedModels1}. Table  \ref{tab:EstimatedModels2} depicts the sparse models that correspond to the estimated dense models. We would like to point out here that the spatial covariates that remain in the sparse models have a strong association with emergency alarm call events. Those spatial covariates that exist in the estimated dense models but not in the estimated sparse models may have weak associations with the events in comparison. To keep things simple, we will interpret the results based on the estimated sparse models identified  by one-standard-error rule in \cite{hastie2017elements}. Accordingly, the one-standard-error rule identified event locations, population age categories (such as 0--5, 6--9, 10--15, 60--64, 65--69), and spatial covariates related to bus stops, main road networks, complete road networks, and densely populated areas as having strong associations with the spatial intensities of emergency alarm call events with priority level 1. With regard to a spatial point pattern with priority level 2, population age categories (such as 0--5, 10--15, 35--39, 55--59, 60--64), as well as spatial covariates related to bus stops, main road networks, complete road networks, and densely populated areas, have played an important role in determining the spatial variation of emergency alarm call events. Population age categories (such as 0--5, 6--9, 10--15, 60--64), event locations, and spatial covariates related to bus stops, main road networks, complete road networks, and densely populated areas are strongly associated with the spatial variation of emergency alarm call events with mark male. The one-standard-error rule has identified spatial covariates such as population age categories (such as 0--5, 10--5, 35--39), event locations, and spatial covariates related to bus stops, main road networks, complete road networks, and densely populated areas as key features in determining the spatial distribution of emergency alarm call events with mark female. In the case of an unmarked spatial point pattern, spatial covariates such as population age categories (such as 0--5, 10--15, 35--39), event locations, and spatial covariates related to bus stops, main road networks, complete road networks, and densely populated areas are strongly related to the spatial variation of emergency alarm call events. In the presence of the remaining spatial covariates, the benchmark intensity is not strongly associated with the emergency alarm call events for each of the marginal spatial point patterns, which may be due to the fact that the spatial locations of the events/event locations are more informative about human mobility than the benchmark intensity. Table \ref{tab:EstimatedModels2} also shows that, in the presence of other spatial covariates, population density is not strongly associated with the spatial variation of events. It could be due to the presence of a densely populated area in the model, which may provide more information about the distribution of the events than population density. Figure  \ref{OrgrgGd3Compare}  compares the estimated spatial intensity models based on the original spatial covariates and the first-order interaction terms to the estimated spatial intensity models based only on the original spatial covariates. The figure clearly shows that the estimated spatial intensity models based solely on the original covariates performed well in capturing the spatial distribution of the events.\\

We evaluated the performance of the estimated dense and sparse models for the unmarked and each of the marginal spatial point patterns in two ways. Firstly, we assessed the performance of the estimated dense and sparse models on undersampled data. To accomplish this, spatial intensity estimates of the unmarked and each of the marginal emergency alarm call events are generated and treated as true spatial intensities of the emergency alarm call events. Each dataset corresponding to the unmarked point pattern and the marginal spatial point patterns is undersampled. That is, from each spatial point pattern, we created one hundred undersampled datasets, accounting for 70\% of each dataset. Using the undersampled datasets, we estimated one hundred dense and sparse models and thus one hundred spatial intensities. The pixel-wise mean absolute errors of the estimated spatial intensity that was treated as the true spatial intensity of the emergency alarm call events and each of the hundred estimated spatial intensities are used to evaluate the stabilities of the estimated dense and sparse models. Furthermore, we assessed the performance of the estimated dense and sparse models using the 5\% and 95 \% quantiles of the  pixel-wise absolute errors of the estimated spatial intensity that was treated as the true spatial intensity and each of the hundred estimated spatial intensities. Figures \ref{Densemae} and \ref{Sparsemae} show the evaluations of the stabilities of the estimated dense and sparse models in estimating the spatial intensities of emergency alarm call events. Secondly, we evaluated the performance of the estimated dense and sparse models for the unmarked point pattern and each of the marginal spatial point patterns, using 70\% of the total data for model fitting and 30\% of the total data for validating or testing the model. The performance of the estimated dense and sparse models in estimating the spatial intensities of emergency alarm call events can be evaluated visually by comparing the estimated spatial intensities and their corresponding patterns of spatial locations in the validation/test spatial point patterns, as shown in Figure \ref{CVDenseSparseData}. The plots in the figure show that the hotspot regions in the estimated spatial intensities of the events and the patterns of the spatial locations in the test spatial point patterns appear to coincide. That is, both the estimated dense and sparse models capture the spatial variations of emergency alarm call events well. Even though the estimated dense models appear to outperform the sparse models, the results show that both models are feasible.\\

The primary challenge in this work is the nature of the spatial data; that is, the locations of the events tend to be on road networks. Traditional kernel smoothing methods are incapable of displaying the spatial variation of emergency alarm call events. To address this issue, semi-parametric modelling of the spatial intensity function of the events has been developed, and it appears that the method is feasible for obtaining the estimated spatial intensity of the events, which will be used to design optimal ambulance dispatching strategies to provide life-threatening emergency medical services as quickly as possible.
Even though the semi-parametric modelling of the spatial intensity function of the events captures the spatial distribution of the emergency alarm call events well, the effect of the nature of the spatial data persists. To address this issue, we will continue to work on modelling the spatial intensity function of road network events. The proposed statistical method is tested with data gathered from Skellefte{\aa}, a municipality in northern Sweden. The spatial intensity function of emergency alarm call events will be modelled for each municipality in northern four counties (Norrbotten, V{\aa}sterbotten,  V{\aa}sternorrland, and J{\aa}mtland) of Sweden using the proposed statistical model.\\

Finally, the study developed a new heuristic algorithm for bandwidth selection and discovered that spatial covariates such as population age categories, spatial location of events, and spatial covariates related to bus stops, main road networks, complete road networks, and densely populated areas play an important role in determining the spatial distribution of emergency alarm call events. The study also demonstrated that semi-parametric modelling of the spatial intensity function of the inhomogeneous Poisson process can handle the spatial variation of emergency alarm call events, and the estimated spatial intensity function of the events can be used as an input in designing optimal ambulance dispatching strategies to provide better emergency medical services for life-threatening health conditions.
\section*{Acknowledgments}
This work was supported by Vinnova [Grant No. 2018-00422] and the regions: V{\aa}sterbotten, Norrbotten, V{\aa}sternorrland and J{\aa}mtland-H{\aa}rjedalen. We are also grateful to our project partners SOS Alarm and the regions: V{\aa}sterbotten Norrbotten,  V{\aa}sternorrland and J{\aa}mtland-H{\aa}rjedalen for fruitful discussions on the Swedish prehospital care.
\section*{Disclosure of Conflicts of Interest}
The authors have no relevant conflicts of interest to disclose.
\bibliographystyle{apalike}
\bibliography{SpatialRef2022}

\end{document}